\documentclass{aa} 
\usepackage{natbib} 
\usepackage{graphicx} 
\usepackage{amsmath}

\def\N{\mathrm{N}}
\def\K{\mathrm{K}}
\def\Lx{\mathrm{L}_{\mathrm{X}}}

\def\T{\mathrm{T}} 
\def\kt{\mathrm{kT}} 
\def\ademi{\frac{a}{2}}
\def\M{\mathrm{M}}

\def\W{\mathrm{W}}
\def\P{\mathrm{P}}

\def\t{\mathrm{T}} \def\t1{\mathrm{T1}}
\def\t2{\mathrm{T2}}          
\def\nh{\mathrm{Nh}}

\def\tx{\mathrm{T}_\mathrm{x}}
\def\sx{\mathrm{\Sigma}_\mathrm{x}}  

\def\ab{\mathrm{Z}}     
    
\def\kev{\mathrm{keV}}

\def\ao{a_\mathrm{0}}

\def\jmax{\mathrm{j}_{\mathrm{max}}} 
\def\z{\mathrm{z}}
\def\l{\mathrm{L}} 
\def\t{\mathrm{T}} 
\def\f{\mathrm{F_{\mathrm{evt}}}}
\def\s{\mathrm{F_{\mathrm{ICM}}}} 
\def\c{\mathrm{F_{\mathrm{CXB}}}}
\def\b{\mathrm{F_{\mathrm{bck}}}} 
\def\o{\mathrm{F_{\mathrm{oot}}}} 
\def\p{\mathrm{F_{\mathrm{p}}}}
 
\def\j{\mathrm{j}} 
\def\w{\mathrm{W}}
\def\nf{\mathrm{n}_\mathrm{evt}} 
\def\ns{\mathrm{n}_\mathrm{ICM}}
\def\nc{\mathrm{n}_\mathrm{CXB}} 
\def\nb{\mathrm{n}_\mathrm{bck}}
\def\np{\mathrm{n}_\mathrm{p}} 
\def\no{\mathrm{n}_\mathrm{oot}}
\def\ea{\mathrm{E}}
\def\vf{\mathrm{a}}

\def\nhunit{10^{24}\mathrm{m}^{-2}}
\def\xmm{XMM-Newton}
\def\fig{Fig.~}
\def\tab{Table~}
\def\equ{Eq.~}
\def\equs{Eqs.~}
\def\part{Sect.~}
\def\parts{Sects.~}

\begin{document}
  
  \title{Temperature structure of the intergalactic medium within seven
    nearby and bright clusters of galaxies observed with XMM-Newton.}
  \titlerunning{ICM temperature structure within seven nearby and
    bright clusters of galaxies}

  \author{H. Bourdin \inst{1} \and P. Mazzotta \inst{1,2}}
  \offprints{H. Bourdin}
  
  \institute{Dipartimento di Fisica, Universit\`a degli Studi di Roma "Tor
    Vergata", \\ via della Ricerca Scientifica, 1, I-00133 Roma, Italy \\
    \email{herve.bourdin@roma2.infn.it, pasquale.mazzotta@roma2.infn.it}
    \and Harvard-Smithsonian Center for Astrophysics, 60 Garden
    Street, Cambridge, MA 02138, USA}

  \date{Received 5 June 2006 / Accepted 31 October 2007}
  
  \abstract{} {Using a newly developed algorithm, we map, to the
  highest angular resolution allowed by the data, the temperature
  structure of the intra-cluster medium (ICM) within a nearly complete
  X-ray flux limited sample of galaxy clusters in the redshift range
  between $\z=0.045$ and $\z=0.096$. Our sample contains seven bright
  clusters of galaxies observed with \xmm: Abell~ 399, Abell~ 401,
  Abell~ 478, Abell~ 1795, Abell~ 2029, Abell~ 2065, Abell~ 2256.}
  {We use a multi-scale spectral mapping algorithm especially designed
  to map spectroscopic observables from X-ray extended emission of the
  ICM. By means of a wavelet analysis, this algorithm couples
  spatially resolved spectroscopy with a structure detection approach.
  Derived from a former algorithm using Haar wavelets, our algorithm
  is now implemented with B-spline wavelets in order to perform a more
  regular analysis of the signal. Compared to other adaptive
  algorithms, our method has the advantage of analysing spatially the
  gas temperature structure itself, instead of being primarily driven
  by the geometry of gas brightness.}  {For the four clusters in our
  sample that are major mergers, we find a rather complex thermal
  structure with strong thermal variations consistent with their
  dynamics. For two of them, A2065 and A2256, we perform a 3-d
  analysis of cold front-like features evidenced from the gas
  temperature and brightness maps. Furthermore, we detect a
  significant non-radial thermal structure outside the cool core
  region of the other 3 more ``regular'' clusters, with relative
  amplitudes of about about 10 $\%$ and typical sizes ranging between
  2 and 3 arcmin. We investigate possible implications of this thermal
  structure on the mass estimates, by extracting the surface
  brightness and temperature profiles from complementary sectors in
  the ``regular'' clusters A1795 and A2029, corresponding to hottest
  and coldest regions in the maps.  For A2029, the temperature and
  surface brightness gradients seem to compensate each other, leading
  to a consistent mass profile. For A1795, however, the temperature
  structure leads to a significant mass discrepancy in the innermost
  cluster region. The third ``regular'' cluster, A478, is located in a
  particular sky region characterised by strong variations of neutral
  hydrogen column density, Nh, even on angular scales smaller than the
  cluster itself. For this cluster, we derive a spectroscopic Nh map
  and investigate the origin of Nh structure by discussing its
  correlation with galactic emission of dust in the infrared.}  {}

   
   \maketitle

  \section{Introduction}

  Clusters of galaxies are thought to form by accretion of less massive
  groups and clusters under the influence of gravity. Following this
  scheme, they successively overcome some transient merging processes
  that alter their energy content and some relaxation phases, which lead
  to the almost virialised systems we observe.

  The thermodynamical state of X-ray emitting intra-cluster medium
  (ICM) depends on both the cluster merger history and some not yet
  perfectly understood physical processes that drives its
  thermalisation, such as heat conduction or viscosity. Revealed by
  X-ray observations, the brightness and temperature structure of the
  ICM testifies to this final state. The brightness structure of the
  ICM has been extensively studied in the past
  \citep[e.g.][]{Forman_jones_82, Slezak_94, Buote_tsai_96,
  Schuecker_01}. The current generation X-ray telescopes (\xmm,
  Chandra) now further allows us to accurately map and investigate the
  temperature structure of the ICM \citep[see e.g.][]{Bauer_05},
  provided that specific algorithms help at optimising a necessary
  compromise between the precision of local temperature measurements
  and the spatial resolution of temperature maps.

  For this reason, a number of spectral-mapping algorithms have been
  recently developed in X-ray astronomy. All of them can be
  essentially subdivided in three categories, depending on the general
  approach used. A first approach combines X-ray imaging in several
  energy bands, and colour analyses obtained from recombination of
  multi-band information. The imaging part of these algorithms usually
  requires de-noising by adaptive smoothing
  \citep[e.g.][]{Markevitch_00}. A second approach proposes performing
  spatially resolved spectroscopy with required signal-to-noise. To do
  so, the field of view is firstly sampled in independent spatial bins
  following a mesh refinement scheme, then spectroscopic estimations
  are performed within each bin. Several binning strategies have been
  proposed in this context: adaptive binning
  \citep[e.g.][]{Sanders_01}, ``contour binning'' \citep{Sanders_06},
  or Voronoi tessellation \citep[e.g.][]{Cappellari_03}. Proposed by
  \citet{Bourdin_04}, a third approach consists in using Haar wavelet
  coefficients in order to couple a multi-scale spectroscopic analysis
  with a structure detection scheme.

  Among these algorithms, only the last one performs a direct
  investigation of structure for the searched parameter itself
  --i.e. the ICM projected temperature-- while the signal analysis is
  essentially driven by the geometry of gas brightness in all the
  other algorithms. Here, we present a new version of the algorithm of
  \citet{Bourdin_04}, now implemented with B-spline wavelets and
  possibly enabling us to apply a more conservative thresholding
  strategy. Other improvements are related to adaptations to more
  recent calibration of \xmm{} instruments, which particularly allow us
  to combine data coming from the three European Photon Imaging
  Cameras (EPIC).

  This algorithm has been used to perform a systematic study of a
  nearly complete X-ray flux selected sample of seven clusters
  observed with \xmm. Along this paper, we first discuss the sample
  selection and data preparation in \parts \ref{sample_selection} and
  \ref{data_preparation}, then expose our data analysis scheme in
  \part \ref{data_analysis}, and finally give a brief description of
  the thermal structure observed in each single cluster in \part
  \ref{thermal_structure}. Before providing general discussions and
  conclusions in \part \ref{conclusion}, we discuss more specific
  issues related to thermal features observed within individual
  objects: isoradial thermal structure in the relaxed clusters
  Abell~1795 and Abell~2029 (\part \ref{tprofs}), cold fronts in the
  merging clusters Abell~2065 and Abell~2256 (\part \ref{cfront}). The
  additional \part \ref{nh_part} is related to the high neutral
  hydrogen column density variations observed across the field of view
  of Abell~478.

  The cluster radii are computed as angular diameter distances,
  assuming a $\lambda$-CDM cosmology with $\mathrm{H}_0=70 ~
  \mathrm{km}.\mathrm{s}^{-1}.\mathrm{Mpc}^{-1}$, $\Omega_0=0.3$,
  $\Omega_\Lambda=0.7$.
  

\section{The cluster sample\label{sample_selection}}

Starting from the X-ray Brightest Cluster Sample (BCS) of
\citet{Ebeling_98}, we selected a flux limited sample of
clusters. In order for the clusters to have an observed angular size
close to the EPIC cameras field of view, we limited the cluster
redshift selection to the range $\delta \z = [0.045, 0.096]$,
corresponding to an angular size from 5 to 10 arcmin.  

Our selection criteria returned eight bright clusters, namely Abell~
2142, Abell~ 2029, Abell~ 478, Abell~ 1795, Abell~ 401, Abell~ 2256,
Abell~ 399, Abell~ 2065, public \xmm{} data being available for each
of them.  However, since the Abell~2142 observation is strongly
altered by solar flares, we decided to remove it from our sample and
focus on the seven remaining clusters.

Previous investigations, based primarily on the analysis of the X-ray
surface brightness, classified 4 of the objects in our sample as major
mergers and the remaining 3 as more ``relaxed'' clusters. The clusters
are listed in \tab\ref{cluster_sample_tab} and are grouped according
to this classification. The cluster redshifts and the ICM brightness
from the BCS survey are reported in the same table, with associated
P4/P0 power ratio calculated by \citet{Buote_tsai_96} when
available. Expected to be an indicator of the dynamical evolution
stage of the cluster, this latter quantity is lower for the 3 most
relaxed clusters than for the clusters classified as major mergers.

\begin{table*}

\caption{Cluster sample with associated redshift, coordinates, total
ICM flux from the BCS survey \citep{Ebeling_98}, and P4/P0 power ratio
reported in \citet{Buote_tsai_96}.  Last column: average neutral
hydrogen column density measured along line of sight of each cluster
\citep{Dickey_lockman_90}.\label{cluster_sample_tab}}

\begin{center}
\begin{tabular}{cccccc}

\hline\hline

Cluster & Redshift & Equatorial coordinates & ICM flux & Power ratio & Neutral hydrogen column \\
name && (J2000)  & ($10^{-15}$ W.m$^{-2}$ / $ 10^{-12}$ erg.s$^{-1}$.cm$^{-2}$) & P4/P0 & density ($\nhunit$) \\

\hline
\multicolumn{6}{c}{Relaxed clusters}\\
\hline

A478  & 0.088 & 04 13 25.0 +10 27 54.0 & 39.9 & 0.025 & 15.1 \\
A1795 & 0.062 & 13 48 53.0 +26 35 32.0 & 68.1 & 0.004 & 1.2  \\
A2029 & 0.077 & 15 10 56.0 +05 44 42.0 & 61.3 & 0.050 & 3.14 \\

\hline
\multicolumn{6}{c}{Major merger clusters}\\
\hline

A399 & 0.072 & 02 57 53.0 +13 02 00.0  & 28.8 & - & 10.9 \\
A401 & 0.074 & 02 58 58.0 +13 34 00.0  & 42.8 & 0.157 & 10.5 \\
A2065 & 0.072 & 15 22 42.0 +27 43 00.0 & 22.2 & - & 2.95 \\
A2256 & 0.058 & 17 03 58.3 +78 38 31.0 & 49.5 & 0.395 & 4.1 \\
\hline

\end{tabular}
\end{center}

\end{table*}

\section{Observations and data preparation\label{data_preparation}}

\subsection{Observations}

All data used for this investigation come from the EPIC \xmm{}
database. We use individual observations of Abell~399, Abell~401,
Abell~478, Abell~1795, Abell~2029, and Abell~2065, and a multiple
pointing observation of Abell~2256. For all of these clusters, data
sets combine observations obtained from the three EPIC instruments:
MOS1, MOS2 and PN. A summary of data sets is provided in \tab
\ref{pointing_params_tab}, with associated exposure times.

\subsection{Data filtering\label{data_filtering}}

In addition to the extended and optically thin emission of ICM, X-ray
observations gather photon impact events related to other sources,
such as spatially resolved X-ray emitting galaxies and the cosmic
X-ray background (CXB). Moreover, observations may be transiently
contaminated by solar flares. While the extended contributions of ICM
and CXB are hardly separable, contributions from point-sources and
solar flares can be isolated and removed from the observed signal
through spatial and temporal wavelet analyses, respectively.

In order to detect high solar flares periods and remove corresponding
data sets, we analyse light curves with associated high energy events
(10-12 keV) and softer events (1-5 keV), respectively. As proposed by
\citet{Nevalainen_05}, this two-step analysis first enables us to
isolate the most prominent flares at high energy, where ICM brightness
is expected to be negligible, then detect some additional contribution
of soft flares only. For each of these curves, we use a B3-spline
``\`a-trous'' wavelet algorithm in order to detect disruptions, and
select the positive irregularities with amplitude overcoming a
$2\sigma$ significance threshold with regard to the light-curve
fluctuation. This ``cleaning'' process has significantly lowered the
effective exposure time of observations, as reported for individual
observations and pointings on \tab\ref{pointing_params_tab}.

In order to identify point-sources, we analyse EPIC-MOS event-lists
which are more suited to imaging, and adopt an object separation
algorithm derived from the multi-scale vision model of
\citet{Bijaoui_95}. After detecting and imaging point-sources, we
associate a mask to regions of the field of view dominated by
point-sources contribution, and isolate events coming from these
regions when analysing the signal.

\subsection{Data sampling -- effective exposure\label{eff_exposure}}

XMM-Newton imaging cameras provide photon impact event lists with
associated energy and position on the detector planes. In order to
analyse our signal using discrete wavelet algorithms, we group events
coming from various observations, if any, and sample them spatially
into sky coordinate grids, with given angular resolution, $\ao$. We also 
sample events in energy so as to perform spectral
analyses, which leads to 3-D event cubes associated with each EPIC
instrument, and sampled in position (k,l) and energy (e).

For imaging and spectral-mapping purposes, we associate to these event
cubes a set of local ``effective exposure'' $\ea(k,l,e)$ and
``background'' cubes, $\nb~\b(k,l,e)$, with similar position-energy
sampling, and use them for modelling the bolometric or spectroscopic
ICM radiation; we report details about the background modelling in
appendix \ref{b_model_app}. Let us define here the ``effective exposure''
$\ea(k,l,e)$ at pixel [k,l] as the linear combination of CCD exposure
times $t_{\mathrm{CCD}}(k,l,p)$ related to individual observations
$p$, with correction for spatial variations of the mirror effective
area or so-called ``vignetting factor'', $\Delta
a_{\mathrm{mirror}}(k,l,e)$, transmission by other focal instrument, 
--i.e. reflection grating spectrometer (RGS)-- $tr_{\mathrm{RGS}}$,
and detector pixel area with correction for gaps and bad pixels,
$a_{\mathrm{CCD}}(k,l)$\footnote{Information about these instrumental
effects are provided in the following \xmm{} Current Calibration Files
(CCF), corresponding to each observation epoch: RGS*\_QUANTUMEF,
XRT*\_XAREAEF, EMOS*\_LINCOORD, EPN\_LINCOORD, EMOS*\_BADPIX,
EPN\_BADPIX}. For $\K$ observations, we get:

\begin{eqnarray}
\ea(k,l,e) = \sum_{p=1}^\K t_{\mathrm{CCD}}(k,l,p) && \times ~\Delta a_{\mathrm{mirror}}(k,l,e,p) \nonumber \\
&& \times ~ tr_{\mathrm{RGS}}(k,l,e,p) \nonumber \\
&& \times ~ a_{\mathrm{CCD}}(k,l,p)  
\label{eff_exposure_equ}
\end{eqnarray}

%
%




\begin{table*}[ht]
\caption{Effective exposure time of each EPIC \xmm{} observation.  In
brackets: fraction of the useful exposure time after solar-flare
``cleaning''. \label{pointing_params_tab}}
\begin{center}
\begin{tabular}{cccc}
\hline\hline
Cluster  & MOS1  effective  &  MOS2  effective &  PN  effective \\
name  & exposure time (ks) & exposure time (ks) & exposure time (ks)  \\
\hline
A399& 8.3 (58.6 \%)& 4.6 (32.8 \%)& 4.2 (45.2 \%) \\
A401&12.0 (92.5 \%)&10.0 (77.4 \%)& 2.8 (35.1 \%) \\
A478&30.3 (44.2 \%)&38.0 (30.5 \%)&47.0 (43.6 \%) \\
A1795&22.5 (45.8 \%)&23.5 (47.8 \%)&19.3 (45.5 \%) \\
A2029& 6.5 (37.5 \%)& 7.7 (44.4 \%)& 6.4 (50.6 \%) \\
A2065&20.4 (61.0 \%)&19.9 (60.3 \%)&12.2 (55.8 \%) \\
\hline
A2256 (1)& 8.7 (62.1 \%)& 8.5 (60.3 \%)& 4.8 (54.8 \%) \\
A2256 (2)& 7.9 (48.5 \%)& 8.6 (53.0 \%)& 5.9 (50.9 \%) \\
A2256 (3)& 8.6 (47.3 \%)& 8.9 (48.6 \%)& 5.3 (26.3 \%) \\
A2256 (4)& 7.8 (47.8 \%)& 7.1 (43.3 \%)& 6.5 (58.0 \%) \\
\hline
\end{tabular}
\end{center}
\end{table*}

\begin{figure*}[ht]
\begin{center}
\begin{tabular}{lll}
\resizebox{.25\hsize}{!}{\includegraphics{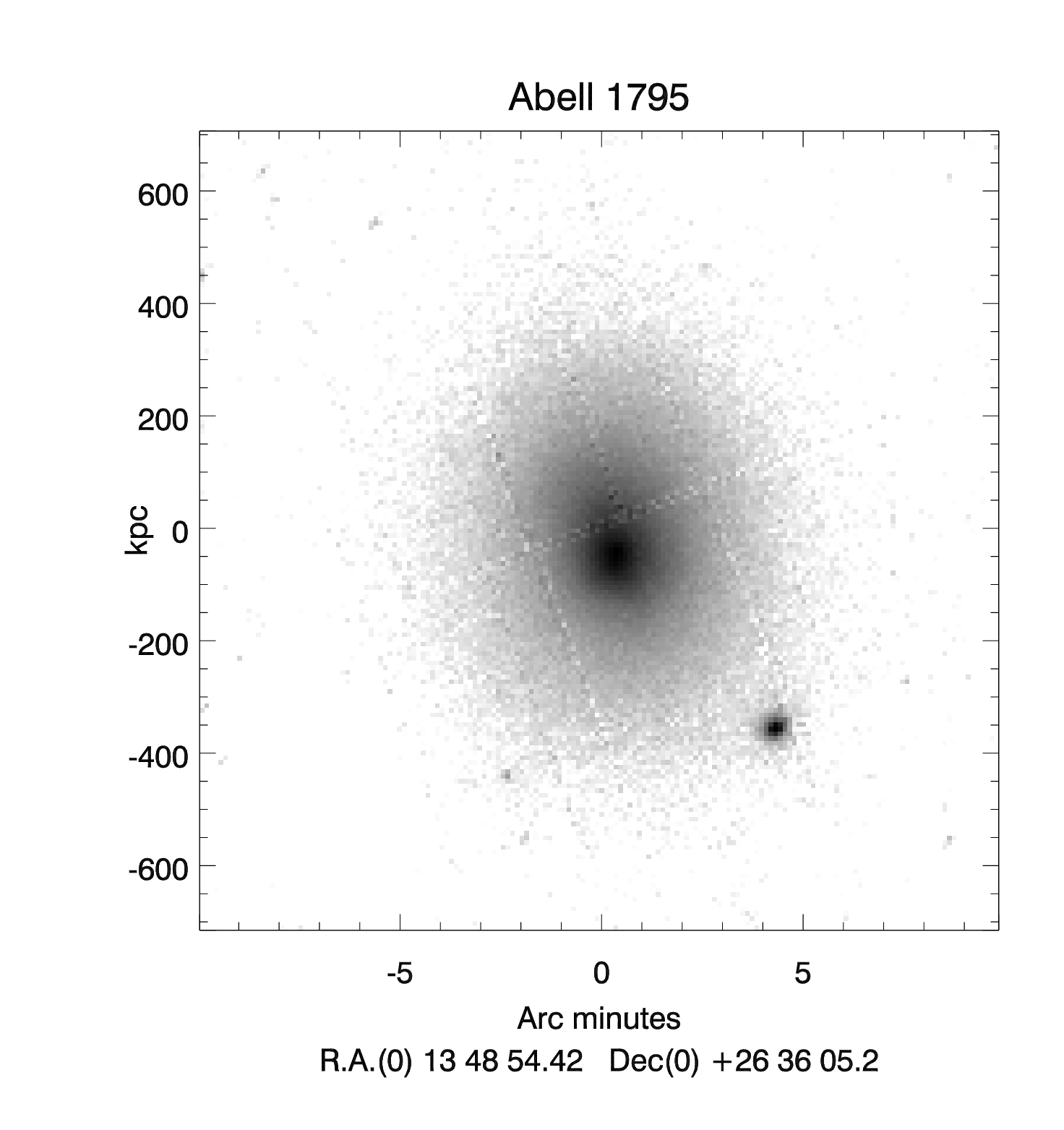}}
&
\resizebox{.25\hsize}{!}{\includegraphics{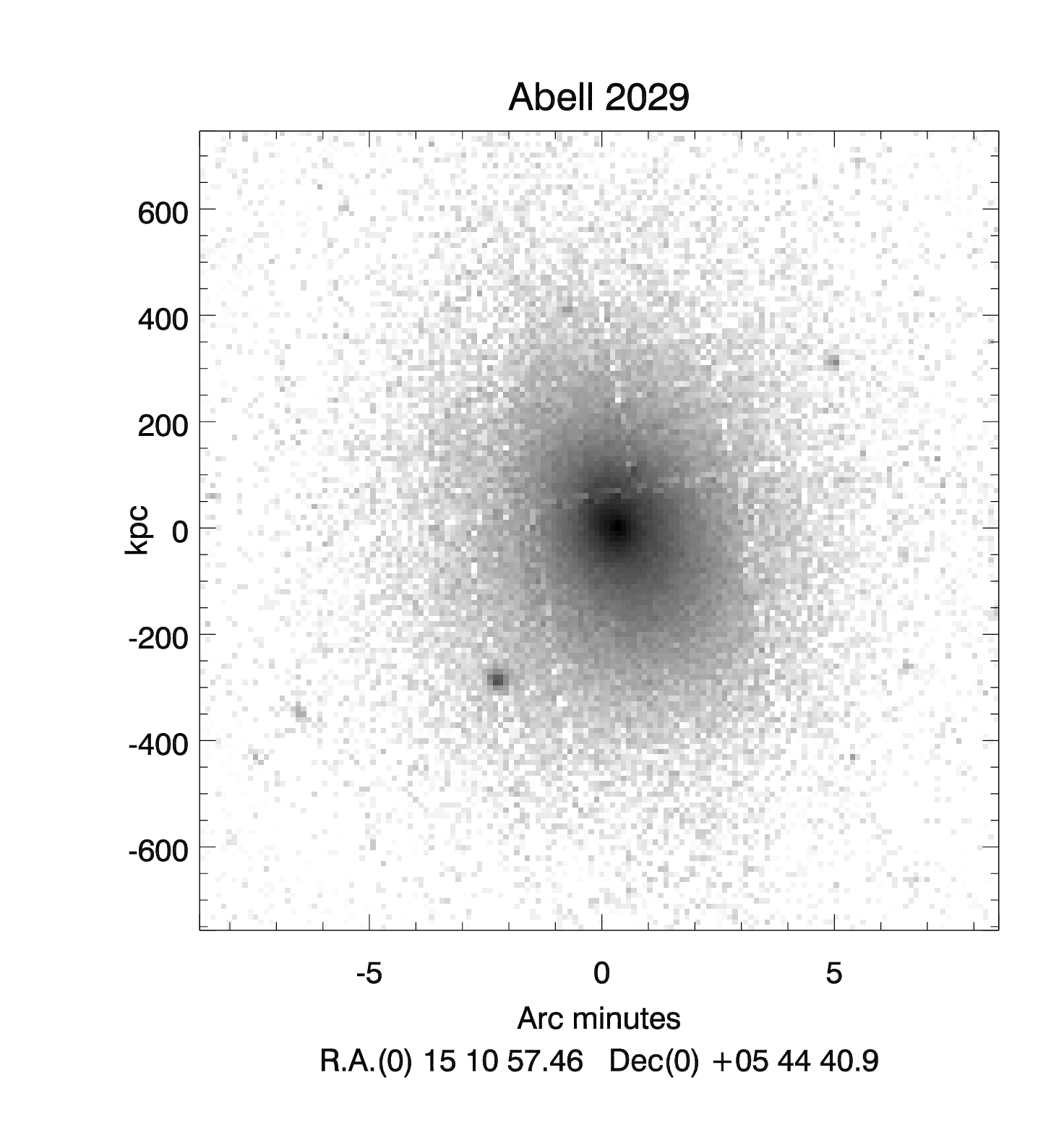}}
&
\resizebox{.25\hsize}{!}{\includegraphics{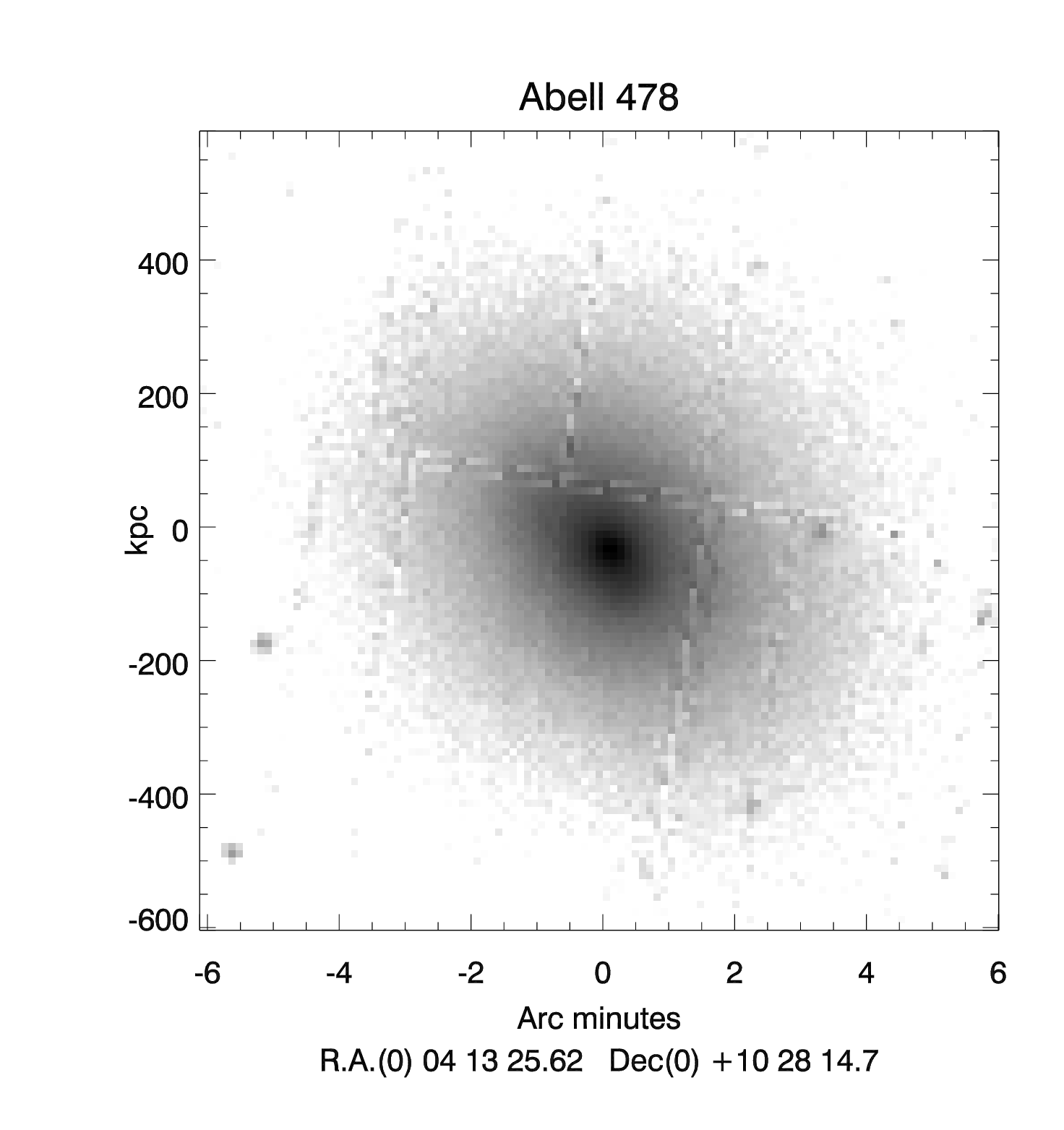}}
\\
\multicolumn{3}{c}{\resizebox{.25\hsize}{!}{\includegraphics{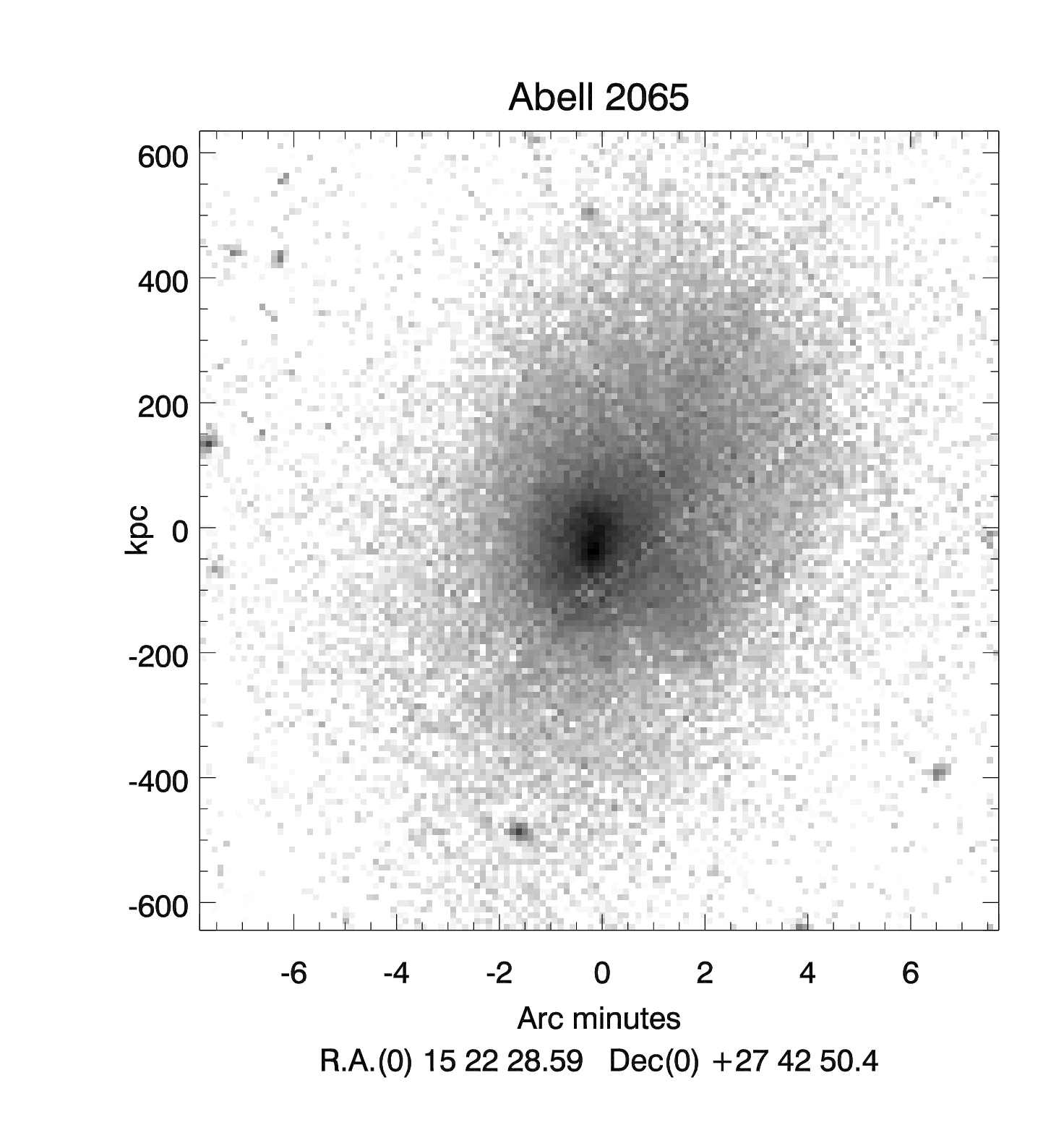}}
  \hspace{.025\hsize}                       
  \resizebox{.25\hsize}{!}{\includegraphics{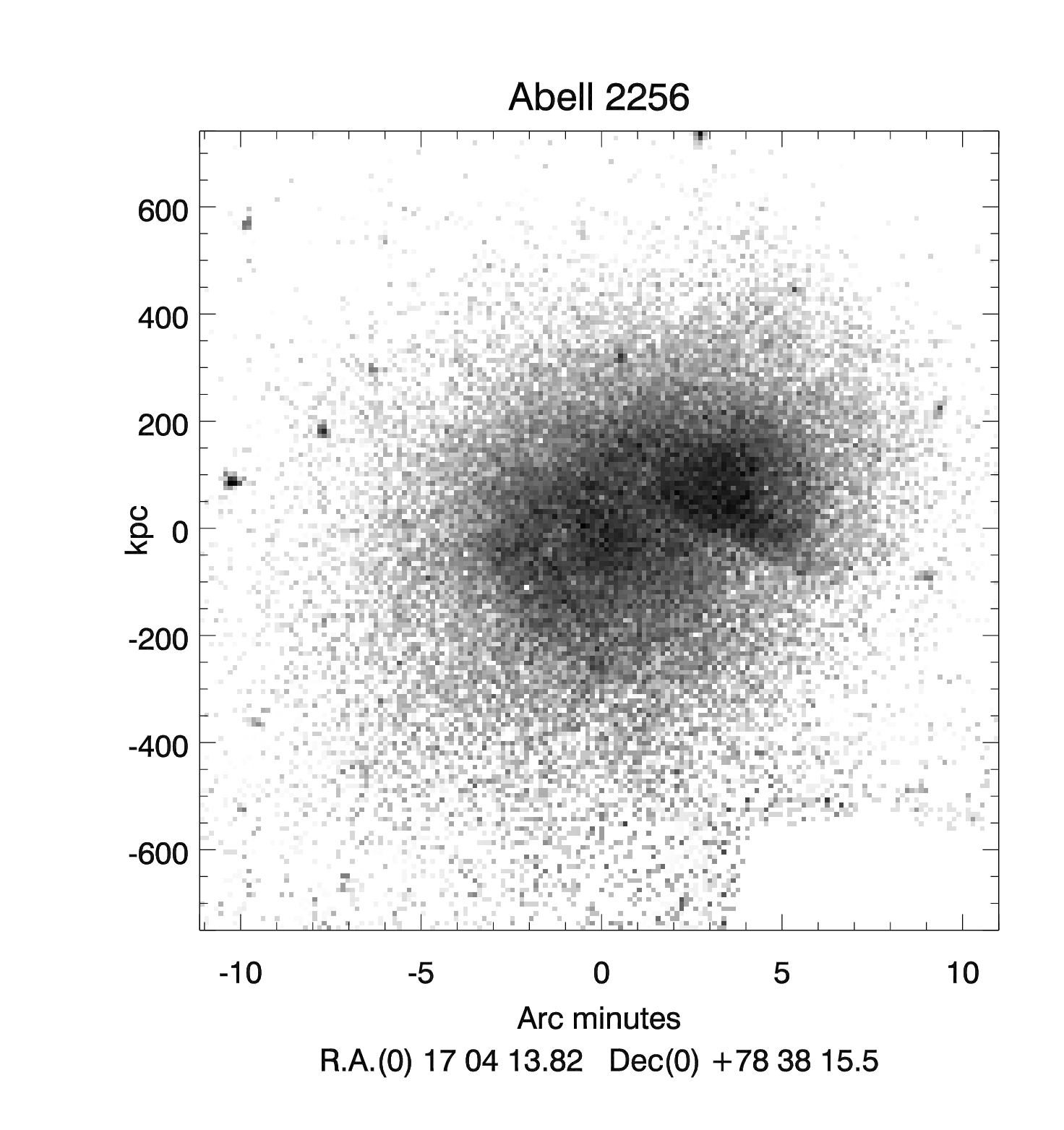}}} \\                      
\multicolumn{3}{c}{\resizebox{.25\hsize}{!}{\includegraphics{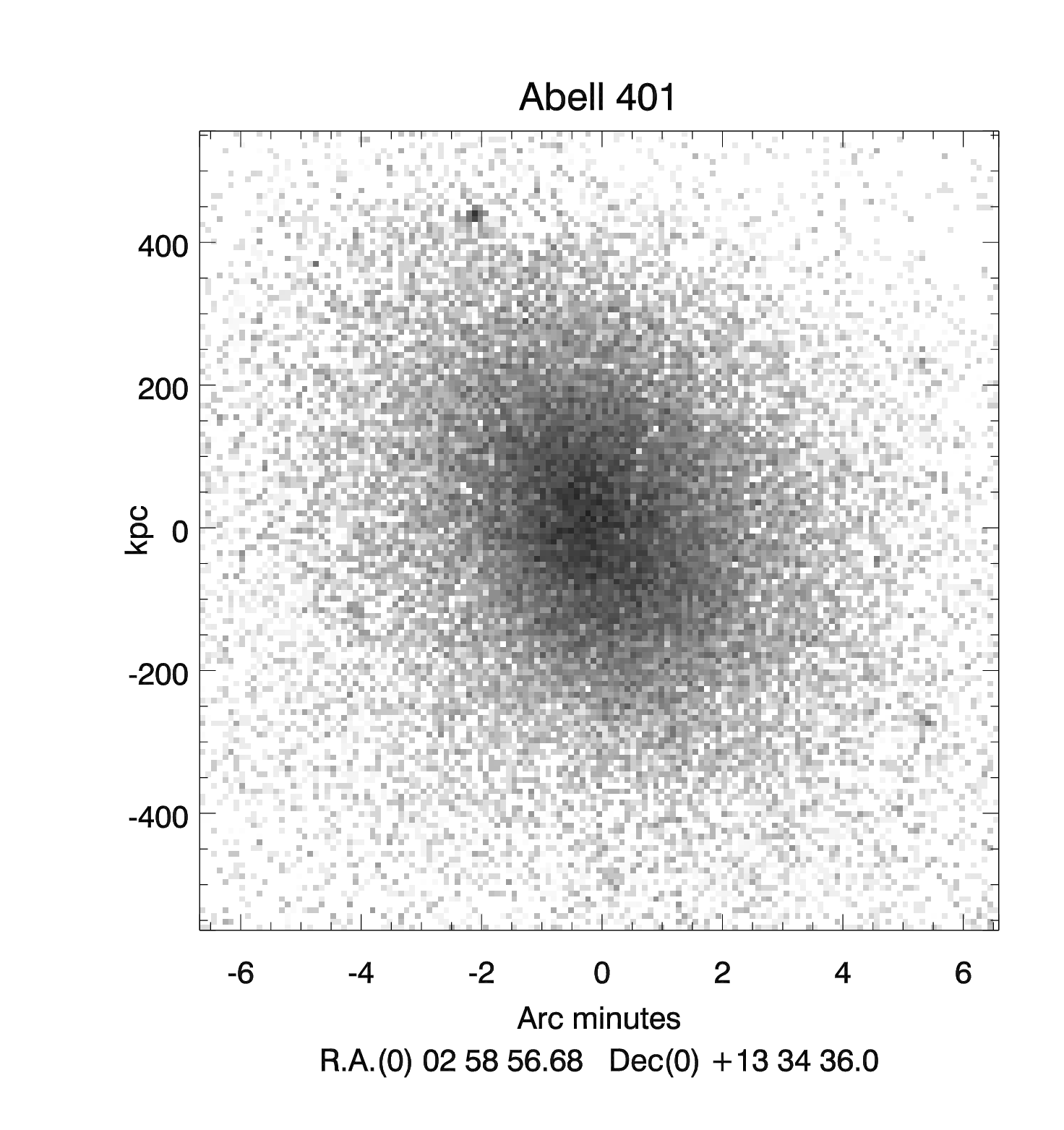}}
  \hspace{.025\hsize}                       
  \resizebox{.25\hsize}{!}{\includegraphics{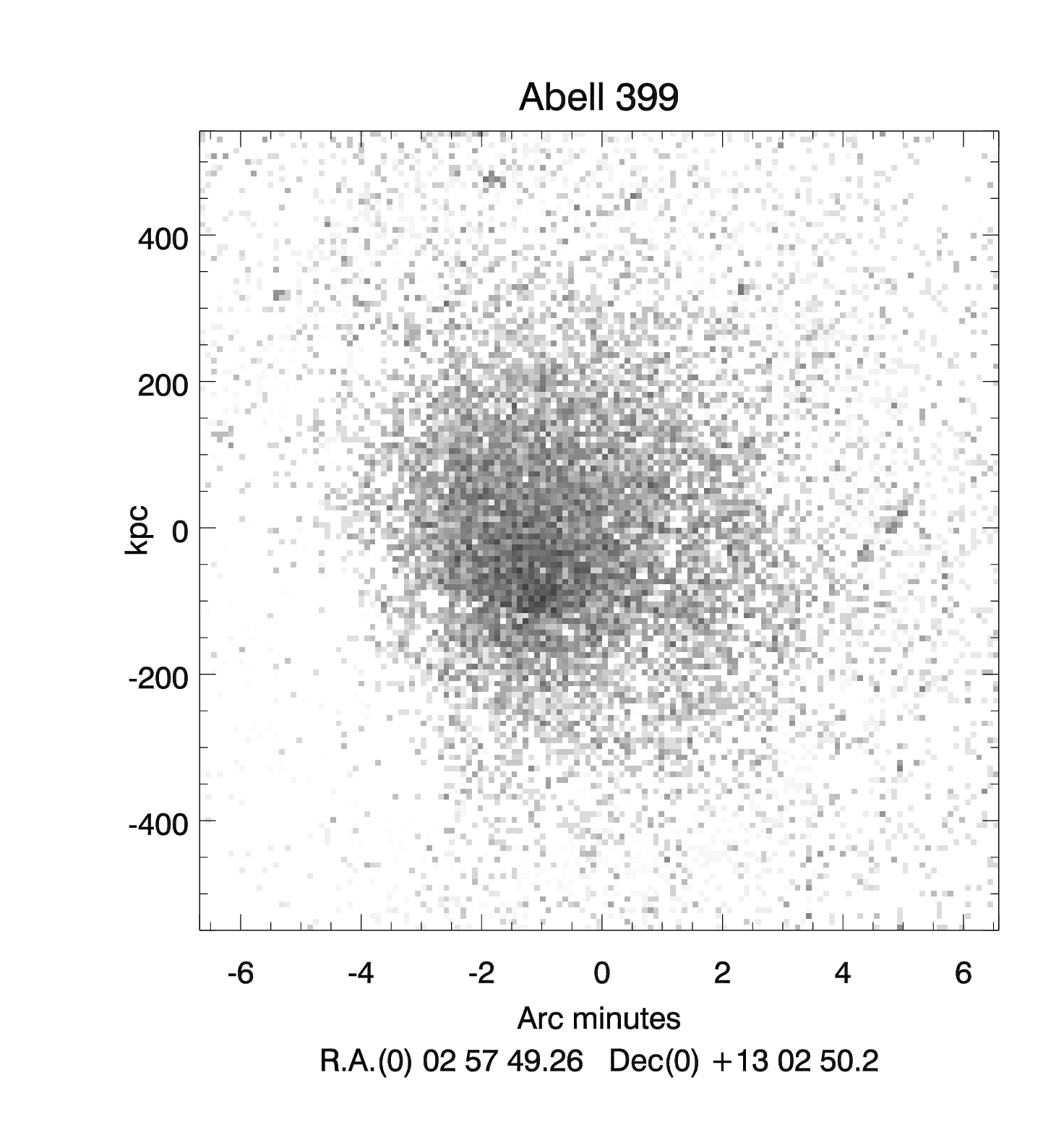}}} \\
\end{tabular}
\caption{EPIC-XMM ``soft'' (0.5-2.5 $\kev$) exposures images of the
clusters in our sample.\label{cxmaps_fig}}
\end{center}
\end{figure*}

\section{Data analysis\label{data_analysis}}

\subsection{ICM temperature mapping}

In order to map the ICM temperature structure, we have built a
spectral mapping algorithm coupling a spectroscopic and multi-scale
analysis of the X-ray signal, to a wavelet mapping of the searched
parameter structure. 

Mainly following the scheme of the \citet{Bourdin_04} algorithm
--hereafter the 2004 algorithm-- we first sample the field of view
using redundant and square grids with typical size $s_\j = 2^{\j} \ao,
\{\j \in [0,\jmax]\}$, according to a dyadic scheme. Then, we locally
estimate the gas temperature T and its fluctuation $\sigma_\t$ by
fitting a spectral model to the data, within each meta-pixel $[k,l,j]$
of the different grids.  A set of temperature maps $\t(k,l,j)$ is
obtained, with associated noise expectation maps
$\sigma_\t(k,l,j)$. Filtering the $\t(k,l,j)$ and $\sigma_\t(k,l,j)$
maps using high-pass analysis filter enables us to code the
temperature variations as wavelet coefficients $\w_\t(k,l,j)$ with
expected noise $\sigma_{\w_\t}(k,l,j)$, and to detect significant
temperature structures as wavelet coefficients with amplitude
overcoming a significance threshold depending on
$\sigma_{\w_\t}(k,l,j)$.  Finally, we map the gas temperature using a
Tikhonov regularised thresholding of the wavelet transform.

More detailed in \citet{Bourdin_04}, this general approach has been
adapted to the present study. In particular, the local spectral
fitting is now implemented using an updated plasma emission code and
now allows a multiple parameter estimation. Furthermore, a B-spline
wavelet transform is now used instead of the Haar wavelet
transform. Below we describe both of these improvements in detail.

\subsubsection{Spectral fitting of local ICM temperature\label{spectral_fitting}}

The local estimation of ICM temperature is performed by fitting a
normalised spectral model, $\f(\t,\ab,\nh,e)$, to the data set
associated with meta-pixel $[k,l,j]$. Combining contributions of ICM
itself, $\ns(k,l)~ \s(\t,\ab,\nh,e)$, and ``overall background'',
$\nb~\b(k,l,e)$, this model is sampled in photon energies, $e$, and
depends on the ICM temperature $\t$, metal abundances, $\ab$, and
neutral hydrogen density column along the line of sight, $\nh$:

\begin{eqnarray}
  && \nf(k,l)~\f(\t,\ab,\nh,e)  \nonumber \\
  &=& \ea(k,l,e) \times \ns(k,l)~\s(\t,\ab,\nh,e) \nonumber \\
  && + \nb~\b(k,l,e),
\label{global_spectra_equ}
\end{eqnarray}

The unnormalised ICM contribution $\s(\t,\ab,\nh,e)$ is modelled from
plasma radiation flux $\phi_{\mathrm{ICM}} (\t,\ab,e)$, following the
Astrophysical Plasma Emission Code \citep[APEC,][]{Smith_01} as a
function of the gas temperature $\t$ and heavy elements abundances
$\ab$, being set according to the solar element composition of
\cite{Grevesse_98}. It is obtained by red shifting and distorting
$\phi (\t,\ab,e)$ in order to take into account X-ray absorption by
the galactic neutral hydrogen $h(\nh,e)$ with given column density
along the line of sight, $\nh$, following absorption parameters of
\citet{Balucinska-Church_92}. Introducing the instrument area $A(e)$,
and convolving ICM radiation flux by the detector response in energy
$R(e)$\footnote{Detector responses are tabulated within the \xmm{}
redistribution matrix files (RMF)}, we get the radiation power
measured by the instrument for a given emitting source at redshift
$z$:

\begin{eqnarray}
  && \s(\t,\ab,\nh,e) \nonumber \\ &=& R(e) \# \left[ A(e) \times h(\nh,e)
    \times \frac {\phi_{\mathrm{ICM}} \left[\t,\ab,(1+z)e\right]}{1+z} \right],
  \label{spectra_equ}
\end{eqnarray}

where instrument area $A(e)$ takes into account the effective
area of mirrors $a_{mirror}(e)$, the filter transmission
${tr}_{filter}(e)$ and detector quantum efficiency
$q_{CCD}(e)$\footnote{Information about these instrumental effects are
provided in the following \xmm{} current calibration files (CCF),
corresponding to each observation epoch: XRT*\_XAREAEF,
EMOS*\_QUANTUMEF, EPN\_QUANTUMEF, EMOS*\_FILTERTRANSX,
EPN\_FILTERTRANSX}:

\begin{equation}
  A(e) = q_{CCD}(e) \times {tr}_{filter}(e) \times a_{mirror}(e).
\end{equation}

In order to get a robust estimation of ICM temperatures, whatever the
local statistics are, the spectral fitting is performed by maximising
the log-likelihood function $\log \l(\t,\ab,\nh,e) = \sum_{i} \log
\f(\t,e_i)$, where our spectral model is summed on all energy
channels, $e_i$.

\subsubsection{Wavelet mapping of ICM temperature structure\label{spectral_mapping}}

\begin{figure}[t]
  \begin{tabular}{ll}
    \resizebox{.43\hsize}{!}{\includegraphics{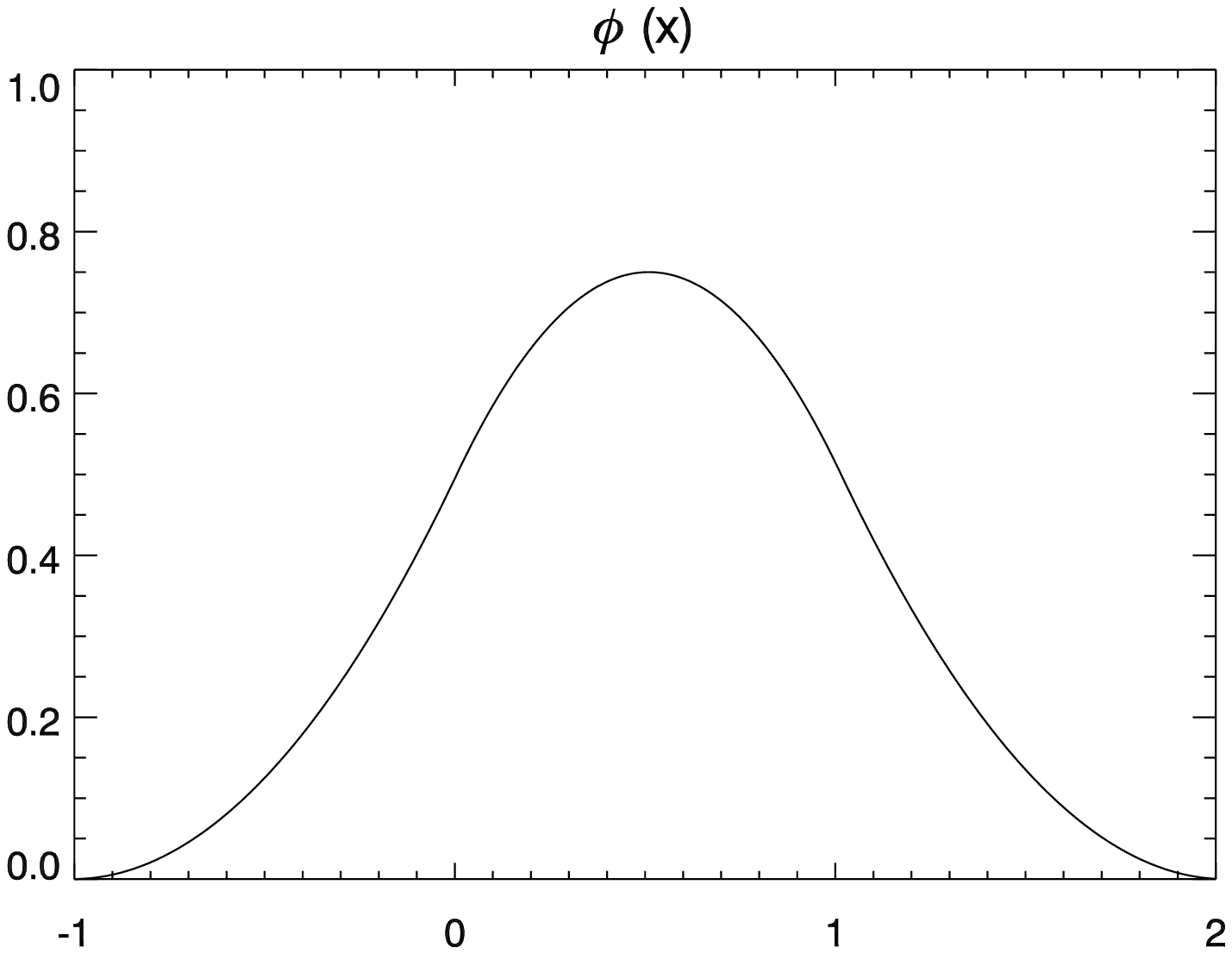}} &
    \resizebox{.43\hsize}{!}{\includegraphics{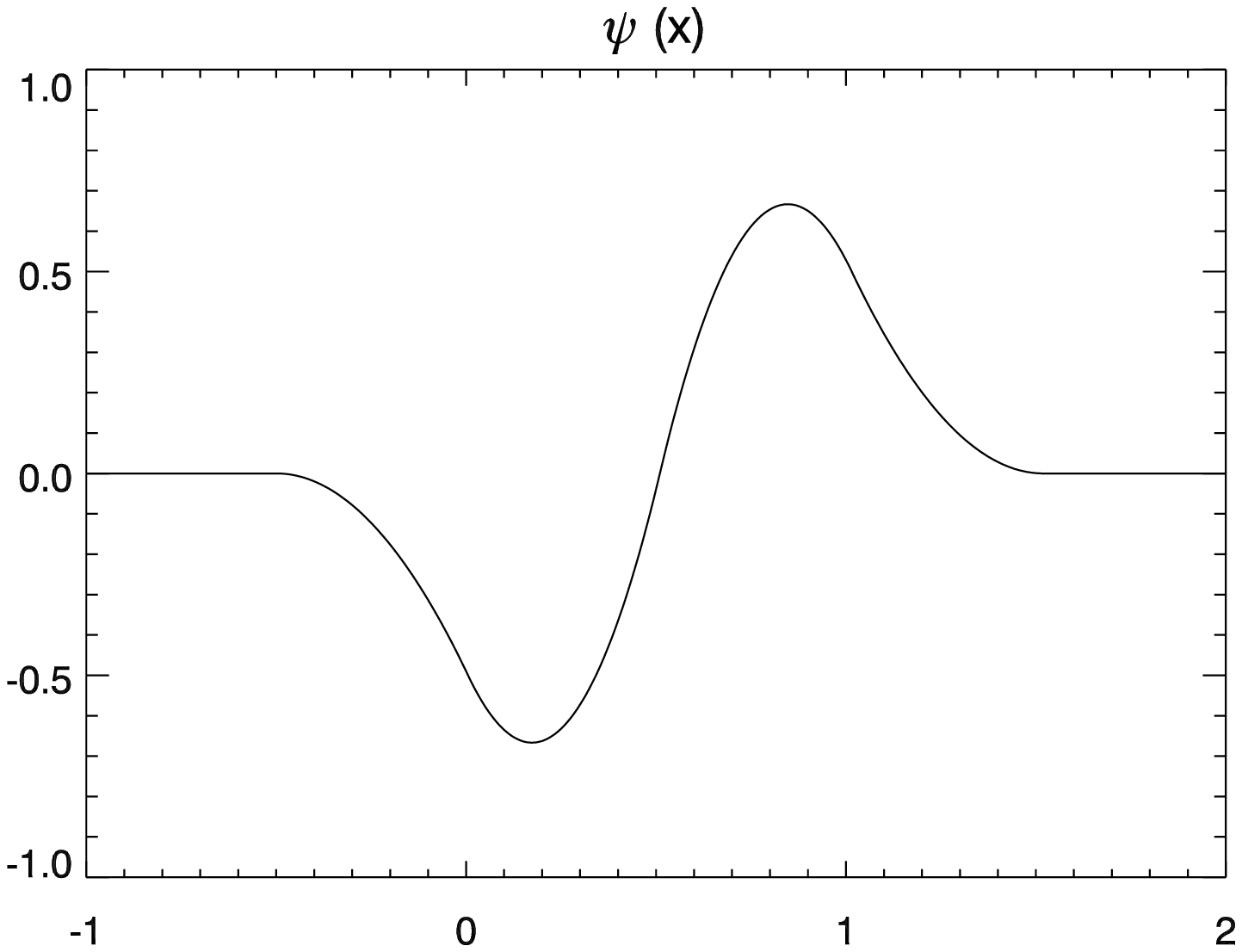}} \\
  \end{tabular} 
  \caption{The quadratic B-spline  scaling function $\phi(x)$ and dual
  wavelet $\psi(x)$ in 1D.\label{b2_spline}}
\end{figure}

\begin{table}[t]
\caption{The quadratic B-spline analysis filters. \label{bspline_filters}}
\begin{tabular}{ccccccc}
\hline\hline
~ & -2 & -1 & 0 & 1 & 2 & 3\\
\hline
$h$ & & .125 & .375 & .375 & .125 & \\ 
$g$ & & & -.5 &  .5 & & \\ 
$\tilde{h}$ & & .125 & .375 & .375 & .125 & \\ 
$\tilde{g}$ &-.03125 & -.21875 & -.6875 & .6875 & .21875 &
  .03125 \\ 
\hline
\end{tabular}
\end{table}

In order to map the ICM temperature structure at scale $j$, we analyse
the spatial correlation of temperature measurements using wavelet
coefficients $\w_{\t,j}(k,l)$, computed from the temperature maps
$\t_j(k,l)$.  A trivial solution to this computation has been proposed
in the 2004 algorithm: filtering the temperature maps $\t_j(k,l)$
using Haar high-pass analysis filters enables one to get a set of Haar
wavelet coefficients $\w_{\t,H,j}(k,l)$.  Indeed, the Haar wavelet is
the dual function of the top-hat smoothing kernel $\Pi_j(k,l)$, which
may be applied to the searched map $\t_0(k,l)$ for computing the maps
$\t_j(k,l)$, at scale $j$.  However, thresholding Haar wavelet transforms
usually generates square artifacts, in particular when analysing
regular signals.

Due to the expected smoothness of our signal, a better analysis of its
spatial correlations can be provided by the more regular B-spline
wavelets; see \citet{Curry_47} for definition and e.g.
\cite{Mallat_98} for application to wavelet bases. Indeed, we do not
expect any strong discontinuities in our signal, due to joint effects
of both the instrument PSF and the 2-D projection of the gas
temperature structure. Since B-spline wavelets are the dual functions
of the m degree B-spline interpolation functions obtained by (m+1)
self-convolving of a top-hat smoothing kernel, they can also be used
in our context; starting from the temperature maps $\t_j(k,l)$ and
convolving them (m) times by a top-hat smoothing kernel $\Pi_j(k,l)$
provides a new set of smoothed maps $\t_{S(m),j}(k,l)$, whose dual
wavelet coefficients are m degree B-spline wavelet coefficients
$\w_{\t,S(m),j}(k,l)$.  Introducing the smoothing length $a = 2^j$ at
scale j, we get:

\begin{eqnarray}
  \Pi_j(k,l)  &=& \mathbf{1}_{\left[k-\ademi,k+\ademi-1\right]} \times
  \mathbf{1}_{\left[l-\ademi,l+\ademi-1\right]},  \\  \t_{S(m)}(k,l,j)
  &=& \Pi_j(k,l)^{(m)} \star \t(k,l,j).
\end{eqnarray}

Like  the Haar  wavelet, the  B-spline  wavelets can  be projected  in
bi-orthogonal    bases.    We therefore adopted the   shift-invariant
\citet{Coifman_Donoho_95}   algorithm   for   computing   the   wavelet
coefficients $\w_{\t,S(m),j}(k,l)$, following  a similar scheme as for
the 2004 algorithm.  To do so, the smoothed maps $\t_{S(m)}(k,l,j)$ are
convolved with  a set of three high-pass  analysis filters $h(k)g(l)$,
$g(k)h(l)$ and $g(k)g(l)$, associated with the B-spline wavelet, which
leads    to     the    three    sets     of    wavelet    coefficients
$\w_{\t,S(m),j,h}(k,l)$,          $\w_{\t,S(m),j,v}(k,l)$          and
$\w_{\t,S(m),j,d}(k,l)$:

\begin{eqnarray}
  \w_{\t,S(m),j,h}(k,l)    &=&   h(k)g(l)    \star   \t_{S(m)}(k,l,j),
\nonumber\\ \w_{\t,S(m),j,v}(k,l) &=& g(k)h(l) \star \t_{S(m)}(k,l,j),
\nonumber\\ \w_{\t,S(m),j,d}(k,l) &=& g(k)g(l) \star \t_{S(m)}(k,l,j).
\end{eqnarray}
  
In  order   to  re-construct  the  signal   $\t_0(k,l)$,  the  wavelet
coefficients  are  first   convolved  using  the  high-pass  inverse
filters,  $\tilde{h}(k)\tilde{g}(l)$,  $\tilde{g}(k)\tilde{h}(l)$  and
$\tilde{g}(k)\tilde{g}(l)$  associated with the  bi-orthogonal wavelet
analysis  and  added  to  each  other, which  allows us to  average  the
redundant information.  Thus,  the risk occurring when re-constructing
the  signal from  the thresholded  wavelet transform  is significantly
lowered.

Here we decided to work with the quadratic B-spline wavelet ($m=2$,
see \fig \ref{b2_spline}), which is regular enough for our purpose but
more compact than B-spline wavelets of higher degree. The quadratic
B-spline analysis filters are reported in \tab \ref{bspline_filters}.

\begin{figure*}[ht]
  \begin{tabular}{lll}
    \resizebox{.3\hsize}{!}{\includegraphics{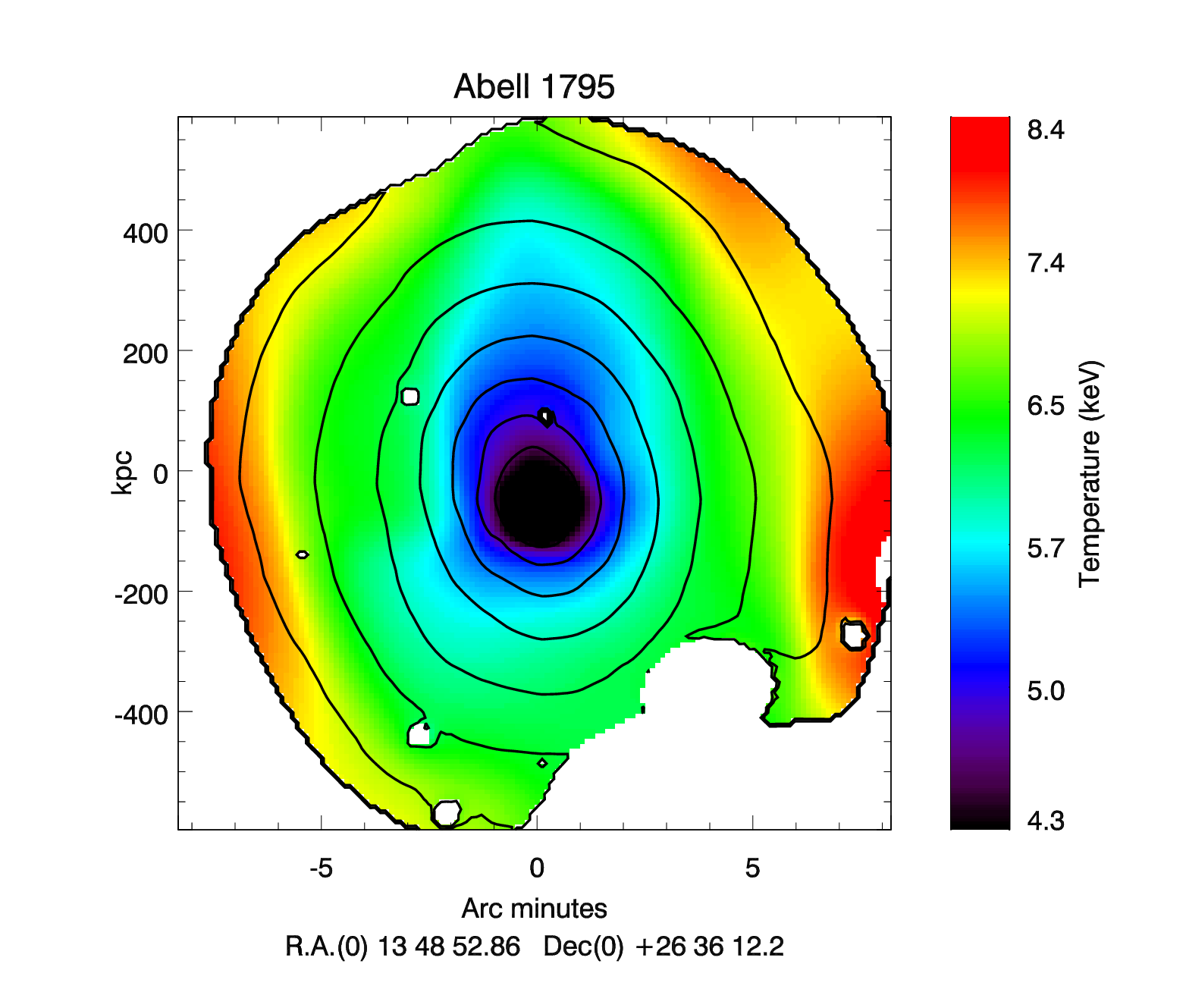}}
    &
    \resizebox{.3\hsize}{!}{\includegraphics{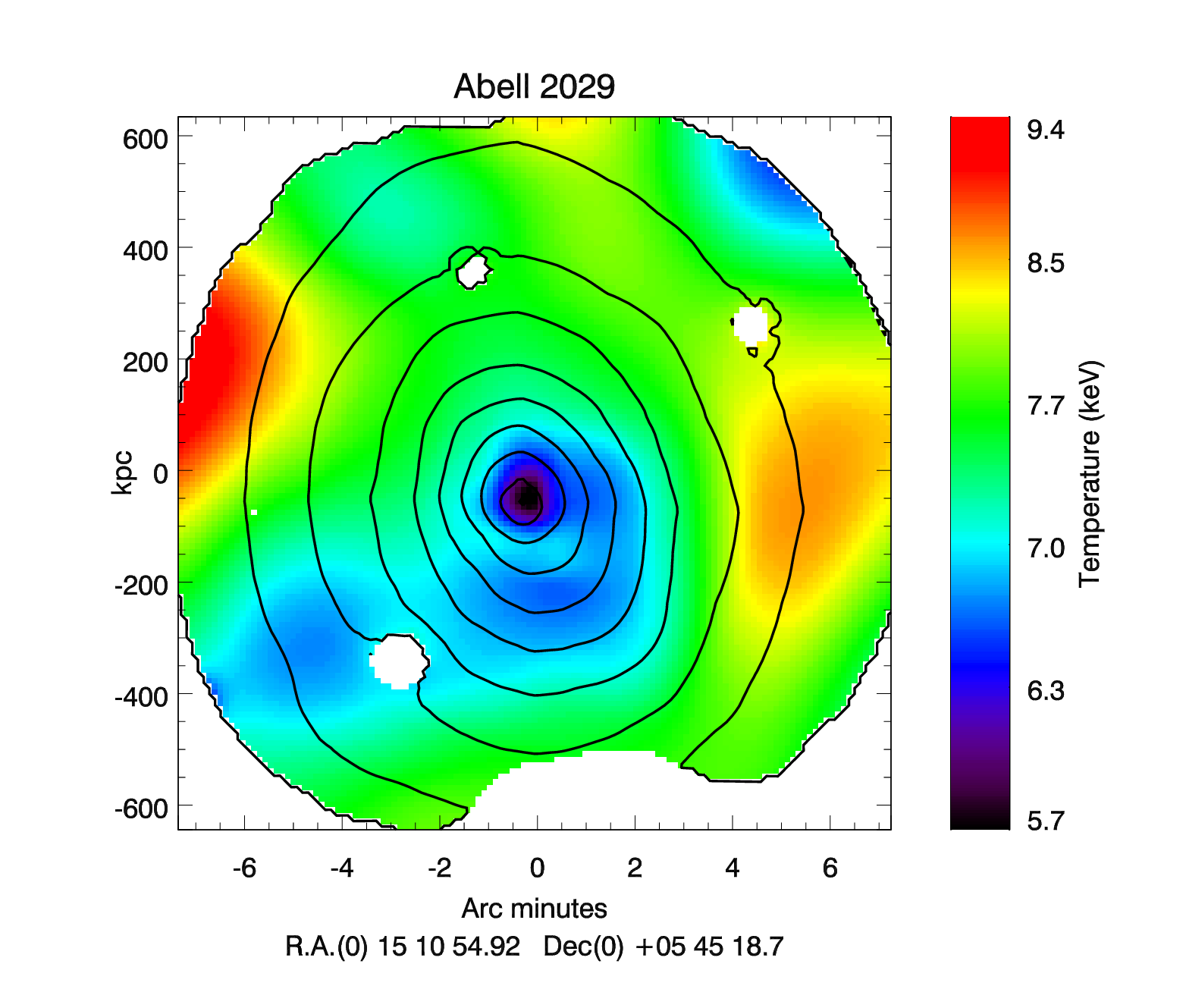}}
    &
    \resizebox{.3\hsize}{!}{\includegraphics{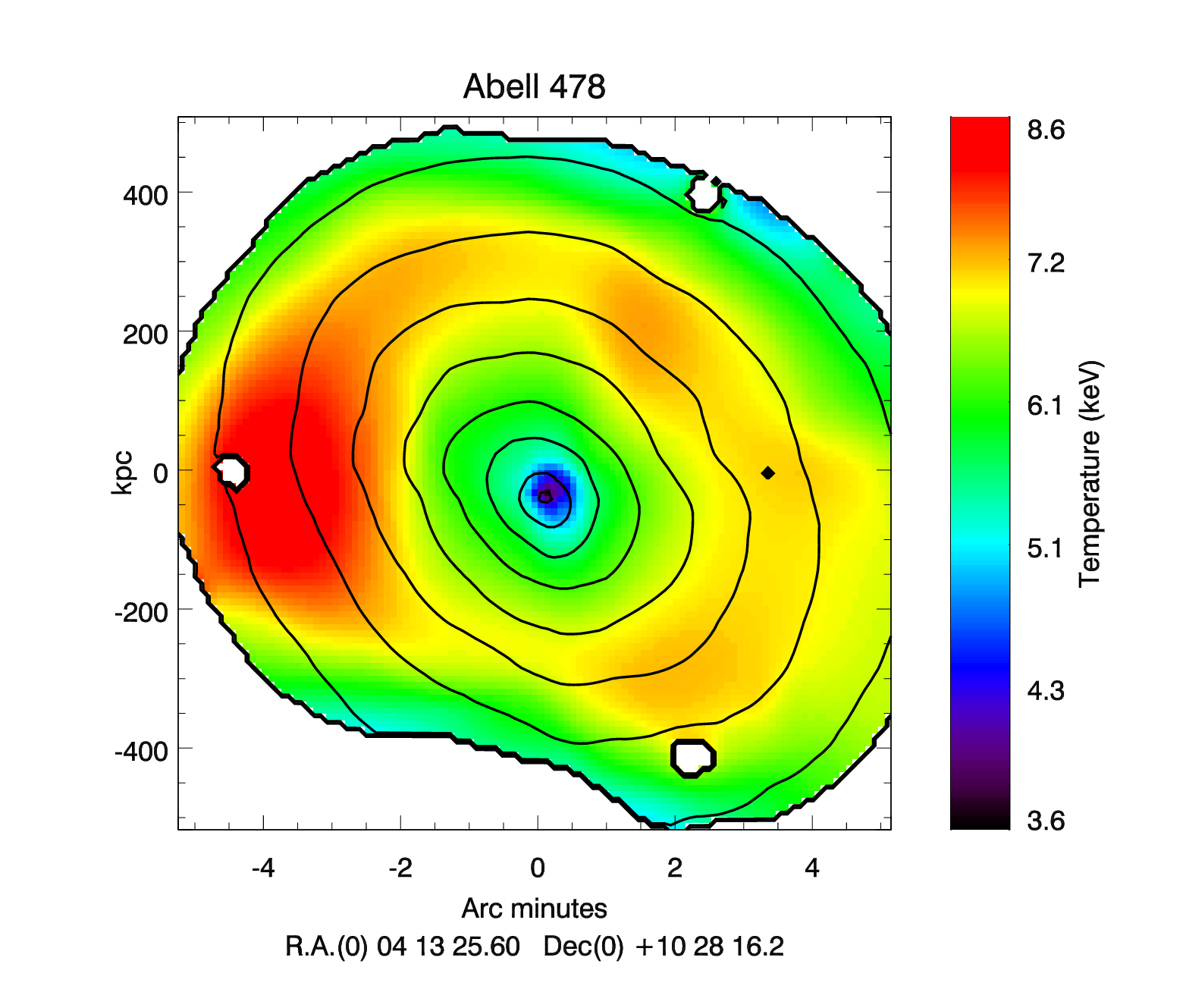}}
    \\
    \multicolumn{3}{c}{\resizebox{.3\hsize}{!}{\includegraphics{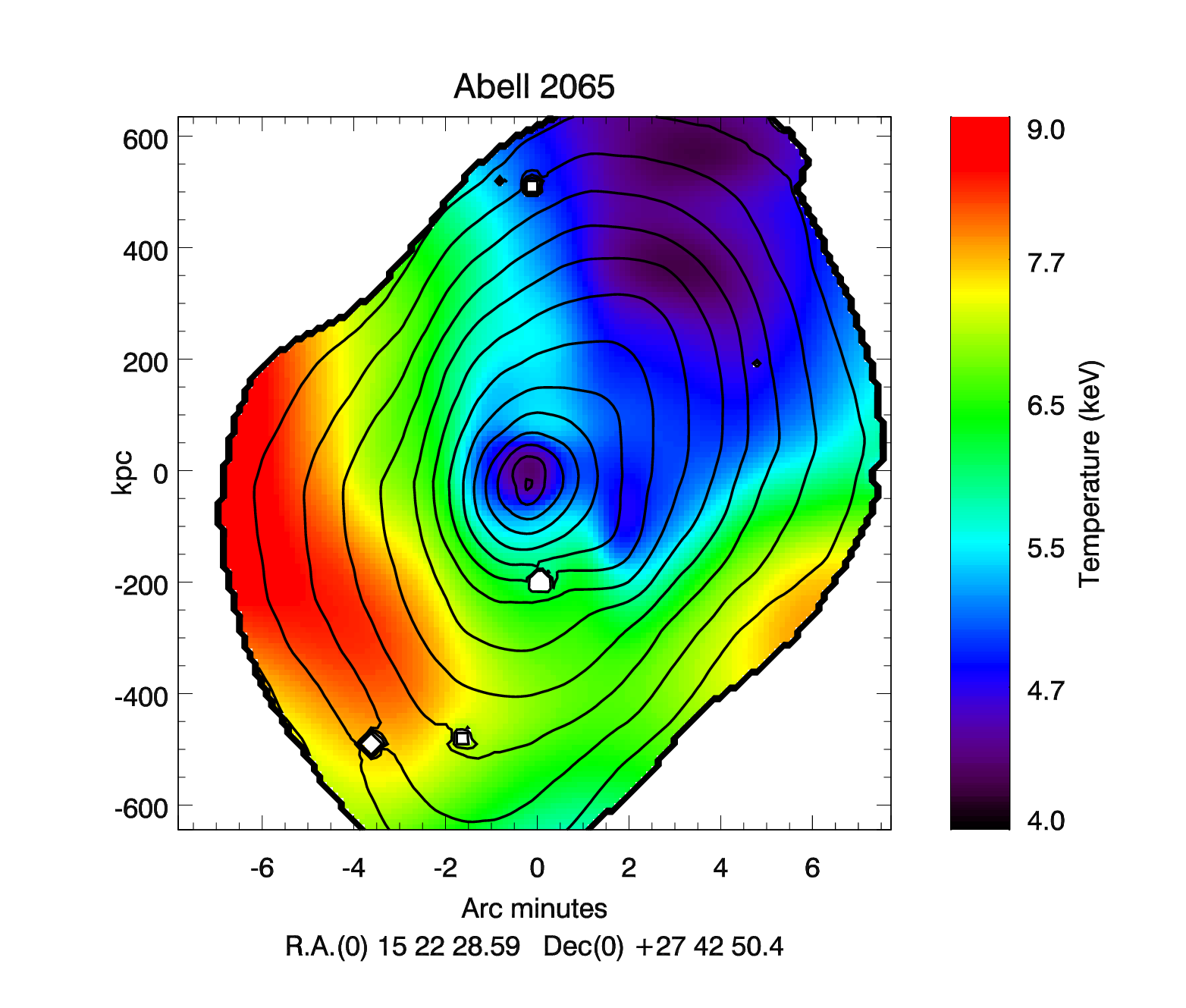}}
      \hspace{.025\hsize}                       
      \resizebox{.3\hsize}{!}{\includegraphics{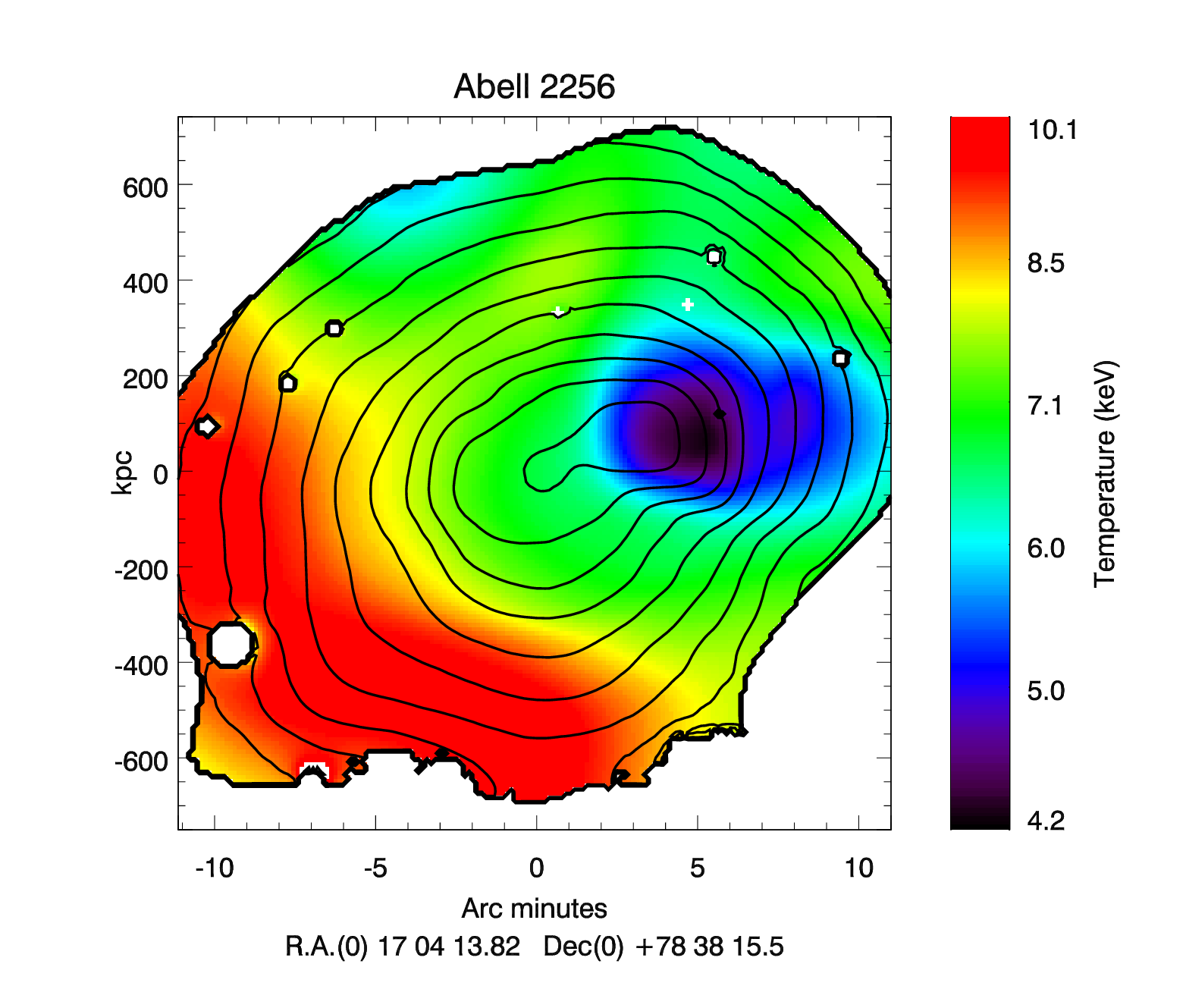}}} \\                       
    \multicolumn{3}{c}{\resizebox{.3\hsize}{!}{\includegraphics{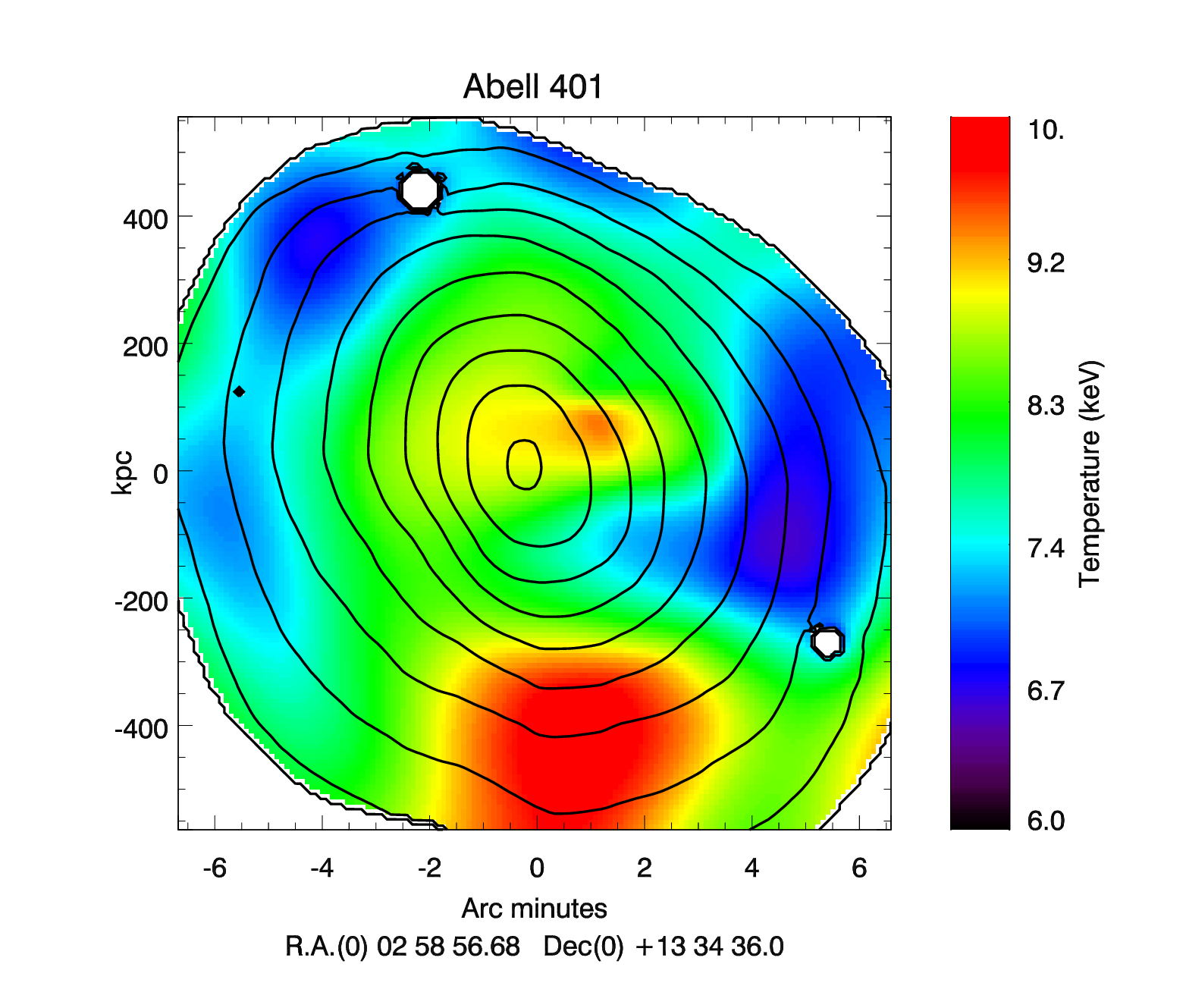}}
      \hspace{.025\hsize}                       
      \resizebox{.3\hsize}{!}{\includegraphics{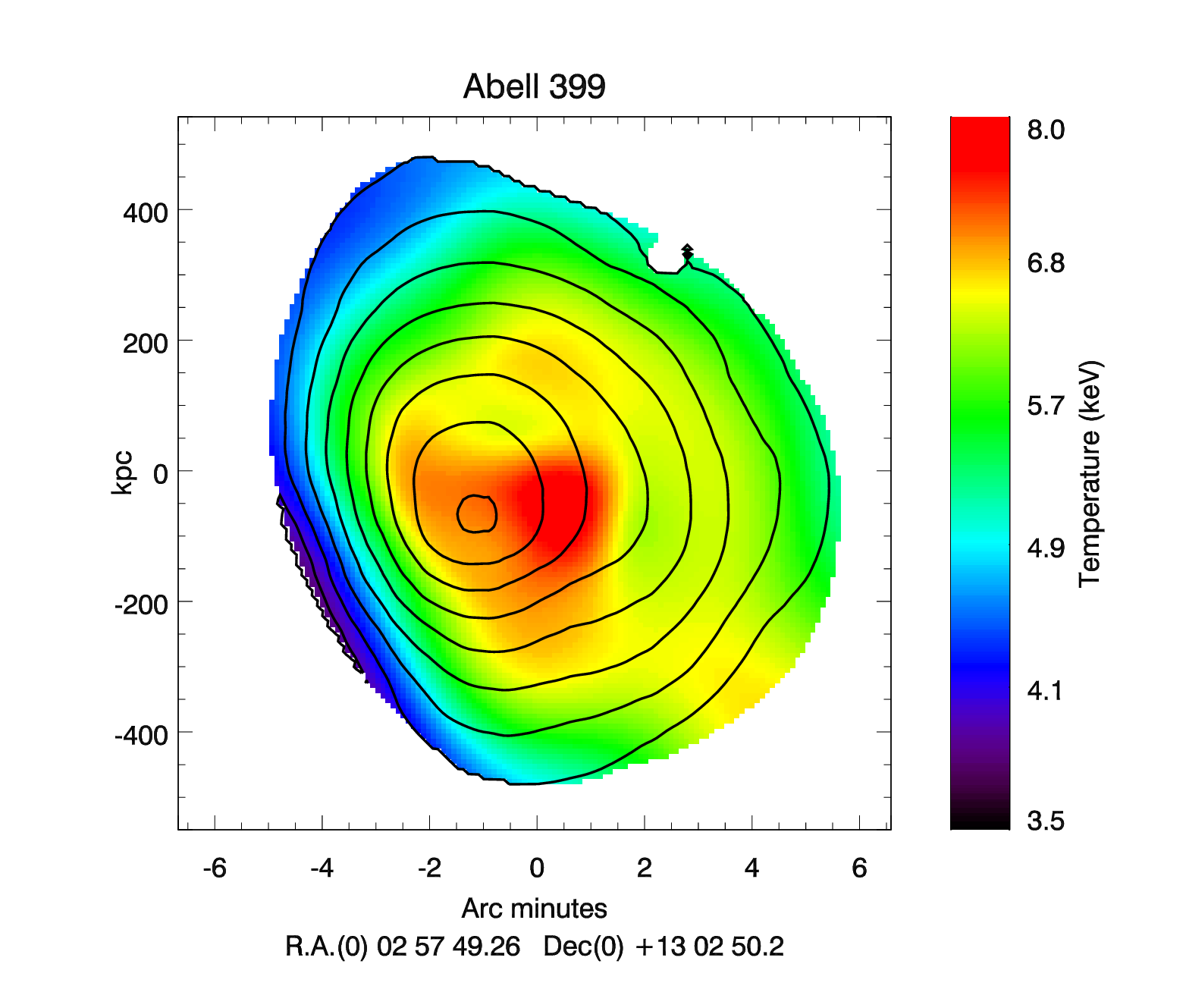}}} \\
  \end{tabular}
  \caption{ICM temperature maps of the clusters in our sample overlaid
    to the soft (0.5-2.5 $\kev$) X-ray brightness isocontours. The
    isocontour levels are logarithmically equispaced by a factor
    $\sqrt{2}$. \label{tmaps_fig}}
\end{figure*}

\subsection{ICM brightness mapping}

In order to map the ICM brightness structure, we have built an imaging
algorithm where the local brightness $\Lx(k,l)$ is first estimated
from a brightness model associated with pixel [k,l], then iteratively
de-noised in the scale-space by means of a discrete Haar wavelet
analysis.

We estimate the local ICM brightness ${\Lx}(k,l)$ for energy band
$\Delta e$, by correcting the number of events $\nf(k,l)$ detected at
pixel (k,l), from the expected ``background'' contribution, $\nb(k,l)$,
and exposure weighted instrument area, $\vf(k,l)$:

\begin{eqnarray}
  \vf(k,l) &=& \int_{\Delta e} \ea(k,l,e) ~ \f(\t_o,\ab_o,\nh_o,e) ~ de \label{weigthed_area_equ}, \\
  \widehat{\Lx}(k,l) &=& \frac{\nf(k,l) - \nb(k,l)}{\vf(k,l)},
  \label{brightness_equ}
\end{eqnarray}

where the weighted instrument area $\vf(k,l)$ is computed for a fixed
ICM emission model, $\f(\t_o,\ab_o,\nh_o,e)$. ``Corrected'' photon maps
obtained from local estimator $\widehat{\Lx}(k,l)$ and for energy band
$0.5-2.5 ~\kev$ are shown in \fig \ref{cxmaps_fig}.

The signal de-noising is then performed by thresholding Haar wavelet
coefficients $\W_{\widehat{\Lx}}[k,l]$, associated with a multi-scale
analysis of $\widehat{\Lx}(k,l)$ using redundant grids. Driven by a
significance criterion of detected structure, this thresholding is set
after estimating the probability density function (PDF) of noise
wavelet coefficients, $\P(\W_{\widehat{\Lx}}[k,l])$. To do so, we
deduce $\P(\W_{\widehat{\Lx}}[k,l])$ from the PDF of noise wavelet
coefficients associated with local event counts, $\nf$:

\begin{equation}
  \P(\W_{\widehat{\Lx}}[k,l]) = \frac{\P(\W_{\nf}[k,l])}{\vf[k,l]},
\end{equation}

which enables us to use an analytical form of $\P(\W_{\nf}[k,l])$,
first proposed by \citet{Bijaoui_01} for Haar wavelet analyses of
uniform Poisson noise. Introducing the modified Bessel function of
order m, $I_m(\nf)$, we get:

\begin{equation}
  \P(\W_{\nf}) = \sum_{m=-\infty}^{\infty} e^{-\nf} I_m(\nf) \delta(\W_{\nf}-m).
\end{equation}

\section{2D temperature structure\label{thermal_structure}}

The ICM temperature maps of each cluster in our sample are presented
in \fig\ref{tmaps_fig}, overlaid to the relative $0.3-2.5 \kev$
brightness isocontours levels obtained by wavelet imaging (see details
above). The emission model $\f(\t_o,\ab_o,\nh_o,e)$ assumed for
computing brightness maps has been obtained by spectral-fitting of an
overall cluster spectrum excluding the core region of relaxed
clusters.

The temperature maps have been computed from spectral-fitting within
the $.7-7.5 \kev$ energy band with fixed values of the redshift, metal
abundances, and neutral hydrogen column densities. The redshift values
have been set to those in \tab\ref{cluster_sample_tab}, while the
metal abundances $\ab_o$ have been deduced from the spectral fitting
of $\f(\t_o,\ab_o,\nh_o,e)$. For each cluster except Abell~478, the
neutral hydrogen column densities, $\nh_o$, have been estimated by
spectral fitting, similarly to the metal abundances. The values
obtained are consistent with measurements by
\citet{Dickey_lockman_90}, in the 1 degree neighbourhood of each
cluster centre (see \tab \ref{cluster_sample_tab}). For A478, the
$\nh$ has been left as free parameter, since \citet{Dickey_lockman_90}
measurements were inconsistent with X-ray spectroscopic estimations
(see \part \ref{nh_part} for details).

The EPIC-XMM field of view has been sampled in $256 \times 256$
pixels, and the wavelet analysis has been unfolded over six spatial
resolutions, allowing detection of features with typical size ranging
from about 12 arcsec to 3.5 arcmin.  The thresholding of the wavelet
coefficients has been performed following the \citet{Donoho_95}
thresholding approach, leading to threshold levels that are always
higher than the noise fluctuation $\sigma$ (typically from $1.75$ to
$2.5~\sigma$ with increasing analysis scales, thresholds lower than $
2 \sigma$ being related to the highest resolution details, namely the
cool core regions of relaxed clusters only).

In the following we give a short description of each cluster separately.
For convenience we group the clusters following the dynamical 
classification, used in previous work and based on the X-ray morphology,  
of relaxed and merging systems. 

\subsection{Relaxed Clusters}

\subsubsection{Abell~ 1795}

As observed in X-rays, Abell~1795 is known as an elliptical and
relaxed cluster, as predicted by the relatively low P4/P0 power ratios
(see \tab\ref{cluster_sample_tab}).  The \xmm{} observation of
Abell~1795 confirms a globally elliptical symmetry for this cluster.
Nevertheless, the \xmm{} X-ray image in \fig\ref{cxmaps_fig} shows the
presence of a sharp surface brightness variation, at about $1 ~{\rm
arcmin}$ to the south with respect to the cluster centre.  From a
Chandra observation, this sharp surface brightness feature has been
identified as a cold front and interpreted as the result of the
sloshing of core gas within the cluster potential well
\citep{Markevitch_01}. This scenario is consistent with what can be
observed on surface brightness contours in \fig\ref{tmaps_fig},
revealing at the same time a strong compression of isophotal lines
across the cold front, and a shift of the cluster brightness peak
towards the south, with regard to large radii isophotes.

The temperature map of Abell~ 1795 in \fig\ref{tmaps_fig} shows a
global elliptical symmetry with a cool core. If we perform a radial
analysis, it is consistent with temperature profiles derived from \xmm{}
\citep{Tamura_01, Arnaud_01, Ikebe_04} and Chandra observations
\citep{Ettori_01, Markevitch_01, Vikhlinin_05}. We notice, however,
some anisotropies with regard to the overall elliptical symmetry of
temperature structure. First of all, the cool core appears as being
shifted towards the south, as coinciding with the cluster brightness
peak. At larger radii ($r > 2$ arcmin) the cluster also presents some
significant non radial thermal structure.  In particular we find that
the gas temperature is colder in a small sector to the north ($\kt
\simeq 5 \kev$) and hotter elsewhere ($\kt \simeq 6 \kev$).

\subsubsection{Abell~ 2029}

Similarly to Abell~1795, Abell~2029 was also known as a regular and
relaxed cluster (see also the P4/P0 power ratios in
\tab\ref{cluster_sample_tab}).  Except for its very central region
where X-ray filaments have been observed \citep[e.g.][]{Clarke_04},
no brightness nor temperature anisotropies have been previously
detected for this cluster.

As already noticed by \citet{Vikhlinin_05}, this cluster is projected
near the galactic region of North Polar Spur, the X-ray spectral
contribution of which has been modelled and added to the background
model of \equ(\ref{background_equ}). This contribution has been fitted
simultaneously with cosmic background, in the same external region of
the field-of-view, which led to a two thermal component model
(kT1=0.22, kT2=0.49, $n_{K1}/n_{J2}=2.05$), consistent with the
analysis of \citet{Vikhlinin_05}\footnote{Normalisation discrepancies
with \citet{Vikhlinin_05} are due to different modellings of the CXB,
our model already including a thermal component at $0.204 ~ \kev$ (see
\equ(\ref{background_equ})).}.

The \xmm{} image in \fig\ref{cxmaps_fig} confirms that Abell~ 2029 is
globally regular and elliptically symmetric (see also
\fig\ref{tmaps_fig}).  The gas temperature map in \fig\ref{tmaps_fig}
shows the presence of a cooler core ($\kt \simeq 5.5 \kev$) and a
positive radial temperature gradient that extends up to 1.5 arcmin from
the cluster centre.  If we perform a radial analysis, the temperature
map is consistent with previously published temperature profiles
\citep{Lewis_03, Vikhlinin_05}, except in the very central region ($r
< 10$ arcsec), where the Chandra data analysis of \citet{Lewis_03}
indicates lower temperatures decreasing to about 2 keV.

Nevertheless, as observed for Abell~1795, even if the temperature
structure is elliptically symmetric in the innermost regions, at
larger radii the temperature map reveals some non radial thermal
features with typical temperature variations of $\simeq 1 \kev$. In
particular, we detect a cold region at 5 arcmin to the southeast of
the cluster centre ($\kt \simeq 6 \kev$).


\subsubsection{Abell~ 478}

Known to be regular and relaxed, Abell~ 478 has been observed recently
 by the Chandra and \xmm{} telescopes \citep{Sun_03,
 Pointecouteau_04}.  While the Chandra observation has revealed
 brightness substructure in the very centre of the cluster
 \citep{Sun_03}, both of these observations have shown a regular
 structure at larger radii. The elliptical symmetry and regularity of
 the surface brightness of Abell~478 is also evident from the X-ray
 cluster image in \fig\ref{cxmaps_fig} and confirmed by the low value
 of the P4/P0 power ratios in \tab\ref{cluster_sample_tab}.

The gas temperature map of Abell~478 is shown in \fig\ref{tmaps_fig}.
Due to the strong hydrogen column density appearing across the field
of view ($\nh = 15.1 \times \nhunit$), the gas temperature map of this
cluster has been computed from a spectral fitting process with a free
$\nh$ parameter.  Using this procedure the temperature map appears
remarkably regular with elliptical symmetry. This map reveals a cool
core ($\kt \simeq 4 \kev$) with 1 arcmin radius, a radial temperature
increase up to a plateau at about 3.5 arcmin from the cluster centre
($\kt \simeq 7.5 \kev$), then a decrease up to 5 arcmin. This overall
structure is in agreement with the shape of Chandra temperature
profile published by \citet{Vikhlinin_05}. Nevertheless, the absolute
temperature values from our temperature map are lower than the values
of \citet{Vikhlinin_05}, and more consistent with the analysis of
\cite{Pointecouteau_04}, using the same data as
here. \citet{Vikhlinin_05} and \cite{Sanderson_05} already noticed
these discrepancies and argued that they are probably related to a
combined effect of differences between instrument PSFs, and the
complex temperature structure of the very central region in this
cluster.

From both the \xmm{} brightness and temperature maps we may conclude
that Abell~478 is the most regular of our sample.  Nevertheless, we
should note that this cluster is located in a particular region of the
sky characterised by strong variations of neutral hydrogen column
density, on angular scales smaller than the cluster size itself.  In
this condition, the gas temperature variations are more difficult to
detect for this cluster than for the other clusters of our sample (see
also \part\ref{nh_part} below). 

\subsection{Merging clusters}

\subsubsection{Abell~399 - Abell~401}

Although, independently selected in our cluster sample, Abell~399 and
Abell~401 form a close pair separated, in projected distance, by
3~Mpc (36 arcmin).  The brightness contours of both Abell~ 399 and
Abell~ 401 (\fig \ref{tmaps_fig}) are elongated along a N-N-E / S-S-W
axis, which is the major direction of the pair. The Abell~401
brightness contours appear mildly disturbed with a centroid shift to
the northeast.  Consistent with the expectation of the P4/P0 power
ratios reported in \tab \ref{cluster_sample_tab}, they are less
regular than the contours of Abell~1795, Abell~ 2029 and Abell
478. Abell~399 looks even more irregular; in particular a sharp edge
can be observed to the southeast of the cluster core.

Using the same data set as used in this work, \citet{Sakelliou_04}
studied the gas brightness and temperature in specific angular sectors
of both clusters and along the major direction of the pair. From this
analysis, they conclude that, currently, the clusters are just starting
to mildly interact and that the sub-features found in their inner
regions are related to the individual merging histories of each
cluster separately, rather than to the remnant of a previous merger of
the two systems.

The temperature maps of Abell~399 and Abell~401 are reported in
\fig\ref{tmaps_fig} to the same spatial scale. We notice a higher
average temperature for Abell~401 ($\kt \simeq 8 \kev$) than for
Abell~ 399 ($\kt \simeq 7 \kev$), consistent with the analysis of
\citet{Sakelliou_04}. The temperature maps of both clusters appear as
strongly irregular, without any elliptical symmetry or large scale
correlation with gas brightness structure, contrary to what is
observed for the relaxed clusters in our sample. Moreover, instead of
cool cores, their central regions host some hot substructure embedded
in the colder ICM. We further confirm the temperature increment
described by \citet{Sakelliou_04} to the south of Abell~401, in the
direction of Abell~399. However, we do not detect any temperature
increment toward the north of Abell~399 along the expected interaction
axis of the cluster pair.  More generally speaking, most of the
temperature irregularities detected here do not have any morphological
link with the interaction axis of the cluster pair, and have typical
sizes more related to the central region of each cluster than to the
overall cluster system.

\subsubsection{Abell~2065}

Abell~2065 has been previously studied by \citet{Markevitch_99} using
ROSAT and ASCA data, and by \citet{Chatzikos_06} using Chandra data.
These observations revealed the asymmetric morphology of this cluster
characterised by a highly elongated inner region that seems to link
together the two cD galaxies located in the cluster centre. The
temperature structure is also highly asymmetric with a hotter region
to the southeast.  Furthermore, the Chandra observation revealed two
cold cores coinciding with the cluster central cD galaxies.

The \xmm{} temperature map of Abell~2065 in \fig\ref{tmaps_fig} shows
the presence of a hot ($\kt \simeq 6.5 \kev$) bow-like region, from
about 3 to 6 arcmin to the southeast of a cold cluster core ($\kt
\simeq 4.5 \kev$). This feature by itself is not isothermal but rather
appears to embed an even hotter sub-feature ($\kt \simeq 8.5 \kev$) to
the eastern cluster outskirts. The map also shows a cold region ($\kt
\simeq 4 \kev$) located at about 7 arcmin to the northwest of the
cluster core, appearing to be linked to this core by an extended tail.

Already reported in previous works \citep{Markevitch_99,Chatzikos_06},
the hot bow-like region is located next to an abrupt variation of gas
surface brightness visible in \fig\ref{cxmaps_fig}. For this reason,
it probably indicates the presence of a cold front, as discussed in
\part \ref{cfront}.

\subsubsection{Abell~2256}

The X-ray image of Abell~2256 in \fig\ref{cxmaps_fig} shows a complex
structure with two X-ray peaks, one of which is coincident with the
cluster central dominant galaxy, while the other is located at 2 arcmin
to the northwest.  This cluster has been observed with ROSAT and ASCA
\citep[e.g.][]{Briel_91, Markevitch_96} and more recently with
Chandra. This latter observation revealed an even more irregular X-ray
morphology, the presence of a third subgroup to the east, and a sharp
edge to the southeast of the northwest peak that has been identified
with a ``cold front''\citep{Sun_02}.
  
The \xmm{} temperature map of Abell~2256 in \fig\ref{tmaps_fig} shows
a bimodal temperature structure along the cluster major elongation
axes. Consistent with \citet{Sun_02}, we find that the gas in the
northwest peak, is the coldest of the cluster with a temperature of
$\kt \simeq 4.5 \kev$.  Unlike previous work, our temperature map also
shows a clear, hot ($\kt \simeq 9 \kev$), bow-like region to the east.
This hot region is located just outside an abrupt variation of the gas
surface brightness, also clearly visible from the compression of the
isocontour levels in \fig\ref{tmaps_fig}. As shown in
\part \ref{cfront}, this new feature is likely to be related to
another cold front in the cluster ICM.

\begin{figure*}[ht]
\begin{center}
\begin{tabular}{ll}
\vspace{.25cm}

\hspace{.1cm} \resizebox{.34\hsize}{!}{\includegraphics{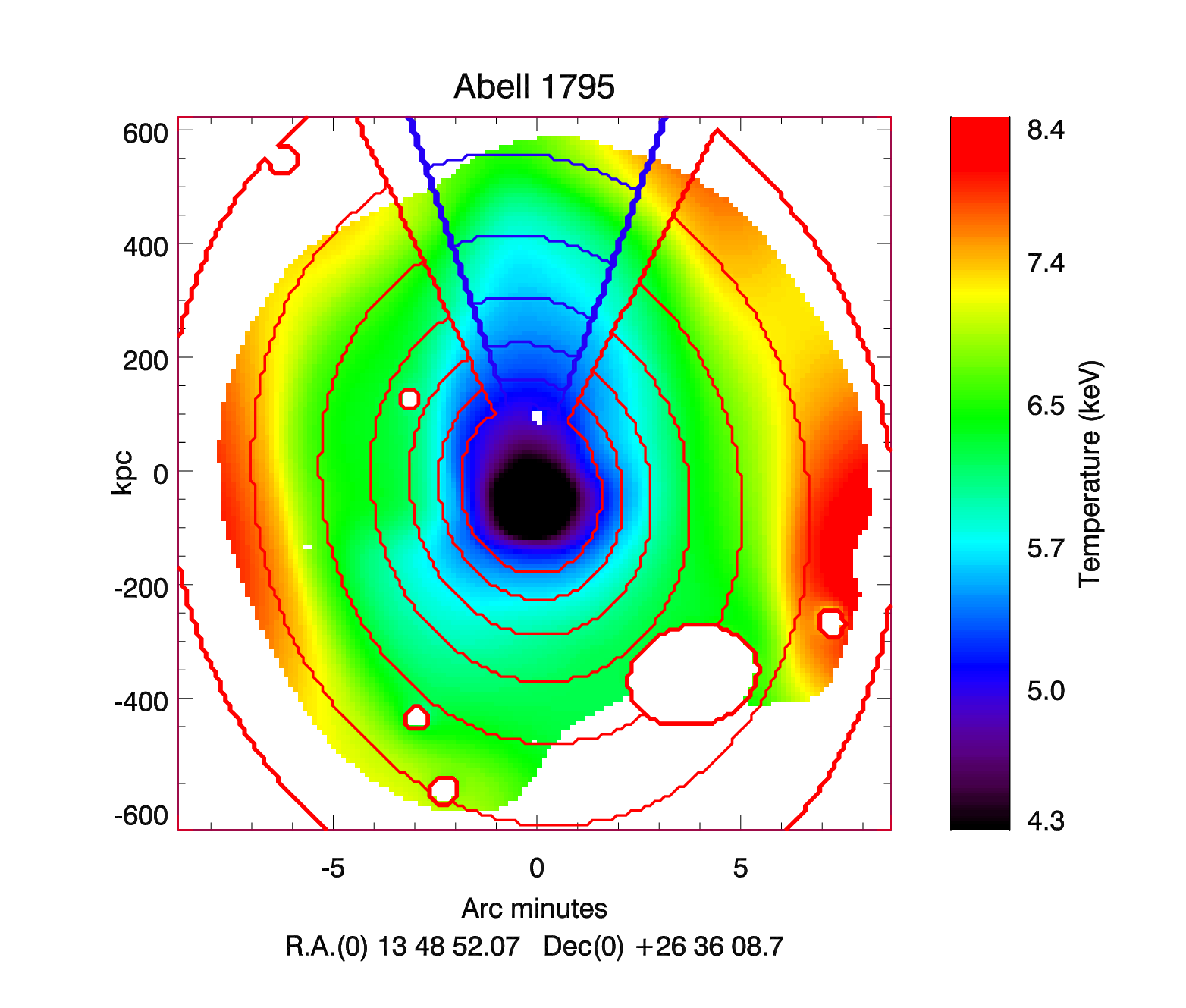}}
& \hspace{.1cm}
\resizebox{.34\hsize}{!}{\includegraphics{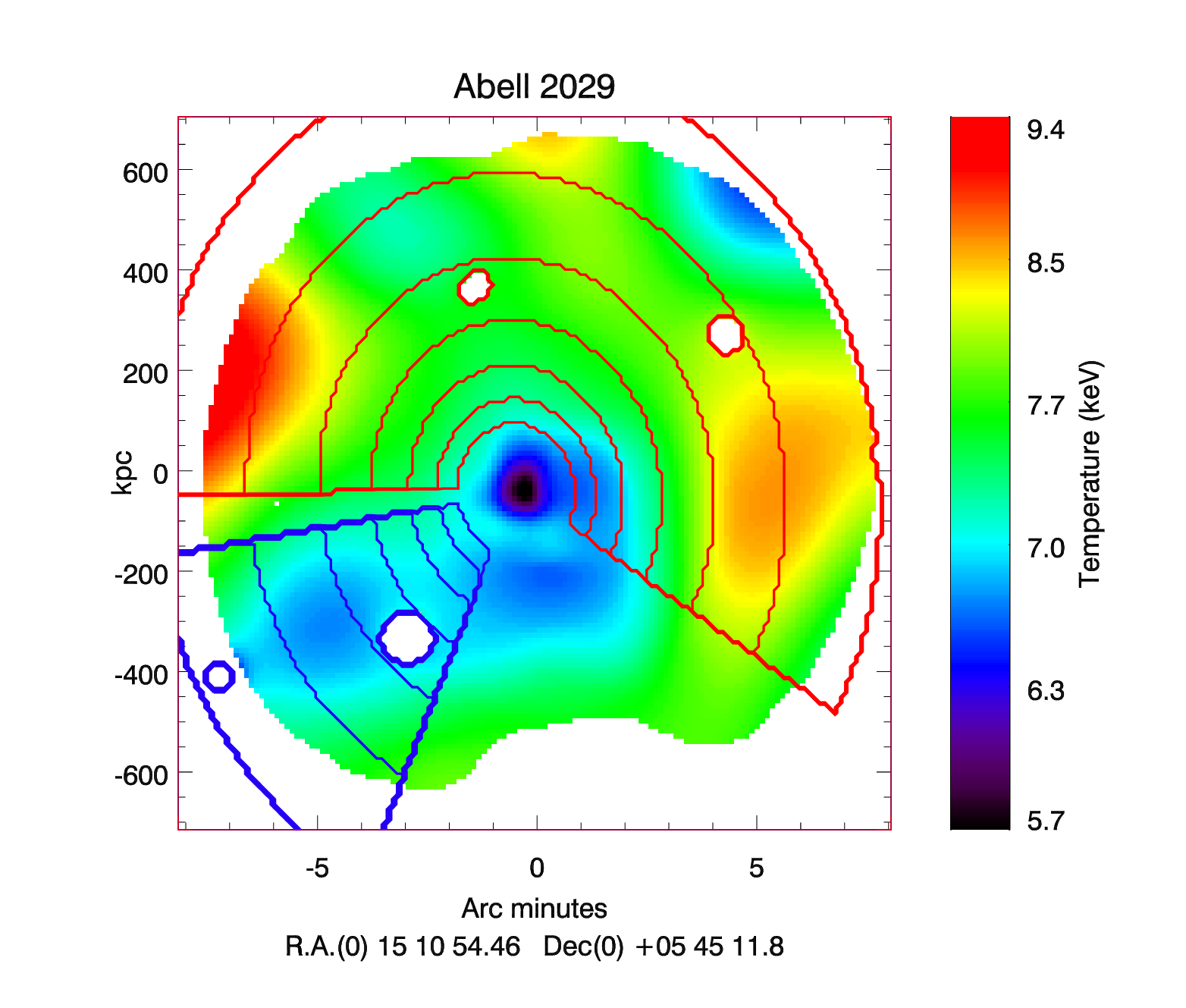}}
\\

\vspace{.25cm}
\resizebox{.34\hsize}{!}{\includegraphics{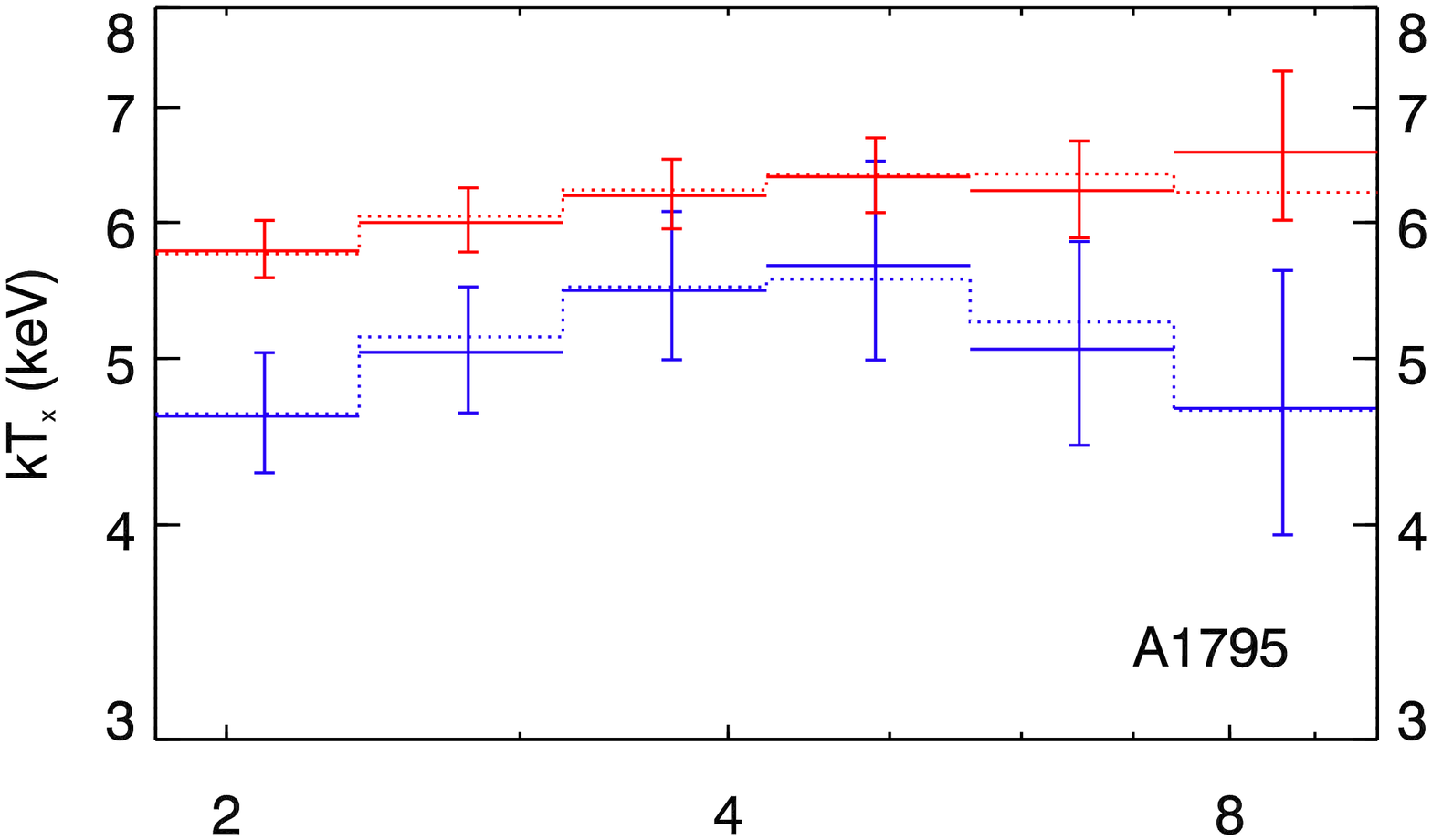}}
&
\resizebox{.34\hsize}{!}{\includegraphics{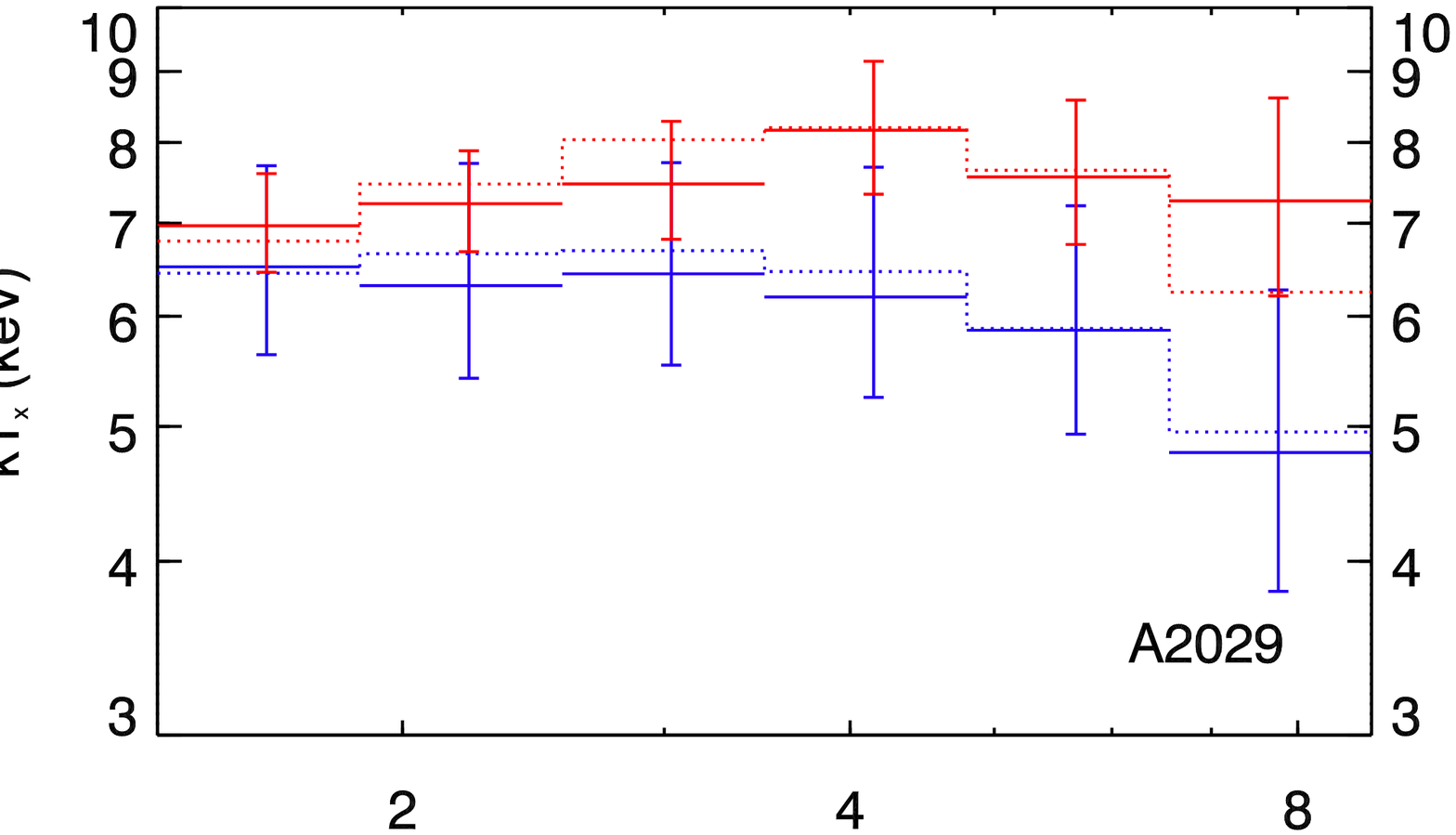}}
\\

\vspace{.5cm}
\resizebox{.34\hsize}{!}{\includegraphics{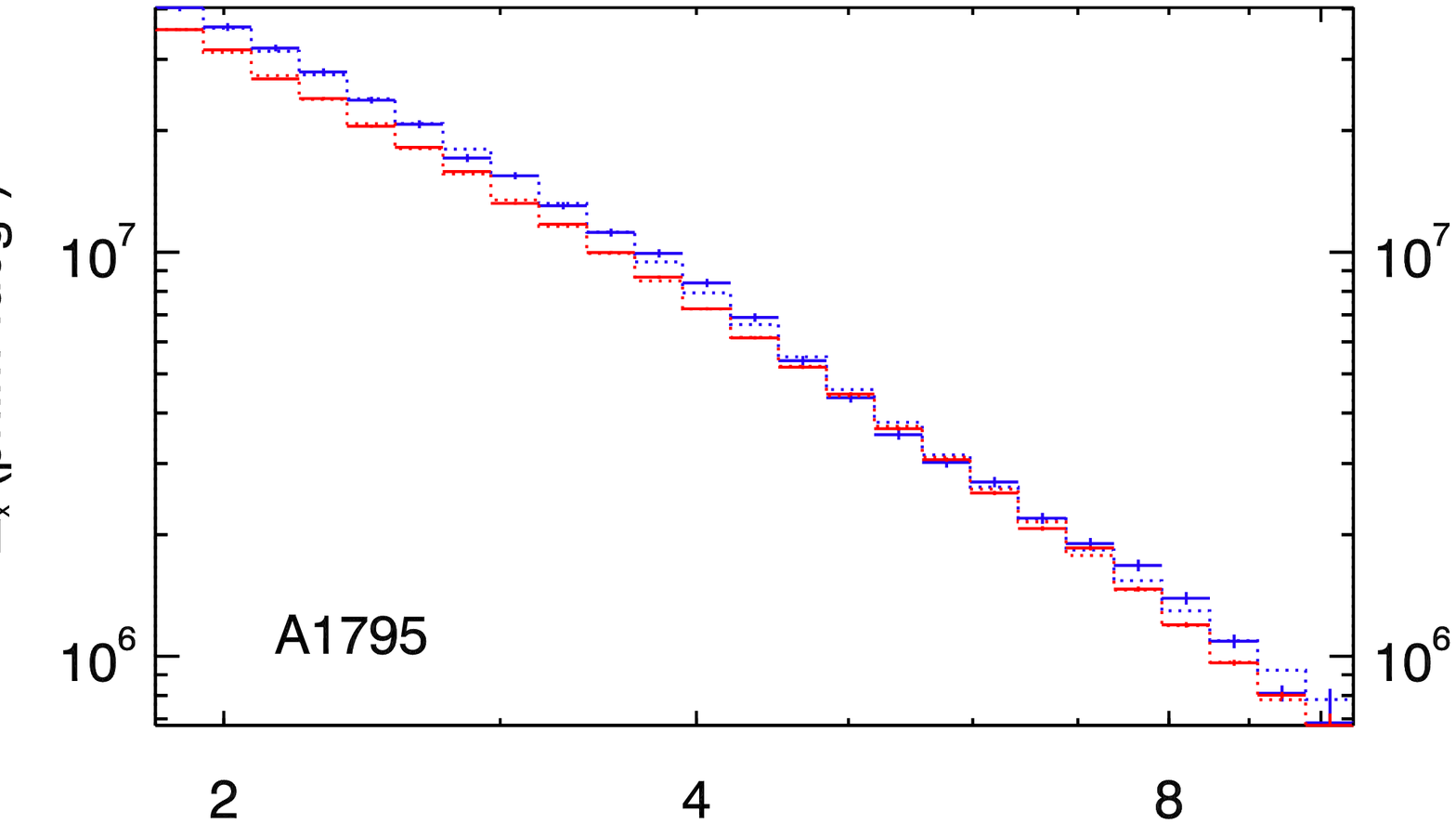}}
&
\resizebox{.34\hsize}{!}{\includegraphics{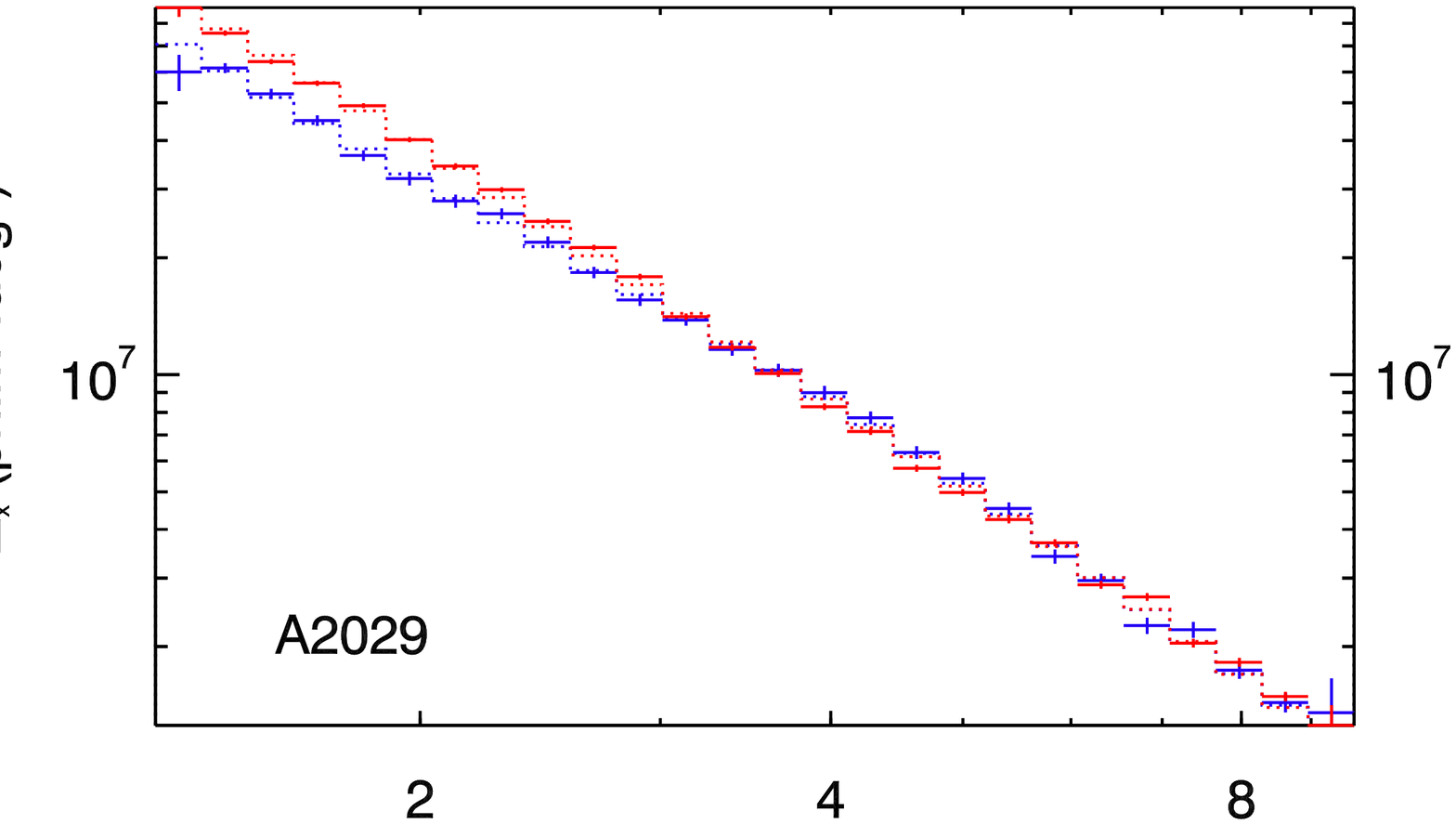}}
\\
\vspace{.2cm}

\resizebox{.34\hsize}{!}{\includegraphics{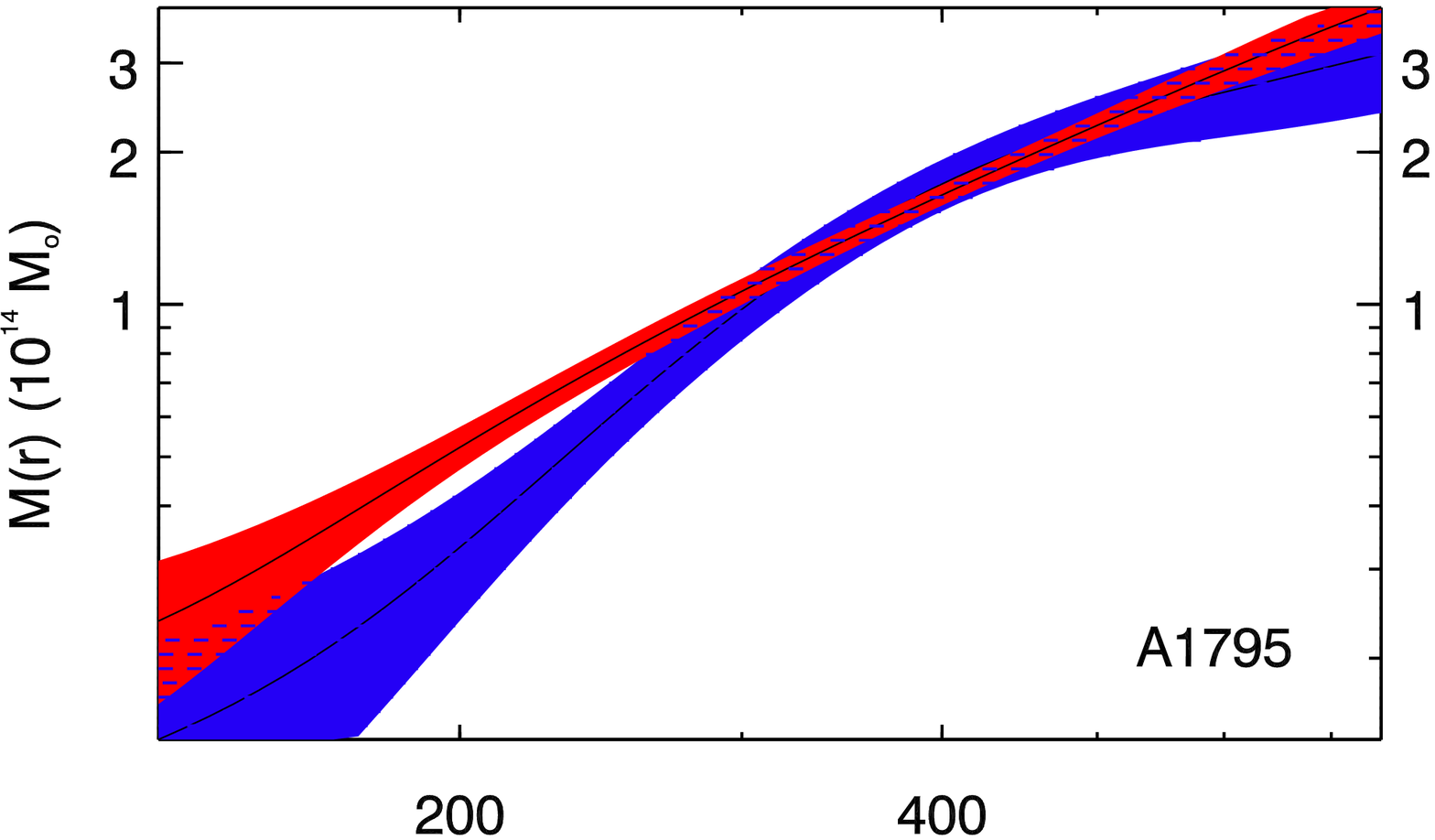}}
&
\resizebox{.34\hsize}{!}{\includegraphics{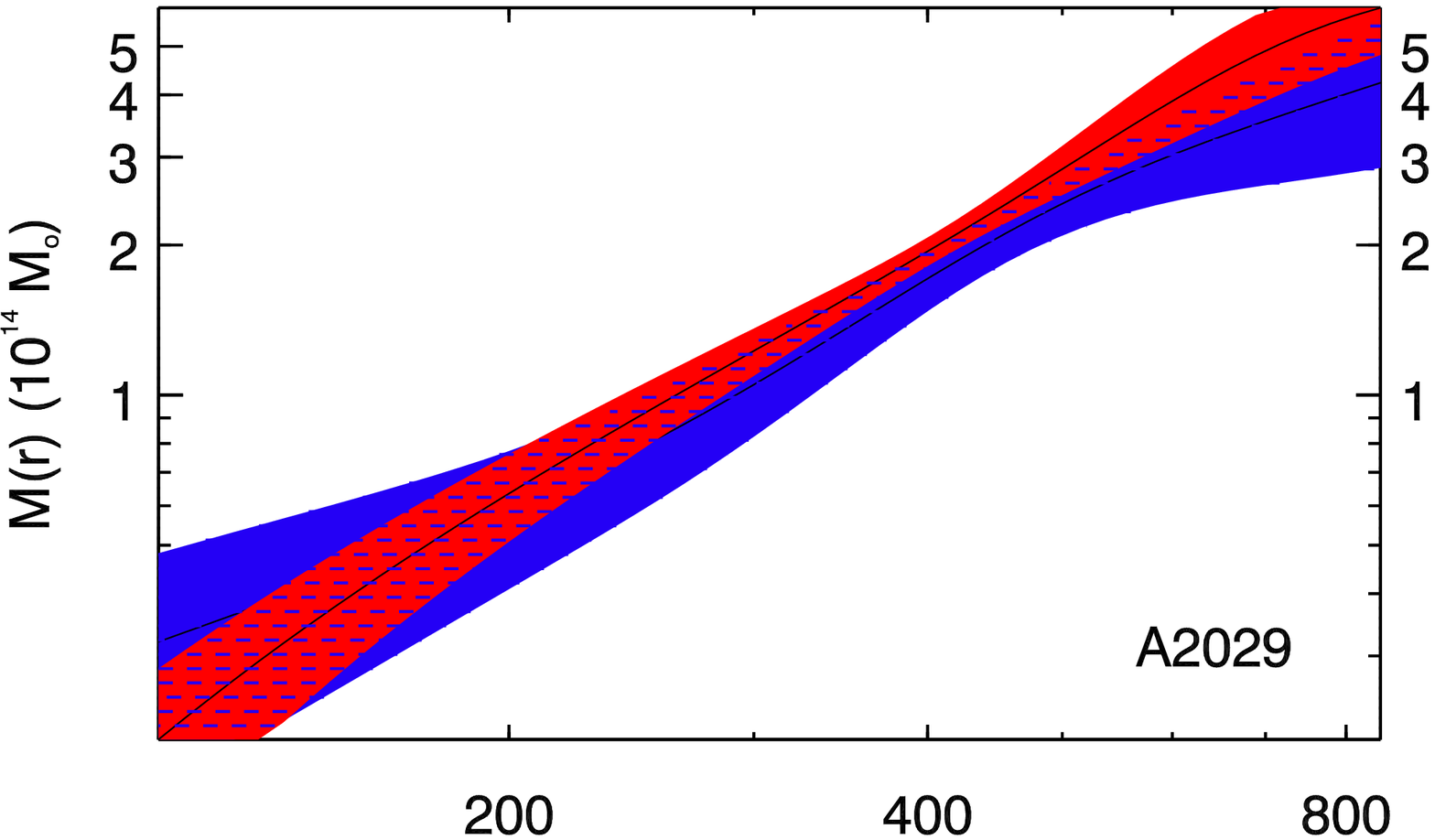}}
\\

\end{tabular}
\caption{Spectroscopic temperature (top), brightness (middle) and
derived mass profiles (bottom) corresponding to the selected red and
blue sectors of Abell~1795 (left) and Abell~2029 (right),
respectively. Dashed lines on temperature and brightness profiles
correspond to fits of the projected functions $\{\tx(r), \sx(r)\}$
used for deriving masses, see \equ(\ref{tsl_equ}) and
(\ref{sigma_x_equ}). Error bars on temperature profiles are 68 $\%$
confidence levels. Dispersions of brightness and mass profiles are
distribution variances. \label{tprofs}}
\end{center}
\end{figure*}

\section{Radial temperature structure and cluster mass\label{tprofs}}

X-ray emission from round relaxed clusters of galaxies is often used
to estimate the cluster mass, assuming hydrostatic equilibrium and
spherical symmetry.  Using hydro N-body simulations, \citet{Rasia_06}
investigated the accuracy of the mass estimate and found that, in the
best of cases, there is at least a 10\% discrepancy between the true
and the estimated mass. They claimed that one of the reasons for such
a discrepancy is related to small, non radial thermal substructure,
which seems to be always present, even in the most relaxed systems
produced by any hydro N-body simulation.

To test whether this problem may also be present in real clusters, we
used our temperature maps and selected specific cluster sectors in
which the gas temperature is either hotter or colder than the gas mean
temperature. From these sectors, we extracted temperature and
brightness profiles, then used these profiles to estimate the relative
cluster mass profiles.

For this test we consider only two of the three more relaxed clusters
in our cluster sample: Abell~2029 and Abell~ 1795. Abell~ 478, the
third round cluster in our sample, was not considered for such a test
because, as explained in \part\ref{nh_part}, it is located in a
particular sky region characterised by strong angular variations of
the neutral hydrogen column density, $\nh$, across the cluster field
of view.  If not treated properly, these strong $\nh$ variations may
introduce spurious thermal features that cannot be easily disentangled
from the real ones.

\subsection*{ICM surface brightness and temperature profiles}

Brightness and temperature profiles associated with hot and cold
regions of temperature maps for Abell~2029 and Abell~ 1795 are shown
in \fig\ref{tprofs} with different colours.

The brightness profiles $\sx(r)$ have been obtained by averaging
the brightness estimator of \equ(\ref{brightness_equ}) within
logarithmically equispaced elliptical annuli of N pixels $\{k,l\}_r$,
with normalisation by the exposure weighted instrument area $\vf(k,l)$
of \equ(\ref{weigthed_area_equ}):

\begin{eqnarray}
  \widehat{\sx}(r) &=& \frac{1}{\N}\sum_{\{k,l\}_r} \widehat{\Lx}(k,l), \nonumber \\
  &=& \frac{1}{\N} \sum_{\{k,l\}_r} {\frac{\nf(k,l) - \nb(k,l)}{\vf(k,l)}},
\end{eqnarray}

We then estimated the brightness variance by considering uncertainties
to be only related to the Poisson fluctuation of $\nf(k,l)$:

\begin{eqnarray}
  \widehat{\sigma_{\sx}}(r) \simeq \frac{1}{\N} 
  \sqrt{\sum_{\{k,l\}_r}{\frac{\widehat{\sigma_{\nf}^2}(k,l)}{\vf^2(k,l)}}}
  = \frac{1}{\N} \sqrt{\sum_{\{k,l\}_r}{\frac{\nf(k,l)}{\vf^2(k,l)}}},
\end{eqnarray}

The spectroscopic temperature profiles $\widehat{\tx}(r)$ have been
estimated by fitting the spectral model $\f_{k,l}(\t,\ab,\nh,e)$ of
\equ(\ref{global_spectra_equ}) to the data associated with a set of
logarithmically equispaced sector annuli. Given the high statistics
available for each annulus, we were able to perform spectral
estimations following a $\chi^2$ minimisation process, and get
straightforward estimations of confidence intervals.

\subsection*{Cluster mass profiles\label{mass_profiles}}

The surface brightness and temperature profiles extracted from the
hottest and coldest sectors of A1795 and A2029 (see \fig\ref{tprofs})
have been used to estimate the cluster mass profile. To do so, we
adopt the approach proposed by \citet{Vikhlinin_06}, that consists in
modelling the 3-d density and 3-d temperature profiles and fitting the
projected quantities to the corresponding data set.

The 3-d gas density is modelled by a double and modified $\beta$-model
with 9 parameters. Introducing $n_p$ and $n_e$, the proton and
electronic density, respectively, we get:

\begin{eqnarray}
  [n_p n_e](r) &=& n_{0}^2
  \frac{(r/r_c)^{-\alpha}}{[1+(r/r_c)^2]^{3\beta - \alpha/2}}
  \frac{1}{[1+(r/r_s)^3]^{\epsilon/3}} \nonumber \\ &&+ \frac
  {n_{02}^2}{[1+(r/r_{c2})^2]^{3\beta_2}}.
  \label{rho3d_equ}
\end{eqnarray}

The projected surface brightness profile, ${\Sigma}(r)$, as observed
within a given energy band $\Delta E$ (here 0.7-2.5 $\kev$), is then
obtained by integrating the ICM brightness,
$\epsilon_{\mathrm{ICM}}(\t)$, along the line of sight:

\begin{equation}
  {\sx}(r) = \frac{1}{d^2 (1+z)^4} \int \epsilon_{\mathrm{ICM}} \left[\t(r_l)\right] ~
  [n_p n_e] (r_l) ~dl,
  \label{sigma_x_equ}
\end{equation}

where $z$ is the redshift of the source located at distance $d$, where
$r_l = \sqrt{r^2+l^2}$, and where $\epsilon_{\mathrm{ICM}}(\t)$ can be
modelled from the source radiation power of \equ(\ref{spectra_equ}),
so as to account for instrument response and effective
area. Normalising with the same average quantities, $\t_o$, $\ab_o$,
$\nh_o$, as used for computing the weighted instrument area in
\equ(\ref{weigthed_area_equ}), we get:

\begin{equation}
  \epsilon_{\mathrm{ICM}}(\t) = \frac{\int_{\Delta E} \s(\t,\ab,\nh,e) ~de}{\int_{\Delta E} \s(\t_o,\ab_o,\nh_o,e) ~de}.
\end{equation}

The 3-d temperature profile is modelled by a 5-parameter broken
power-law, with transition region:

\begin{equation}
  \T(r) = \T_o \frac{(r/r_t)^{-a}}{[1+(r/r_t)^b]^{c/b}},
  \label{t3d_equ}
\end{equation}

then integrated along the line of sight to get a projected profile of
 ``spectroscopic-like'' temperatures, as defined in \equs (14) and (15)
 of \cite{Mazzotta_04}:

\begin{equation}
  {\tx}(r) = \frac{1}{\int w(r_l)~dl} \int w(r_l) \T(r_l) ~dl,
  \label{tsl_equ}
\end{equation}

with weighting factor $w(l)=\frac{n^2(l)}{\T^{3/4}(l)}$.

After fitting the projected ICM brightness ${\sx}(r)$, and
temperature profile ${\T_{s}}(r)$ to the observable set
$\{\widehat{\sx}(r), \widehat{\tx}(r)\}$, and estimating the
related 3-d density $\rho(r)$, and temperature $\T(r)$, we use
 hydrostatic equilibrium of the ICM to derive mass profiles
(see e.g. \citet{Sarazin_88}:

\begin{equation}
  \M(r) = -3.68 \times 10^{13} \M_{\odot} \T_(r) r \left[ \frac{d \log
  \rho}{d \log r} + \frac{d \log \T}{d \log r}\right]
\end{equation}

Best fits of the projected ICM brightness and temperature profiles
with associated cluster mass profiles $\M(r)$, are shown in
\fig\ref{tprofs}. The confidence intervals on mass profiles have been
estimated by minimising the distance between the projected models and
a set of random realisations of observed profiles. These profiles have
been obtained assuming Gaussian statistics around observed values
$\{\widehat{\sx}(r), \widehat{\tx}(r)\}$, with the constraint of
rejecting realisations leading to the non-physical solution of
non-monotonically increasing mass profiles.

\subsection{Abell~1795}

The Abell~1795 ICM temperature map reveals a cold region to the north
of the cluster, while the gas temperature is observed as hotter
elsewhere in the 2 to 5.5 arcmin range of cluster radii. We selected
two complementary ``cold'' and ``hot'' cluster sectors in order to
compute surface brightness and temperature profiles in
\fig\ref{tprofs}. As we want to ignore effects of gas thermal
variations near the cluster core, we excluded this region from our
analysis and computed profiles within a radii range of $1.5-8$
arcmin. We further centred sectors so as to fit the cluster
brightness isophotes at large radii, therefore the cool core appears
as being shifted with regard to our sectors in \fig\ref{tprofs}.

Consistent with what is observed on the temperature map, we notice
that the absolute temperatures in the two sectors are significantly
different.  Furthermore, we observe that the two temperature profiles
show different shapes: the profile corresponding to the coldest sector
peaks at approx 4 arcmin, while the complementary hottest profile is
much flatter. The two complementary brightness profiles also show
different shapes.

It is worth noticing that the resulting mass profiles show
significantly different shapes (see \fig\ref{tprofs}).  In particular,
as for the temperature profiles, the mass profile of the coldest
sector shows stronger gradient variation than the hottest one. These
different shapes result in the observation that while the relative
mass estimates are consistent with each other at large radii ($r > 250
$ kpc) --and also consistent with previously published profiles of
e.g. \citet{Ikebe_04} or \citet{Vikhlinin_06}-- they become
significantly different towards the innermost part of the cluster ($r
< 250 $ kpc).

We should conclude that, within the radii range investigated, the
assumption of hydrostatic equilibrium for this cluster may be not valid.

\subsection{Abell~2029}

The Abell~2029 ICM temperature map reveals a non symmetric temperature
structure within a region ranging from 2 to 5 arcmin cluster radii.
We selected the coldest region to the southeast for computing a first
set of brightness and temperature profiles, and a complementary hot
sector for computing the second set.

As with Abell~1795, the discrepancy observed between temperature
values on profiles in \fig\ref{tprofs} is consistent with the radial
thermal structure of the temperature map. In particular, the profile
corresponding to the cold sector is almost flat ($\kt \simeq 6.5
\kev$), while the profile corresponding to the hot sector has a
positive gradient up to approx 4 arcmin radius (where it reaches $\kt
\simeq 8.5 \kev$), then decreases down to $\kt \simeq 6 \kev$ at
larger radii.  It is important to say, however, that due to the
limited statistics available for fitting temperatures in the coldest
region of A2029, the temperature discrepancies we find are only
marginal for most of the radial bins. However, as with A1795, we find
that the surface brightness profiles of the two complementary sectors
have different shapes.

Despite this, and in contrast with what we find for Abell~1795, the
mass profiles derived for the two complementary cluster sector seem to
have remarkably similar shapes that lead to a consistent mass estimate
over the entire radial range considered ($r \in [200, 800]$ kpc). This
estimate is also consistent with previous works of
e.g. \citet{Lewis_03} and \citet{Vikhlinin_06}.

For this cluster, we can say that the temperature and surface
brightness gradients compensate each other very well, leading to the
same mass estimates regardless of the sector used.  Thus, in this
specific case, we may conclude that the hydrostatic equilibrium
assumption is well satisfied.

\section{Cold fronts\label{cfront}}

The temperature maps of both merging clusters Abell~2065 and
Abell~2256 show two hot bow-like regions to the southeast of each
cluster centre, which seem to be located next to abrupt variations of
the gas surface brightness, as visible on photon maps in
\fig\ref{cxmaps_fig} or isocontour levels in \fig\ref{tmaps_fig}. For
this reason, they are likely to be related to cold fronts separating
the dense and moving cluster cores from the hotter ICM of the cluster
outskirts.

\begin{figure*}
\begin{center}
\begin{tabular}{ll}
\vspace{.25cm}

\hspace{.1cm} \resizebox{.34\hsize}{!}{\includegraphics{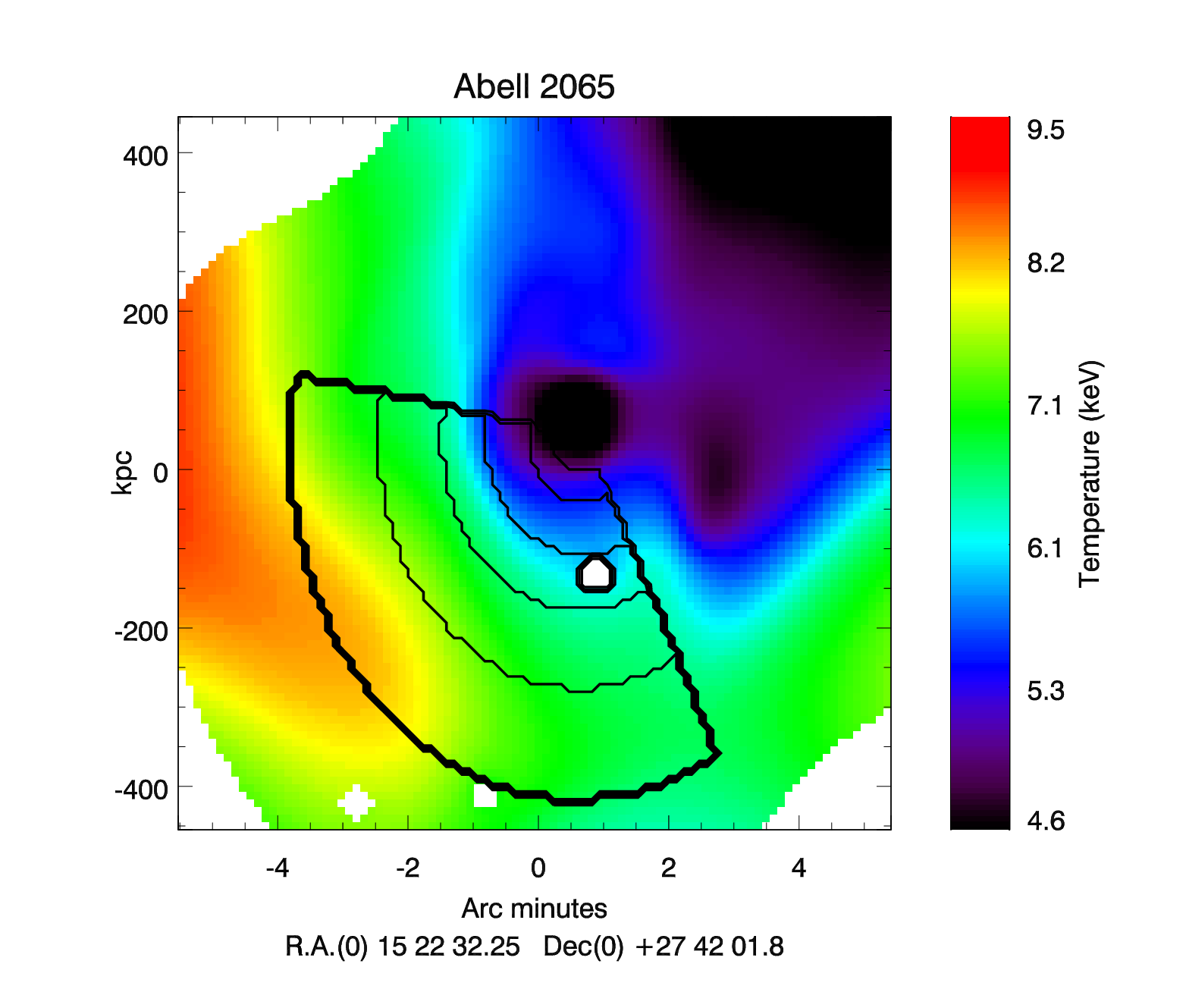}}
& \hspace{.1cm}
\resizebox{.34\hsize}{!}{\includegraphics{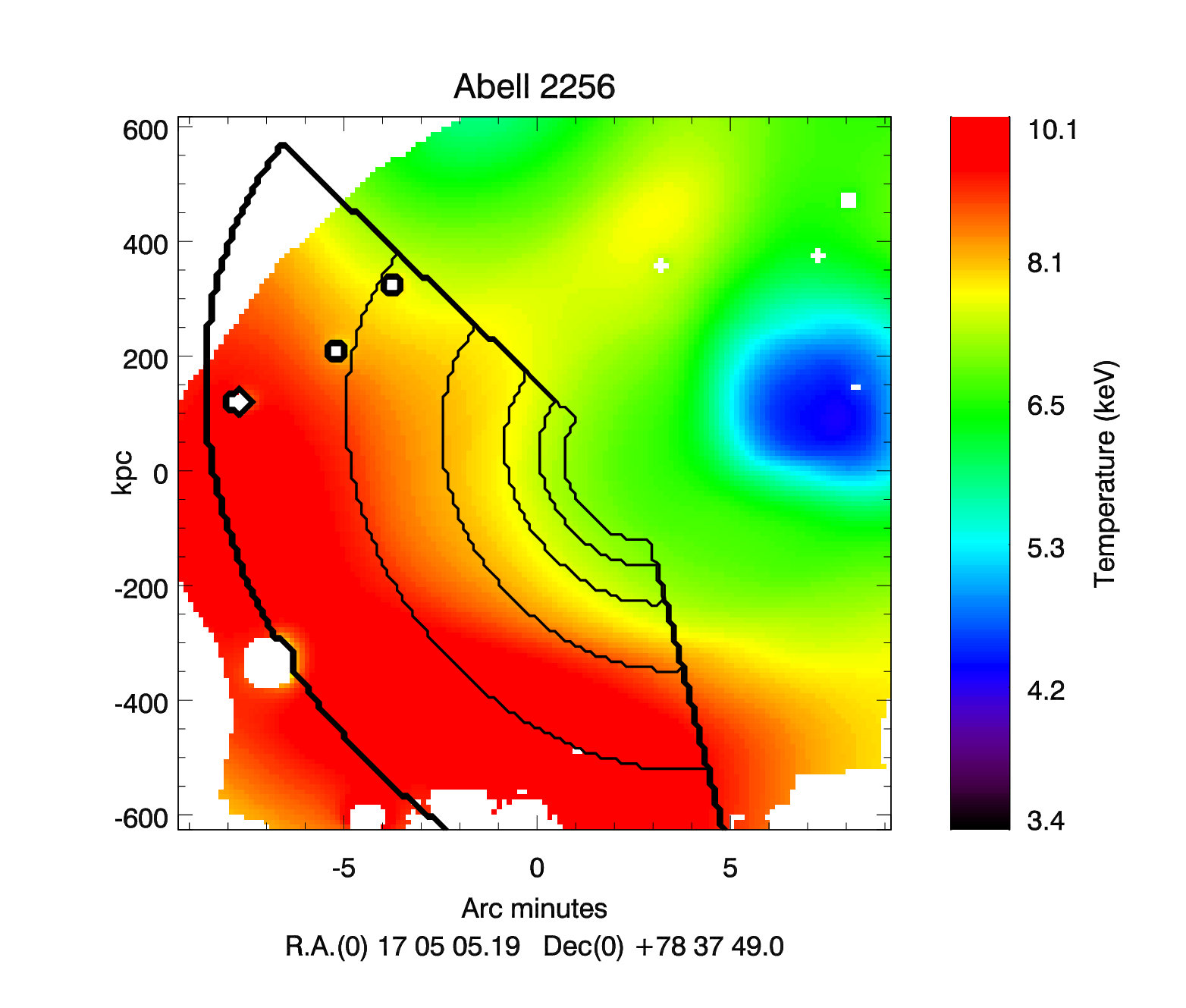}}
\\
\vspace{.25cm}
\resizebox{.34\hsize}{!}{\includegraphics{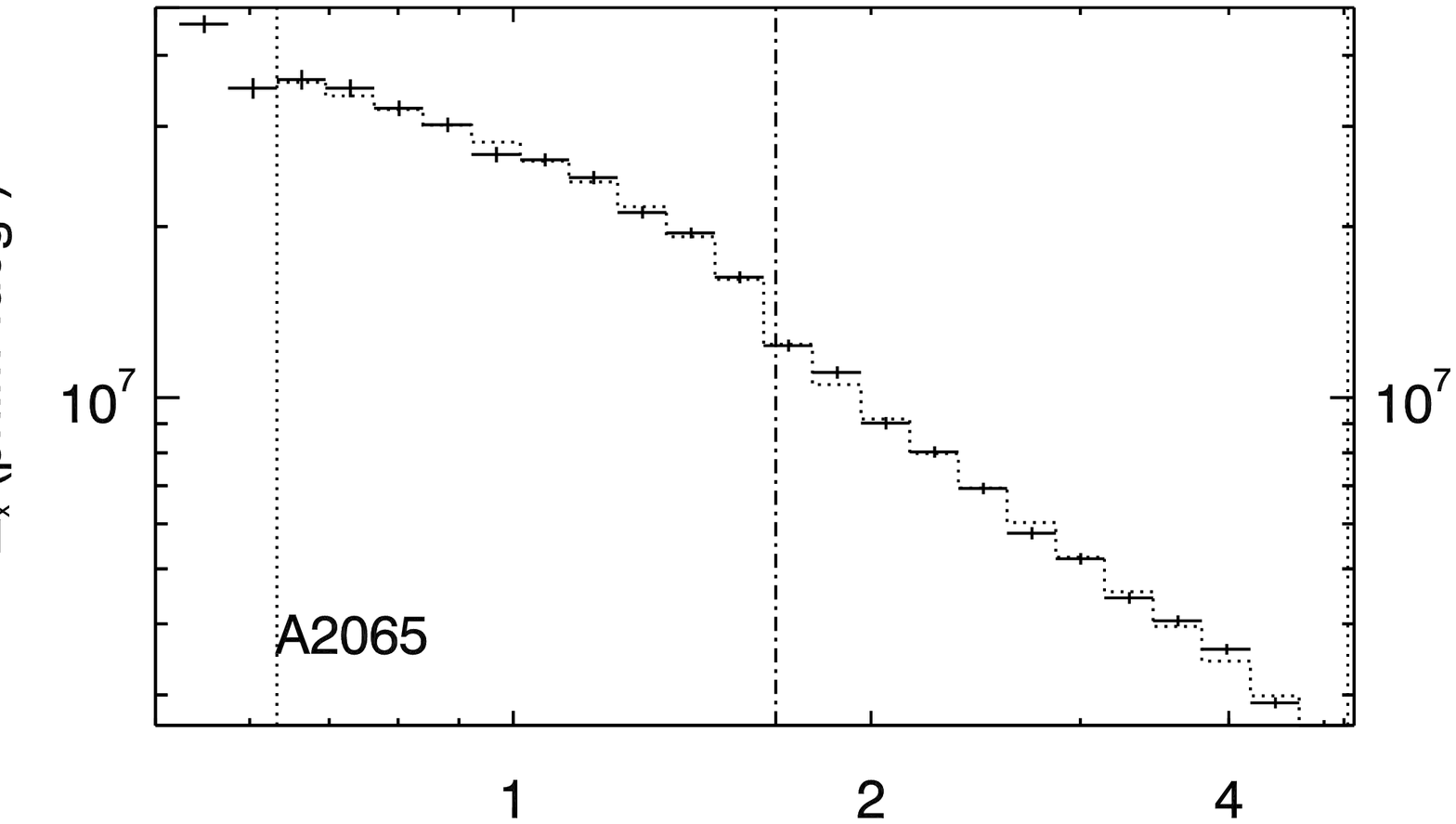}}
&
\resizebox{.34\hsize}{!}{\includegraphics{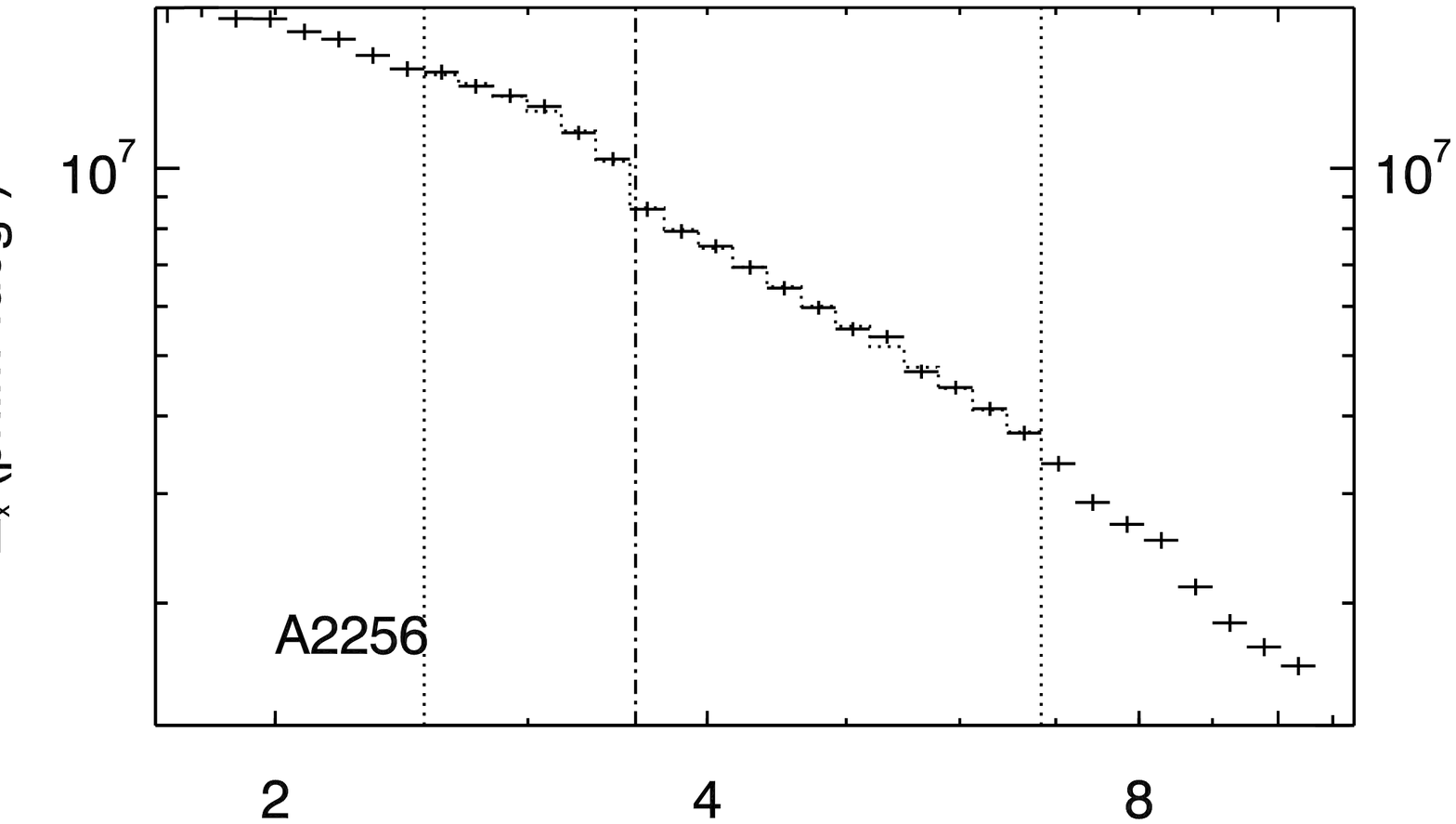}}
\\
\vspace{.5cm}
\resizebox{.34\hsize}{!}{\includegraphics{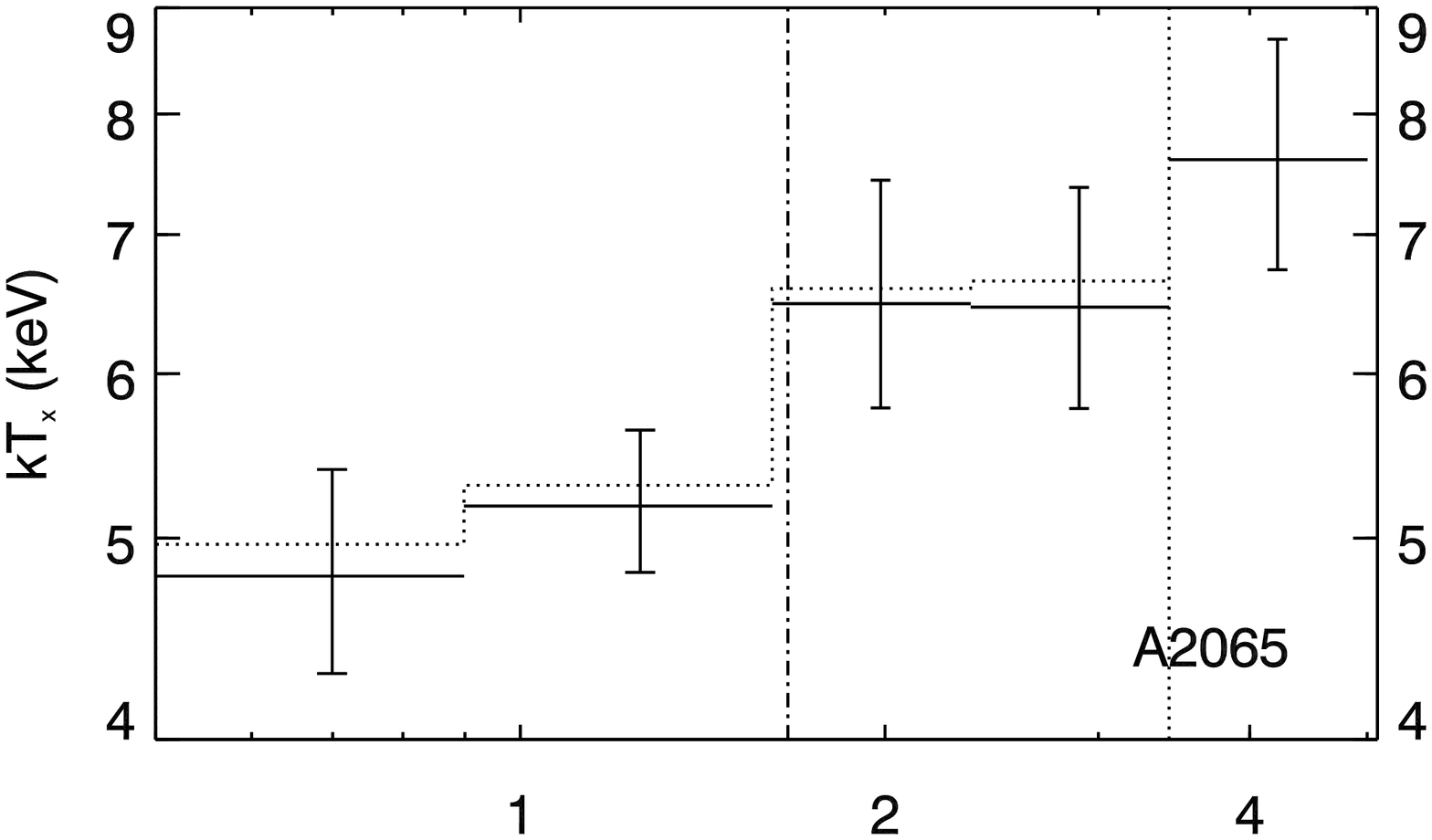}}
&
\resizebox{.34\hsize}{!}{\includegraphics{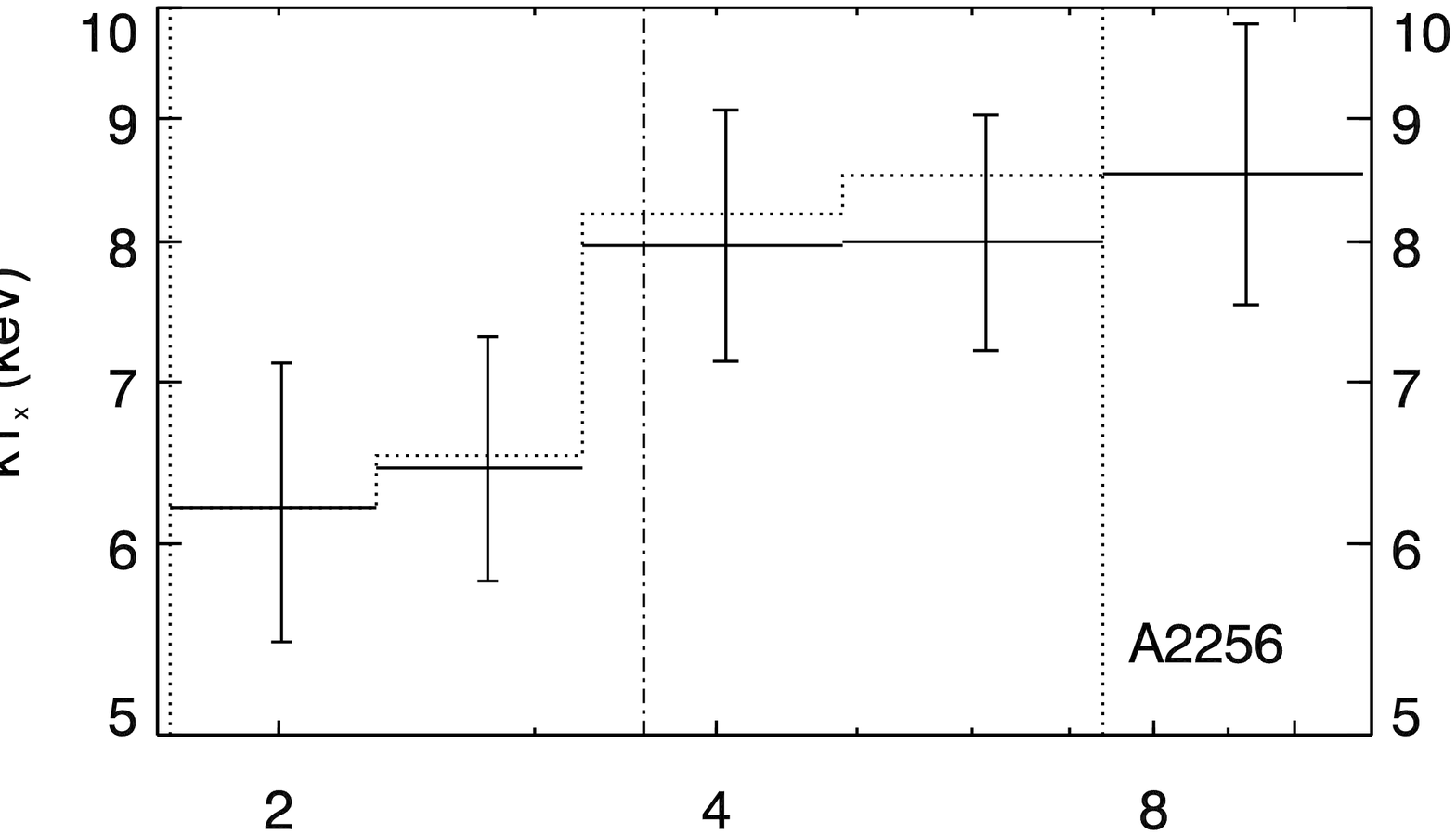}}
\\
\vspace{.25cm}
\resizebox{.34\hsize}{!}{\includegraphics{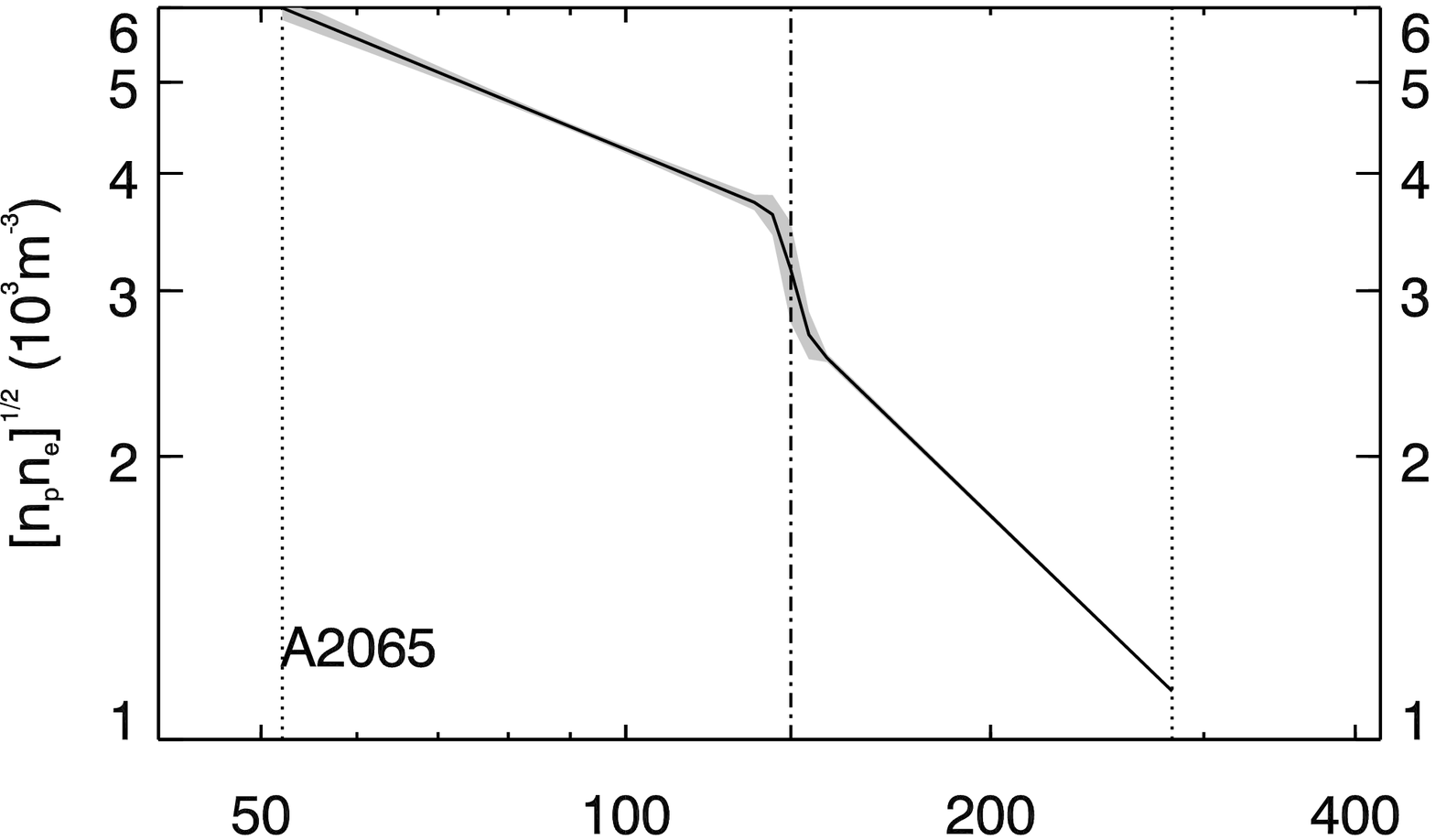}}
&
\resizebox{.34\hsize}{!}{\includegraphics{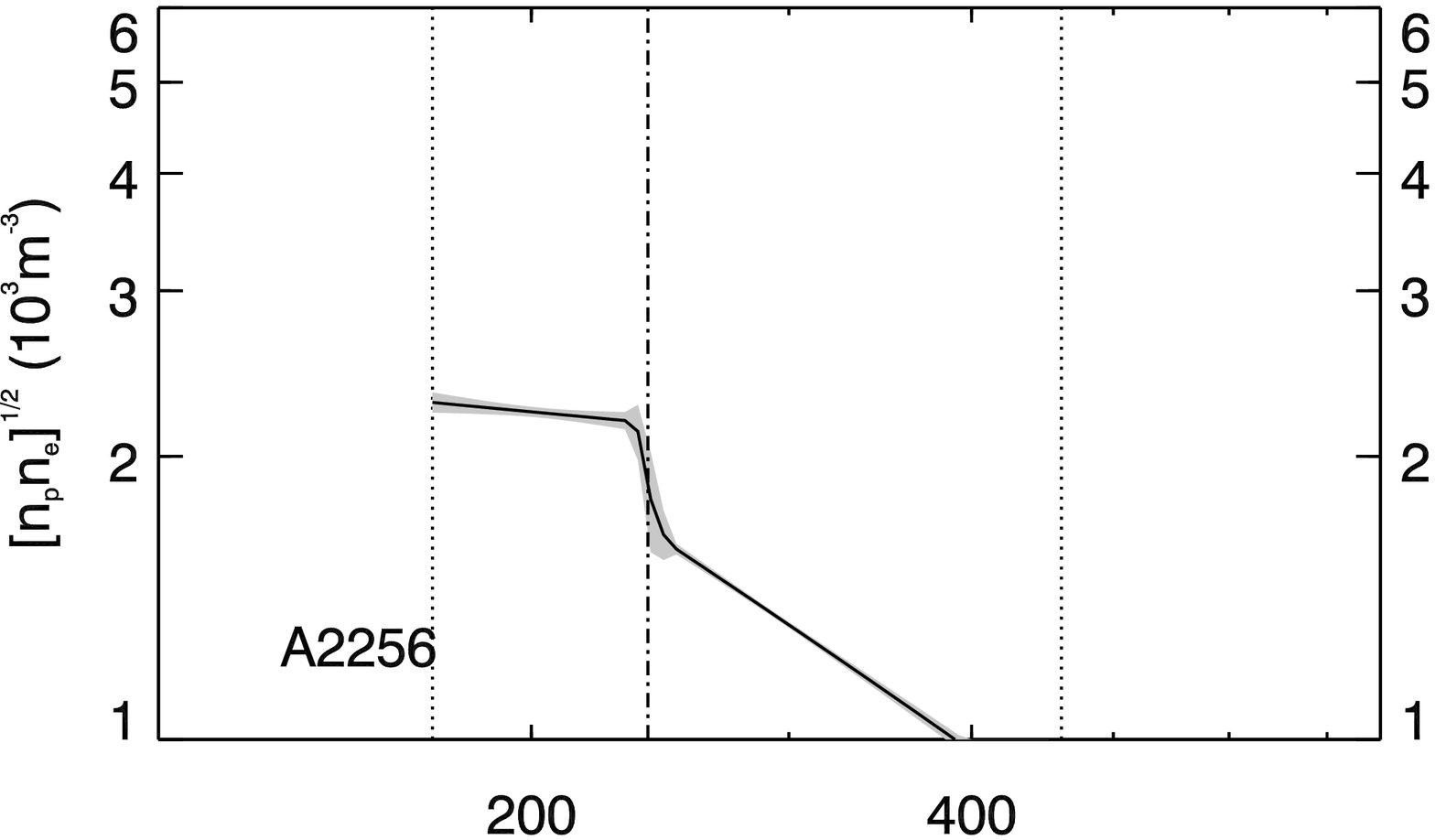}}
\\
\vspace{.25cm}
\resizebox{.34\hsize}{!}{\includegraphics{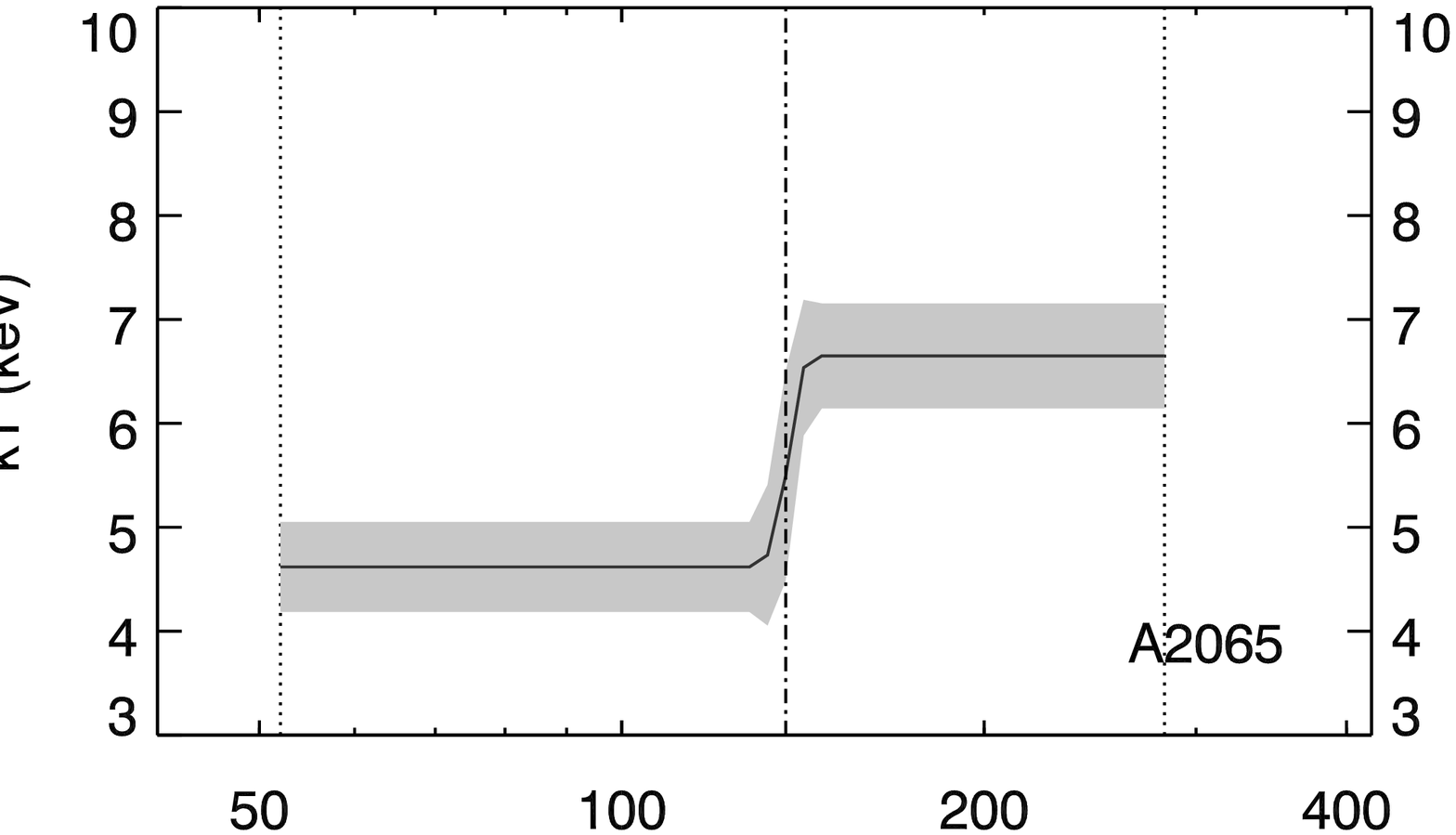}}
&
\resizebox{.34\hsize}{!}{\includegraphics{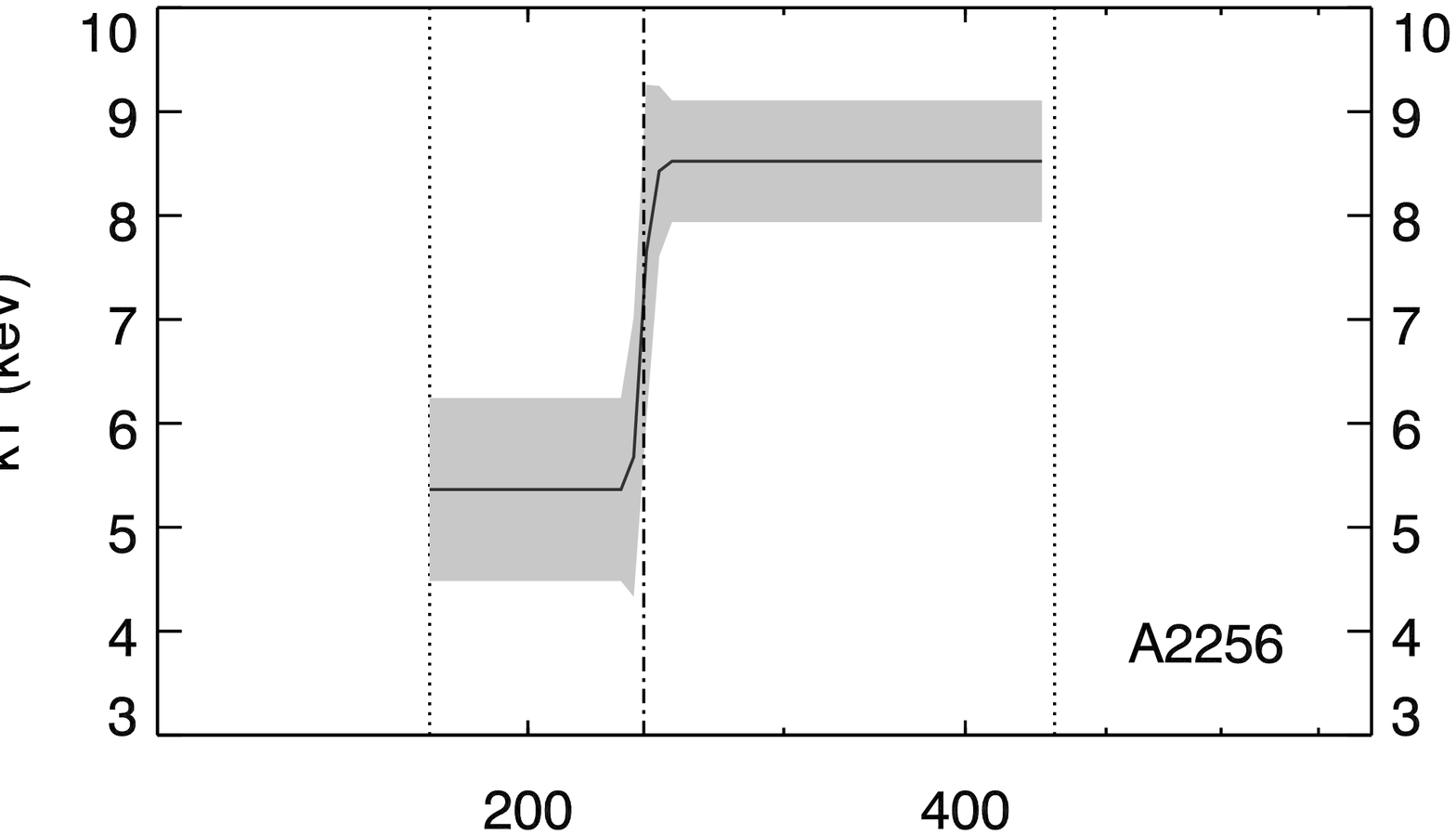}}
\\

\end{tabular}
\caption{From top to bottom: gas brightness, spectroscopic
temperature, and derived density and temperature profiles
corresponding to sectors shown on the top maps for Abell~2065 (left)
and Abell~2256 (right), respectively. Dashed lines on temperature and
brightness profiles corresponds to fits of the projected functions
$\{\tx(r), \sx(r)\}$ (see \part \ref{cfront}). The front modelling and
fitting region is bounded by vertical dashed lines on the gas
brightness and spectroscopic temperature profiles, while the fitted
front position $r_j$ is reported by a vertical dot-dashed line. Error
bars on the temperature profiles are 68 $\%$ confidence
levels. Dispersions reported on the 3d profiles correspond to
variances on each distribution. \label{cf_profs}}
\end{center}
\end{figure*}

In order to investigate this hypothesis, for clusters Abell~2065 and
Abell~2256 we extracted the ICM surface brightness and temperature
profiles corresponding to sectors shown in \fig
\ref{cf_profs}. Located across the brightness front regions, these
sectors follow the brightness isophotes and match regions with almost
uniform isoradial thermal structure.


As expected from temperature and brightness maps, we clearly see two
temperature jumps on the profiles located at a jump radius $r_j$, also
corresponding to a change of slope in surface brightness profiles. In
order to check whether these features can be related to jumps in the 3-d
distributions of the gas density and temperature, we modelled these
distributions by two disrupted functions.

The gas density profile is modelled by two independent $\beta$-models
corresponding to regions located inside and outside the jump radius
$r_j$, respectively:

\begin{equation}
  [n_p n_e]_{cf}(r) = \left\{ \begin{matrix}
    D_{j}^2 n_o^2 (r/r_{j})^{-2\eta_1}
    \left[ \frac {1+\left(r_{j}/r_{c1}\right)^2} {1+\left(r/r_{c1}\right)^2} \right]^{3\beta_1}, 
    ~\forall r \in [0,r_{j}[ \\
    n_o^2 (r/r_{j})^{-2\eta_2}
    \left[ \frac {1+\left(r_{j}/r_{c2}\right)^2} {1+\left(r/r_{c2}\right)^2} \right]^{3\beta_2}, 
    ~\forall r \in [r_{j},\infty[
  \end{matrix} \right. ,
  \label{npne_cf_equ}
\end{equation}

while the gas temperature is modelled by a step function:

\begin{equation}
  \T_{cf}(r) = \left\{ \begin{matrix}
    \T_o, ~~\forall r \in [0,r_{j}[ \\
	D_{\T} \T_o, ~~\forall r \in [r_{j},\infty[ \end{matrix} \right. .
  \label{t3d_cf_equ}
\end{equation}

The 3-d distributions $[n_p n_e]_{cf}(r)$ and $\T_{cf}(r)$ are
projected according to \equ(\ref{sigma_x_equ}) and (\ref{tsl_equ}) in
order to derive X-ray brightness and ``spectroscopic-like''
temperature profiles, $\sx(r)$ and $\tx(r)$, which enables us to
estimate all free parameters of \equ (\ref{npne_cf_equ}) and
(\ref{t3d_cf_equ}), including $r_j$, by fitting the projected models
to the data. Best fits of X-ray brightness and ``spectroscopic-like''
temperature profiles are shown in \fig\ref{cf_profs}, with associated
3-d distributions of gas density and temperature. Similarly to the
derivation of mass profiles in \part \ref{mass_profiles}, the
confidence intervals on 3-d profiles have been estimated by minimising
the distance between the projected models and a set of random
realisations of the observed profiles.

\subsection{Abell~2065}

The Abell~2065 sector is located within 5 arcminutes to the southeast
of the cool elongated central cluster region visible in
\fig\ref{cxmaps_fig}.

Fitting the disrupted density and temperature profile of \equs
(\ref{npne_cf_equ}) and (\ref{t3d_cf_equ}) for the southeastern
sector of Abell~2065 leads to a jump radius of $r_j \simeq
0.137^{+0.002}_{-0.002}$ Mpc, a density jump factor between regions
located immediately above and below $r_j$, of $D_{j} \simeq
1.29^{+0.01}_{-0.01}$, and a temperature jump factor of $D_{\T} \simeq
1.49^{+0.09}_{-0.14}$. These values yield an almost continuous gas
pressure across the front. Interestingly, we notice that the value of
$r_j$ is consistent with the location of a density jump already
reported by \citet{Chatzikos_06}, who performed a sector analysis of
the same cluster region using Chandra data. The additional detection
of a temperature jump using \xmm{} enables us to identify the feature as
a cold front.

\subsection{Abell~2256}


The profiles shown in \fig\ref{cf_profs} correspond to a sector
located within 8 arcminutes to the southeast of the eastern cluster
peak. As with Abell~2065, we detect a cold front feature on the
profiles. The front is located at $r_j \simeq 0.240^{+0.002}_{-0.002}$
Mpc from the sector centre, with density and temperature jump factor
of $D_{j}=1.30^{+0.02}_{-0.02}$ and $ D_{\T}= 1.61^{+0.19}_{-0.15}$,
respectively. Also here, the values of jump factors are consistent
with continuous gas pressure across the front.

\subsection{Possible origin of the fronts}

As shown by \citet{Chatzikos_06} using Chandra data, the cold front
feature seen in Abell~2065 is located to the southeast of a dual
cluster core, which is likely the remnant of a binary interacting
system, now merged. Following this scheme, the cold front may be
related to the motion of the residual cool core of one of the former
interacting clusters, with regard to the hot shocked and mixed gas
of the present cluster.

In the same way, it is worth noticing that the front in Abell~2256 is
located around a complex --at least dual-- cluster core, as revealed
by wavelet analyses of ROSAT PSPC \citep{Slezak_94} and Chandra
\citep{Sun_02} images. Also in this case, we may infer that the cold
front indicates the accretion of a subgroup, the core of which is now
part of the eastern complex system in the cluster.

\section{Neutral hydrogen column density variations across the Abell~478 field of view\label{nh_part}}

\begin{figure*}[ht]
  \begin{tabular}{lll}
    \resizebox{.48\hsize}{!}{\includegraphics{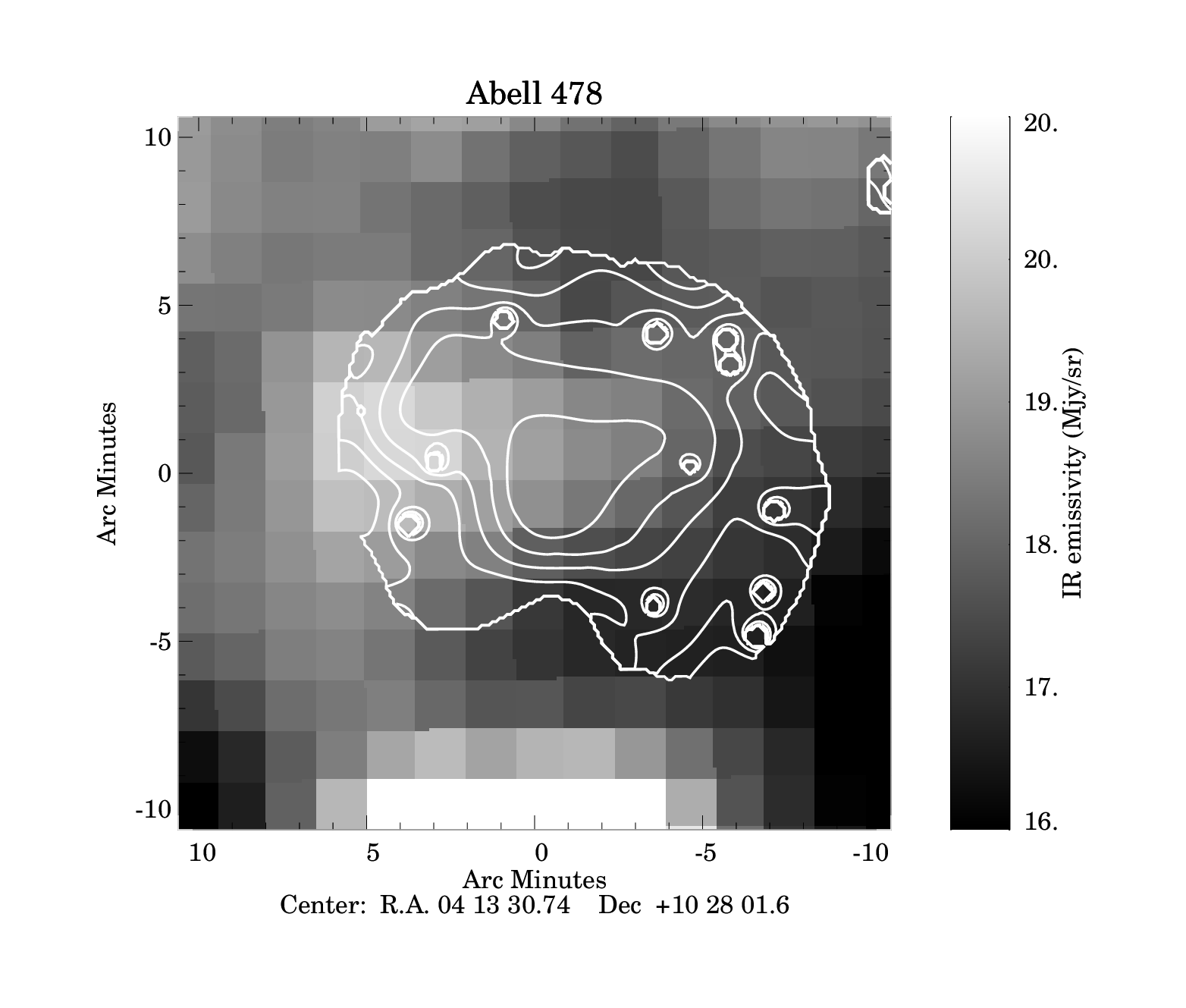}}
    &
    \resizebox{.48\hsize}{!}{\includegraphics{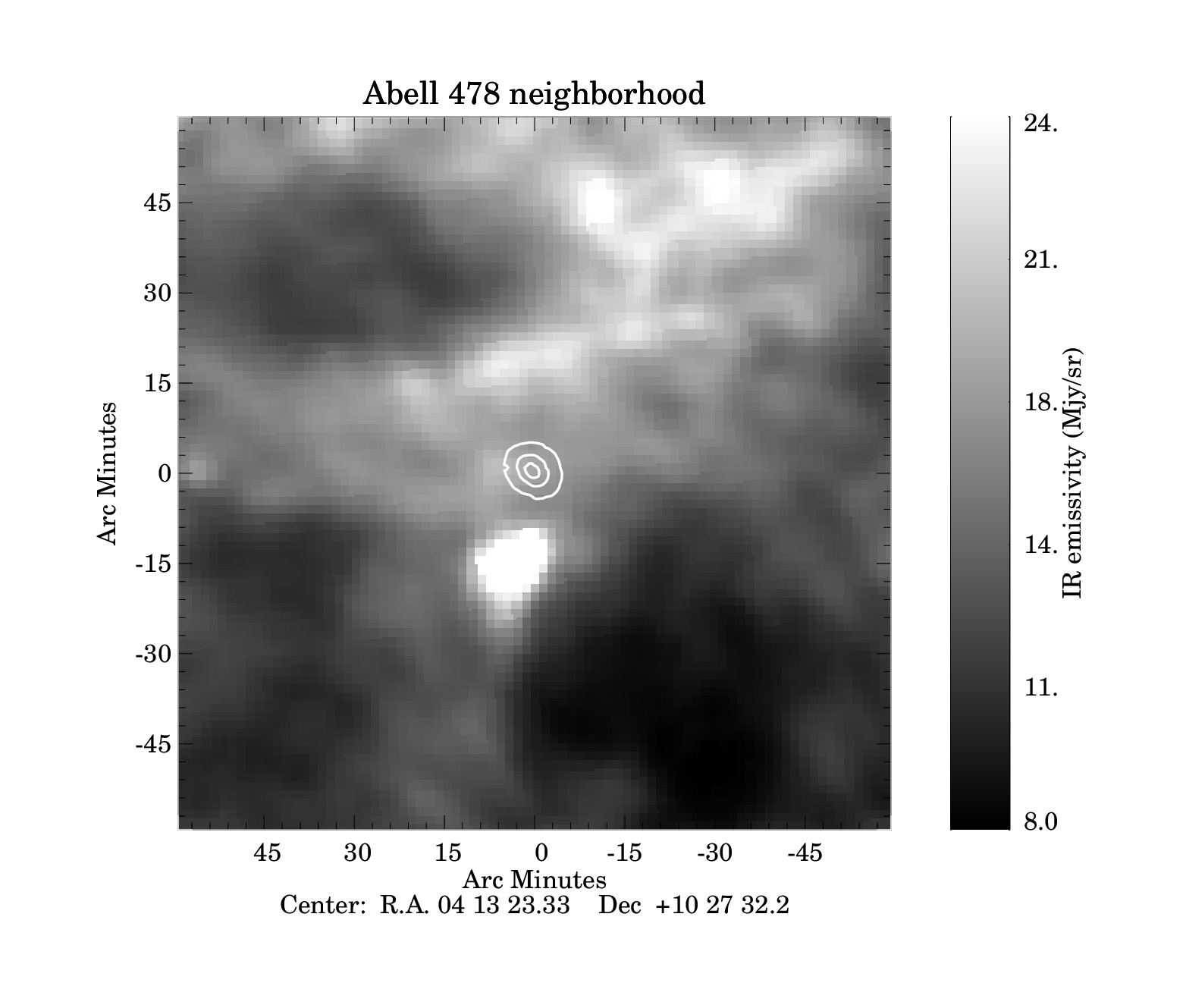}}
    \\
  \end{tabular}
  \caption{Left image: IRAS/IRIS 100 $\mu m$ galactic dust emission
    map across the field of view of Abell~478 overlaid to the neutral
    hydrogen column density estimated by X-ray spectroscopy ($\nh$
    isocontours are equispaced by $2 \times \nhunit$ and decrease from
    a central value of $30 \times \nhunit$). Right image: IRAS/IRIS
    100 $\mu m$ galactic dust emission map of the 2 degree
    neighbourhood of A478, with A478 ICM brightness contours
    overlaid. The IR emissivity of the black region to the south of
    the cluster is of about 50 Mjy/sr. \label{nhmap_fig}}
\end{figure*}

The X-ray emission spectra of extragalactic sources are distorted by
the neutral hydrogen absorption along the line of sight, whose origin
is mostly galactic. We modelled this effect by introducing an
absorption law which is a function of the neutral hydrogen column
density, $\nh$, to the ICM emission models $\nf(k,l) \f(\t,\nh,[e])$,
of \equ(\ref{global_spectra_equ}).  Depending on the knowledge of the
average $\nh$ value in the direction of the source observed, and on
the spatial variation across the source field of view, this parameter
may be fixed a priori or locally estimated from the observed ICM
emission spectra. The average $\nh$ across the field of view of each
cluster in our sample was first estimated by fitting $\nh$ absorbed
emission models to the overall ICM emission spectra. For all clusters
except Abell~478, these overall spectroscopic measurements were
consistent with $\nh$ values measured by \citet{Dickey_lockman_90};
for A478, the spectroscopic $\nh$ was found to be about twice as
high. Due to the inconsistency seen between both of these values, we
left the $\nh$ as a free parameter for X-ray spectroscopy. This
$\t-\nh$ dual parameter spectral-fitting process has enabled us to map
the estimated $\nh$ across the field of view of A478, using a similar
wavelet algorithm to the ICM temperature mapping algorithm described
in
\part\ref{spectral_mapping}.

A map of the spatial distribution of $\nh$ measured by X-ray
spectroscopy across the field of view of A478 is shown by white
isocontours on the left panel of \fig \ref{nhmap_fig}. In order to
reduce a possible spectroscopic bias with brightness gradient, the map
has been obtained from a wavelet analysis limited to scales smaller
than 2 arcmin. Significant wavelet coefficients have been selected
according to a hard thresholding at $2 \times \sigma$. The $\nh$
distribution appears irregular with an east-west elongation, an
extended excess to the centre of the field of view, and a peak to the
to the east. Starting from the central excess ($\nh \simeq 30 \times
\nhunit$), the $\nh$ decreases strongly to the north and south of the
field of view, up to values consistent with the average galactic value
of \citet{Dickey_lockman_90} ($\nh < 20 \times \nhunit$) at a distance
of 5 arcmin from the centre.


The $\nh$ central excess shown on the map of \fig \ref{nhmap_fig} has
already been reported by \citet{Pointecouteau_04}, who performed a
spectroscopic analysis within sectors of the same data set as here. As
already proposed by \citet{Pointecouteau_04}, its origin can be
investigated using the IRAS far-infrared survey of galactic dust
emission at 100 $\mu m$, with angular resolution of about 4
arcmin. Indeed, the far-infrared galactic dust emission can be used as
a tracer of the galactic neutral hydrogen column density, since dust
emission is expected to be correlated to the galactic neutral hydrogen
at high galactic latitude ($|b| > 10 \deg$), as shown by
e.g. \citet{Boulanger_96} and \citet{Schlegel_98}. Notice however that
this correlation is expected to be better constrained within low-$\nh$
regions of the sky ($\nh < 5 \times \nhunit$) than within high-$\nh$
regions like the Abell~478 neighbourhood, since the scatter of the
IR/$\nh$ correlation is higher in such regions, possibly due to the
presence of molecular hydrogen, as discussed by
\citet{Boulanger_96}. A galactic dust emission map at 100 $\mu m$ of
the 2 degree neighbourhood of Abell~478 is shown in the right panel of
\fig \ref{nhmap_fig}. Coming from the last generation of IRAS maps
obtained from the IRIS processing \citep{Miville_Deschenes_05}, this
map reveals the structure of the galactic filament where Abell~478 is
located, to a resolution that cannot be reached by the radio survey of
\citet{Dickey_lockman_90}, with angular resolution of 1 degree. The IR
brightness of this map appears to be strongly structured in this sky
region, with relative flux variations of more than 100 \%. Within a 5
arcmin region centred on the brightness peak of A478, the IR
emissivity is of about 18 Mjy/sr. This value corresponds to a $\nh$
density column of $33 \times \nhunit$ following the IR/$\nh$
correlation factor of \citet{Boulanger_96}. It is consistent with
density column estimations obtained by X-ray spectroscopy, and
encourages us to use the dust emission at 100 $\mu m$ as a tracer of
the galactic neutral hydrogen distribution within this region of the
sky.

We investigated the accuracy of this tracer by superimposing our $\nh$
map to a portion of the IRAS map in the left panel of \fig
\ref{nhmap_fig}. Interestingly, we notice some spatial correlation
between the IR and X-ray estimated $\nh$ map. Indeed, both the central
extended excess and the eastern peak shown on the $\nh$ map correspond
to similar structures in IR. However, both maps are not fully
correlated in flux, since the maximum of the $\nh$ map is located to
the centre of the field of view, while the maximum of the IR map is
located at the eastern peak. Observations of this sky region using
infrared data of higher resolution may help to show conclusively
whether the $\nh$ excess detected near the centre of Abell~478 corresponds
to a real feature, or if it is due to a bias of X-ray spectroscopic
estimations.

\section{Discussions and Conclusions\label{conclusion}}

We present a multi-scale algorithm based on wavelets for mapping the
ICM temperature structure within clusters of galaxies. Following the
approach proposed by \citet{Bourdin_04}, the basic scheme of this
algorithm is i) estimate the searched parameter within square
resolution elements, sampling the field of view at different scales,
ii) filter the analysed signal in order to get wavelet coefficients,
iii) threshold the wavelet transform obtained, in order to de-noise the
signal and map its 2-D structure. In order to perform a more regular
reconstruction of structures and to reduce thresholding artifacts, we
improved the algorithm so that, to filter the signal, we use a
B2-spline wavelet instead of the Haar wavelet used in
\citet{Bourdin_04} (see paragraph {\ref{spectral_mapping}} for
details). This algorithm was used to map the X-ray and temperature
structure of a nearly complete X-ray flux limited cluster sample
containing the eight clusters that, to date, have useful \xmm{}
observations.  For self-consistency of the results, a similar
thresholding approach was adopted for each target.

From previous work, three of the clusters in our sample --namely A2029,
A1795, and A478-- have been identified as relaxed systems, while the
remaining clusters have been identified as major mergers.  This
classification is supported by the power ratios analysis of the
surface brightness of these clusters, obtained from ROSAT observations
\citep{Buote_tsai_96}. The \xmm{} photon images in \fig\ref{cxmaps_fig}
and the surface brightness contours in \fig\ref{tmaps_fig} confirm
this previous classification: the X-ray morphology of the relaxed
systems is quite regular and elliptical symmetric while the morphology
of the merging systems is more complex with the presence of multiple
X-ray peaks.

In \fig\ref{tmaps_fig} we show the temperature maps for all clusters
in our sample overlaid to the soft (0.3-2.5 $\kev$) X-ray brightness
isocontours.  Even if with a very different degree of complexity, we
notice that all clusters, including the most relaxed ones, show non
radial thermal structure. Let us recall that, since for this work we
are only interested in detecting highly significant thermal structure,
we adopted a quite conservative approach which is based on the
\citet{Donoho_95} wavelet shrinkage.  For this reason the angular
resolution of the detected structure appears to be lower than the
angular resolution expected from EPIC \xmm{} data.  A wavelet
shrinkage with constant threshold, as tested in \citet{Bourdin_04} on
simulated observations, or used in \citet{Belsole_04,Belsole_05} and
\citet{Sauvageot_05} on real data, would reveal non radial thermal
structure even on smaller angular scales.

Despite the relatively low angular resolution we find that,
consistent with predictions from numerical hydro N-body simulations,
the complexity of the thermal structure of clusters strongly depends
on the dynamical status of clusters themselves.  From
\fig\ref{tmaps_fig} we can see that the temperature maps of the
clusters in our sample may be classified into three different types,
characterised by an increasing regularity of the thermal structure: 

\begin{itemize}

  \item``irregular'', as for A399 and A401. While selected
  independently in our sample, these clusters actually form a binary
  and mildly interacting system. It has been proposed that the strong
  irregularity of their temperature structure is the result of
  previous merger activities within each cluster of the system,
  independently from the present interaction \citep{Sakelliou_04}.

  \item ``bimodal temperature'', as for A2065 and A2256. These
  clusters are known to show some strong temperature anisotropies and
  to be presently overcoming a late stage of merging. They show an
  elongated geometry of surface brightness, with temperature maps
  characterised by a colder and hotter regions on opposite sides along
  the major direction of elongation. This thermal structure is
  dominated by the contrast between cool features to the one side,
  probably associated with accreted material, and a hotter region to
  the other side, possibly associated with shocked gas --see
  e.g. \citet{Markevitch_99} and \citet{Sun_02}, for A2065 and A2256,
  respectively-- and separated from the cluster core by a cold front.

  \item ``regular'', as for the relaxed clusters, A2029, A1795, and
  A478. These clusters have a characteristic cool core region
  surrounded by a hotter gas annulus region peaking at about 300-400
  kpc from the cluster centre.  As with the surface brightness, the
  temperature map of these clusters is almost regular and elliptically
  symmetric, indicating a good relaxed status for these systems. It is
  worth noting, however, that beyond the overall elliptical symmetry, we
  also detect a number of significant non radial thermal structures
  outside their cores. This result seems to be consistent with what is
  predicted by hydro N-body simulations of cluster formation, where
  clusters continuously accrete small galaxy groups during and between
  major mergers, which lead to temperature irregularities near the
  region where groups are accreted.

\end{itemize}

To conclude, we would like to stress that, despite the large
dispersion of intrinsic brightness and exposure times of the clusters
in our sample, the multi-scale algorithm presented here enabled us to
reveal the thermal structure of the ICM of each cluster. Apart the
detection of strong features mainly present in merging systems
--e.g. cold fronts--, we were also able to detect some mild non-radial
temperature structure outside the core of relaxed clusters. We notice
that the relative temperature variation associated with these
irregularities at fixed radii is about 10 $\%$. We show that these non
radial thermal variations affect the measured radial temperature
profiles (see paragraph \ref{tprofs}), and lead, at least for one of
our clusters --A1795--, to an inconsistency of mass derivations
obtained using hydrostatic equilibrium assumption. Nevertheless, for
another cluster, A2029, the hydrostatic equilibrium assumption appears
to be valid despite the presence of temperature irregularities.

These results enlighten the possibility of constraining the
thermalisation status of the ICM and the departure from gas
hydrostatic equilibrium within bright clusters, using data obtained
with current X-ray observatories. It invites us to further
investigations of ICM temperature anisotropies within larger samples
of relaxed clusters --possibly at even larger cluster radii than here
by assuming specifically an elliptical geometry of the wavelet
analysis-- and to a more accurate evaluation of the dispersions
implied by these anisotropies on cluster mass estimations.

\begin{acknowledgements}      
We thank Albert Bijaoui and Eric Slezak for their contribution to the
conception of the wavelet imaging and spectral-mapping algorithms, and
Jean-Luc Sauvageot for his help in reducing XMM-Newton data. We
further thank the anonymous referee for suggestions and comments that
improved the paper significantly. H.B. acknowledges the financial
support from contract ASI--INAF I/023/05/0. This work is based on
observations obtained with XMM-Newton, and ESA science mission funded
by ESA Member States and the USA (NASA).
\end{acknowledgements}

\appendix

\section{Background modelling for the EPIC cameras 
on board of the \xmm{} satellite \label{b_model_app}}

A typical list of photon impact detection provided by the EPIC CCD
cameras gathers events associated with various processes that have to
be taken into account for modelling the observed spectra of extended
sources. In the case of ICM observations, the detected photons are at
least related to two extended contributions: the observed source
itself and the cosmic X-ray background (CXB). Furthermore, a
significant fraction of the events-list is actually associated with
false detections due to cosmic-ray induced particles interacting with
the detector. Eventually, a known fraction of events is registered
during readout periods of detectors, which leads to an additive noise
due to wrong position registration of these so-called ``out-of-time''
events.

We model, as an overall ``background contribution'' to the event
spectrum $\nf(k,l)~\f(\t,\ab,\nh,e)$ of
\equ(\ref{global_spectra_equ}), a combination of normalised spectral
contributions associated with CXB, $\c(e)$, induced particles,
$\p(e)$, and readout noise, $\o(k,l,e)$:

\begin{eqnarray}
  \nb \b(k,l,e) &=& ~ \ea(k,l,e) \times \nc~\c(e) \nonumber \\ 
  &&+ \no(k,l)~\o(k,l,e) \nonumber \\
  &&+ \np~\p(e),
  \label{background_equ}
\end{eqnarray}

where $\nc$, $\np$ and $\no$ are normalisation terms, and where
$\c(e)$ is corrected for spatially variable exposure $\ea(k,l,e)$ (see
equation \ref{eff_exposure_equ}), as it is related to a physical
observation. The CXB spectrum $\c(e)$ is modelled as the combination
of a soft radiation associated with foreground galactic gas and a
broad band contribution accounting for extragalactic background of
unresolved AGNs. Following \citet{Lumb_02} and \citet{Kuntz_00}, we
use a two temperature thermal radiation ($k\t1 = 0.074 ~ \kev$, $k\t2
= 0.204 ~ \kev$) and a power-law ($\gamma = 1.42$) for modelling the
galactic and extragalactic contributions, respectively. The cosmic-ray
induced particle background $\p(e)$ has been modelled from cumulated
expositions of telescopes to the particles, during in-flight
calibration phases. The readout noise contribution $\o(k,l,e)$ is
estimated by integrating and normalising the overall signal along CCD
columns. While normalisation of the readout noise is known a priori
for a given observing mode, normalisations of both the CXB spectrum
and particle background must be set by fitting our background model,
$\nb \b(e)$, to an observed ``background spectrum''. To extract this
spectrum, we select events from an external ring of the field-of-view
with radius range of 10.5-12 arcmin, which is a region where
background emissivity is already dominant but where effective area
uncertainties are lower than they would be for even larger radii.


\clearpage
\bibliography{5758}

\begin{thebibliography}{51}
\expandafter\ifx\csname natexlab\endcsname\relax\def\natexlab#1{#1}\fi

\bibitem[{{Arnaud} {et~al.}(2001){Arnaud}, {Neumann}, {Aghanim}, {Gastaud},
  {Majerowicz}, \& {Hughes}}]{Arnaud_01}
{Arnaud}, M., {Neumann}, D.~M., {Aghanim}, N., {et~al.} 2001, \aap, 365, L80

\bibitem[{{Balucinska-Church} \& {McCammon}(1992)}]{Balucinska-Church_92}
{Balucinska-Church}, M. \& {McCammon}, D. 1992, \apj, 400, 699

\bibitem[{{Bauer} {et~al.}(2005){Bauer}, {Fabian}, {Sanders}, {Allen}, \&
  {Johnstone}}]{Bauer_05}
{Bauer}, F.~E., {Fabian}, A.~C., {Sanders}, J.~S., {Allen}, S.~W., \&
  {Johnstone}, R.~M. 2005, \mnras, 359, 1481

\bibitem[{{Belsole} {et~al.}(2004){Belsole}, {Pratt}, {Sauvageot}, \&
  {Bourdin}}]{Belsole_04}
{Belsole}, E., {Pratt}, G.~W., {Sauvageot}, J.-L., \& {Bourdin}, H. 2004, \aap,
  415, 821

\bibitem[{{Belsole} {et~al.}(2005){Belsole}, {Sauvageot}, {Pratt}, \&
  {Bourdin}}]{Belsole_05}
{Belsole}, E., {Sauvageot}, J.-L., {Pratt}, G.~W., \& {Bourdin}, H. 2005, \aap,
  430, 385

\bibitem[{{Bijaoui} \& {Jammal}(2001)}]{Bijaoui_01}
{Bijaoui}, A. \& {Jammal}. 2001, Signal Processing, 81, 1789

\bibitem[{{Bijaoui} \& {Rue}(1995)}]{Bijaoui_95}
{Bijaoui}, A. \& {Rue}. 1995, Signal Processing, 46, 345

\bibitem[{{Boulanger} {et~al.}(1996){Boulanger}, {Abergel}, {Bernard},
  {Burton}, {Desert}, {Hartmann}, {Lagache}, \& {Puget}}]{Boulanger_96}
{Boulanger}, F., {Abergel}, A., {Bernard}, J.-P., {et~al.} 1996, \aap, 312, 256

\bibitem[{{Bourdin} {et~al.}(2004){Bourdin}, {Sauvageot}, {Slezak}, {Bijaoui},
  \& {Teyssier}}]{Bourdin_04}
{Bourdin}, H., {Sauvageot}, J.-L., {Slezak}, E., {Bijaoui}, A., \& {Teyssier},
  R. 2004, \aap, 414, 429

\bibitem[{{Briel} {et~al.}(1991){Briel}, {Henry}, {Schwarz}, {Bohringer},
  {Ebeling}, {Edge}, {Hartner}, {Schindler}, {Trumper}, \& {Voges}}]{Briel_91}
{Briel}, U.~G., {Henry}, J.~P., {Schwarz}, R.~A., {et~al.} 1991, \aap, 246, L10

\bibitem[{{Buote} \& {Tsai}(1996)}]{Buote_tsai_96}
{Buote}, D.~A. \& {Tsai}, J.~C. 1996, \apj, 458, 27

\bibitem[{{Cappellari} \& {Copin}(2003)}]{Cappellari_03}
{Cappellari}, M. \& {Copin}, Y. 2003, \mnras, 342, 345

\bibitem[{{Chatzikos} {et~al.}(2006){Chatzikos}, {Sarazin}, \&
  {Kempner}}]{Chatzikos_06}
{Chatzikos}, M., {Sarazin}, C.~L., \& {Kempner}, J.~C. 2006, \apj, 643, 751

\bibitem[{{Clarke} {et~al.}(2004){Clarke}, {Blanton}, \& {Sarazin}}]{Clarke_04}
{Clarke}, T.~E., {Blanton}, E.~L., \& {Sarazin}, C.~L. 2004, \apj, 616, 178

\bibitem[{{Coifman} \& {Donoho}(1995)}]{Coifman_Donoho_95}
{Coifman}, R. \& {Donoho}, D.~L. 1995, in Lecture Notes in Statistics: Wavelets
  and Statistics, Vol. 103 (Springer-Verlag), 125--150

\bibitem[{{Curry} \& {Schoenberg}(1947)}]{Curry_47}
{Curry}, H.~B. \& {Schoenberg}, I.~J. 1947, Bull. Amer. Math. Soc, 53, 1114

\bibitem[{{Dickey} \& {Lockman}(1990)}]{Dickey_lockman_90}
{Dickey}, J.~M. \& {Lockman}, F.~J. 1990, \araa, 28, 215

\bibitem[{Donoho(1995)}]{Donoho_95}
Donoho, D.~L. 1995, IEEE Transactions on Information Theory, 41, 613

\bibitem[{{Ebeling} {et~al.}(1998){Ebeling}, {Edge}, {Bohringer}, {Allen},
  {Crawford}, {Fabian}, {Voges}, \& {Huchra}}]{Ebeling_98}
{Ebeling}, H., {Edge}, A.~C., {Bohringer}, H., {et~al.} 1998, \mnras, 301, 881

\bibitem[{{Ettori} {et~al.}(2002){Ettori}, {Fabian}, {Allen}, \&
  {Johnstone}}]{Ettori_01}
{Ettori}, S., {Fabian}, A.~C., {Allen}, S.~W., \& {Johnstone}, R.~M. 2002,
  \mnras, 331, 635

\bibitem[{{Forman} \& {Jones}(1982)}]{Forman_jones_82}
{Forman}, W. \& {Jones}, C. 1982, \araa, 20, 547

\bibitem[{{Grevesse} \& {Sauval}(1998)}]{Grevesse_98}
{Grevesse}, N. \& {Sauval}, A.~J. 1998, Space Science Reviews, 85, 161

\bibitem[{{Ikebe} {et~al.}(2004){Ikebe}, {B{\" o}hringer}, \&
  {Kitayama}}]{Ikebe_04}
{Ikebe}, Y., {B{\" o}hringer}, H., \& {Kitayama}, T. 2004, \apj, 611, 175

\bibitem[{{Kuntz} \& {Snowden}(2000)}]{Kuntz_00}
{Kuntz}, K.~D. \& {Snowden}, S.~L. 2000, \apj, 543, 195

\bibitem[{{Lewis} {et~al.}(2003){Lewis}, {Buote}, \& {Stocke}}]{Lewis_03}
{Lewis}, A.~D., {Buote}, D.~A., \& {Stocke}, J.~T. 2003, \apj, 586, 135

\bibitem[{{Lumb} {et~al.}(2002){Lumb}, {Warwick}, {Page}, \& {De
  Luca}}]{Lumb_02}
{Lumb}, D.~H., {Warwick}, R.~S., {Page}, M., \& {De Luca}, A. 2002, \aap, 389,
  93

\bibitem[{Mallat(1998)}]{Mallat_98}
Mallat, S. 1998, A wavelet tour of signal processing (Academic Press)

\bibitem[{{Markevitch}(1996)}]{Markevitch_96}
{Markevitch}, M. 1996, \apjl, 465, L1+

\bibitem[{{Markevitch} {et~al.}(2000){Markevitch}, {Ponman}, {Nulsen}, {Bautz},
  {Burke}, {David}, {Davis}, {Donnelly}, {Forman}, {Jones}, {Kaastra},
  {Kellogg}, {Kim}, {Kolodziejczak}, {Mazzotta}, {Pagliaro}, {Patel}, {Van
  Speybroeck}, {Vikhlinin}, {Vrtilek}, {Wise}, \& {Zhao}}]{Markevitch_00}
{Markevitch}, M., {Ponman}, T.~J., {Nulsen}, P.~E.~J., {et~al.} 2000, \apj,
  541, 542

\bibitem[{{Markevitch} {et~al.}(1999){Markevitch}, {Sarazin}, \&
  {Vikhlinin}}]{Markevitch_99}
{Markevitch}, M., {Sarazin}, C.~L., \& {Vikhlinin}, A. 1999, \apj, 521, 526

\bibitem[{{Markevitch} {et~al.}(2001){Markevitch}, {Vikhlinin}, \&
  {Mazzotta}}]{Markevitch_01}
{Markevitch}, M., {Vikhlinin}, A., \& {Mazzotta}, P. 2001, \apjl, 562, L153

\bibitem[{{Mazzotta} {et~al.}(2004){Mazzotta}, {Rasia}, {Moscardini}, \&
  {Tormen}}]{Mazzotta_04}
{Mazzotta}, P., {Rasia}, E., {Moscardini}, L., \& {Tormen}, G. 2004, \mnras,
  354, 10

\bibitem[{{Miville-Desch{\^e}nes} \& {Lagache}(2005)}]{Miville_Deschenes_05}
{Miville-Desch{\^e}nes}, M.-A. \& {Lagache}, G. 2005, \apjs, 157, 302

\bibitem[{{Nevalainen} {et~al.}(2005){Nevalainen}, {Markevitch}, \&
  {Lumb}}]{Nevalainen_05}
{Nevalainen}, J., {Markevitch}, M., \& {Lumb}, D. 2005, \apj, 629, 172

\bibitem[{{Pointecouteau} {et~al.}(2004){Pointecouteau}, {Arnaud}, {Kaastra},
  \& {de Plaa}}]{Pointecouteau_04}
{Pointecouteau}, E., {Arnaud}, M., {Kaastra}, J., \& {de Plaa}, J. 2004, \aap,
  423, 33

\bibitem[{{Rasia} {et~al.}(2006){Rasia}, {Ettori}, {Moscardini}, {Mazzotta},
  {Borgani}, {Dolag}, {Tormen}, {Cheng}, \& {Diaferio}}]{Rasia_06}
{Rasia}, E., {Ettori}, S., {Moscardini}, L., {et~al.} 2006, \mnras, 369, 2013

\bibitem[{{Sakelliou} \& {Ponman}(2004)}]{Sakelliou_04}
{Sakelliou}, I. \& {Ponman}, T.~J. 2004, \mnras, 351, 1439

\bibitem[{{Sanders}(2006)}]{Sanders_06}
{Sanders}, J.~S. 2006, \mnras, 371, 829

\bibitem[{{Sanders} \& {Fabian}(2001)}]{Sanders_01}
{Sanders}, J.~S. \& {Fabian}, A.~C. 2001, \mnras, 325, 178

\bibitem[{{Sanderson} {et~al.}(2005){Sanderson}, {Finoguenov}, \&
  {Mohr}}]{Sanderson_05}
{Sanderson}, A.~J.~R., {Finoguenov}, A., \& {Mohr}, J.~J. 2005, \apj, 630, 191

\bibitem[{{Sarazin}(1988)}]{Sarazin_88}
{Sarazin}, C.~L. 1988, {X-ray emission from clusters of galaxies} (Cambridge
  Astrophysics Series, Cambridge: Cambridge University Press, 1988)

\bibitem[{{Sauvageot} {et~al.}(2005){Sauvageot}, {Belsole}, \&
  {Pratt}}]{Sauvageot_05}
{Sauvageot}, J.~L., {Belsole}, E., \& {Pratt}, G.~W. 2005, \aap, 444, 673

\bibitem[{{Schlegel} {et~al.}(1998){Schlegel}, {Finkbeiner}, \&
  {Davis}}]{Schlegel_98}
{Schlegel}, D.~J., {Finkbeiner}, D.~P., \& {Davis}, M. 1998, \apj, 500, 525

\bibitem[{{Schuecker} {et~al.}(2001){Schuecker}, {B{\"o}hringer}, {Reiprich},
  \& {Feretti}}]{Schuecker_01}
{Schuecker}, P., {B{\"o}hringer}, H., {Reiprich}, T.~H., \& {Feretti}, L. 2001,
  \aap, 378, 408

\bibitem[{{Slezak} {et~al.}(1994){Slezak}, {Durret}, \& {Gerbal}}]{Slezak_94}
{Slezak}, E., {Durret}, F., \& {Gerbal}, D. 1994, \aj, 108, 1996

\bibitem[{{Smith} {et~al.}(2001){Smith}, {Brickhouse}, {Liedahl}, \&
  {Raymond}}]{Smith_01}
{Smith}, R.~K., {Brickhouse}, N.~S., {Liedahl}, D.~A., \& {Raymond}, J.~C.
  2001, \apjl, 556, L91

\bibitem[{{Sun} {et~al.}(2003){Sun}, {Jones}, {Murray}, {Allen}, {Fabian}, \&
  {Edge}}]{Sun_03}
{Sun}, M., {Jones}, C., {Murray}, S.~S., {et~al.} 2003, \apj, 587, 619

\bibitem[{{Sun} {et~al.}(2002){Sun}, {Murray}, {Markevitch}, \&
  {Vikhlinin}}]{Sun_02}
{Sun}, M., {Murray}, S.~S., {Markevitch}, M., \& {Vikhlinin}, A. 2002, \apj,
  565, 867

\bibitem[{{Tamura} {et~al.}(2001){Tamura}, {Kaastra}, {Peterson}, {Paerels},
  {Mittaz}, {Trudolyubov}, {Stewart}, {Fabian}, {Mushotzky}, {Lumb}, \&
  {Ikebe}}]{Tamura_01}
{Tamura}, T., {Kaastra}, J.~S., {Peterson}, J.~R., {et~al.} 2001, \aap, 365,
  L87

\bibitem[{{Vikhlinin} {et~al.}(2006){Vikhlinin}, {Kravtsov}, {Forman}, {Jones},
  {Markevitch}, {Murray}, \& {Van Speybroeck}}]{Vikhlinin_06}
{Vikhlinin}, A., {Kravtsov}, A., {Forman}, W., {et~al.} 2006, \apj, 640, 691

\bibitem[{{Vikhlinin} {et~al.}(2005){Vikhlinin}, {Markevitch}, {Murray},
  {Jones}, {Forman}, \& {Van Speybroeck}}]{Vikhlinin_05}
{Vikhlinin}, A., {Markevitch}, M., {Murray}, S.~S., {et~al.} 2005, \apj, 628,
  655

\end{thebibliography}

\end{document}